\pdfoutput=1

\documentclass[a4paper,12pt,reqno]{amsart}

\usepackage[english]{babel}
\usepackage[utf8x]{inputenc}
\usepackage[T1]{fontenc}

\usepackage{newtxtext,newtxmath}    

\usepackage{enumerate}        
\usepackage{url}              
\usepackage[usenames]{color}  
\usepackage{xspace}           
\usepackage{datetime}         
\usepackage{hyperref}
\usepackage{a4wide}           

\usepackage{graphicx}      
\graphicspath{{./figs/}}
\usepackage{epstopdf}

\usepackage{multirow}    

\usepackage[text={130mm,200mm}]{geometry}

\theoremstyle{plain}
\newtheorem{theorem}{Theorem}
\newtheorem*{theorem*}{Theorem}

\newtheorem*{corollary*}{Corollary}
\newtheorem{lemma}{Lemma}
\newtheorem*{lemma*}{Lemma}

\newtheorem*{proposition*}{Proposition}

\newtheorem*{conjecture*}{Conjecture}
\theoremstyle{definition}

\newtheorem*{definition*}{Definition}
\theoremstyle{remark}

\newtheorem*{remark*}{Remark}

\providecommand{\surname}[1]{#1}
\providecommand{\country}[1]{#1}

\newcommand{\sgn}{\textsf{sign} }
\newcommand{\RR}{\mathbb{R}\xspace}

\newcommand{\EE}{\mathbb{E}\xspace}

\newcommand*{\ie}{i.e.\@\xspace}

\newcommand{\inv}[1]{#1^{\star}}
\newcommand{\yinv}[1]{\hat{#1}}         

\newcommand{\zspace}{$\mathcal{Z}$-space\xspace}
\newcommand{\wspace}{$\mathcal{W}$-space\xspace}
\newcommand{\yspace}{$\mathcal{Y}$-space\xspace}

\newcommand{\overbar}[1]%
 {\mkern 1.5mu\overline{\mkern-1.5mu#1\mkern-1.5mu}\mkern 1.5mu}
\newcommand{\ov}[1]{\overrightarrow{#1}}

\newcommand{\oo}{o_{ijk}}               
\newcommand{\HH}{\mathcal{H}\xspace}    
\newcommand{\OO}{\mathcal{O}}           
\newcommand{\sh}[1]{\mathcal{SR}(#1)}   

\newcommand{\tts}[1]{\mathcal{T}(#1)}   
\newcommand{\itp}[1]{\inv{\Pi}(#1)}     
\newcommand{\ill}[1]{\yinv{\ell}(#1)}   
\newcommand{\cone}{\mathcal{K}}         
\newcommand{\wcone}{\inv{\mathcal{K}}}  
\newcommand{\ycone}{{\mathcal{K}}'}     
\newcommand{\tri}[1]{\tau_{#1}}         
\newcommand{\edge}[1]{e_{#1}}           
\newcommand{\map}[1]{\zeta(#1)}         
\newcommand{\invmap}[1]{\zeta^{-1}(#1)} 
\newcommand{\arc}[1]{\psi(#1)}           

\newcommand{\orient}{{\sc{}Orient3D}\xspace}
\newcommand{\incone}{{\sc{}InCone}\xspace}
\newcommand{\insphere}{{\sc{}InSphere}\xspace}
\newcommand{\exist}{{\sc{}Existence}\xspace}
\newcommand{\shadow}{{\sc{}Shadow}\xspace}
\newcommand{\conflict}{{\sc{}EdgeConflict}\xspace}

\newcommand{\order}{{\sc{}Order}\xspace}
\newcommand{\tritype}{{\sc{}Trisector}\xspace}
\newcommand{\distance}{{\sc{}Distance}\xspace}

\newcommand{\inside}{\textsf{Inside}\xspace}
\newcommand{\outside}{\textsf{Outside}\xspace}
\newcommand{\ptouch}{\textsf{OnePointTouch}\xspace}
\newcommand{\ctouch}{\textsf{CircleTouch}\xspace}

\newcommand{\noconflict}{\textsf{NoConflict}\xspace}
\newcommand{\fullconflict}{\textsf{EntireEdge}\xspace}
\newcommand{\leftvertex}{\textsf{LeftVertex}\xspace}
\newcommand{\rightvertex}{\textsf{RightVertex}\xspace}
\newcommand{\verticesconflict}{\textsf{BothVertices}\xspace}
\newcommand{\middleconflict}{\textsf{Interior}\xspace}

\newcommand{\answer}{OrderCase\xspace}

\begin{document}

\title[The EdgeConflict Predicate in the 3D Apollonius Diagram]{The EdgeConflict Predicate \protect\\ in the 3D Apollonius Diagram}

\author{Manos \surname{kamarianakis}}
\address{Department of Mathematics \& Applied Mathematics\\
  University of Crete\\
  Voutes University Campus\\
  Heraklion\\
  GR-70013\\
  \country{Greece}}
\email{m.kamarianakis@gmail.com}
\urladdr{https://www.tem.uoc.gr/~manosk/}


\thanks{  This work is part of author's Ph.D. thesis and was 
supported by Onassis Foundation and Project Thales.  A simpler version 
of this paper was accepted at \emph{The Sixth International Conference on Analytic Number Theory and Spatial Tessellations} held in Kyiv, Ukraine
at September 2018. This work was supported by Onassis Foundation and Project Thales.}



\date{November 15, 2018}

\begin{abstract}
  In this paper we study one of the fundamental predicates required for
  the construction of the 3D Apollonius diagram (also known as the 3D
  Additively Weighted Voronoi diagram), namely the \conflict predicate:
  given five sites $S_i, S_j,S_k,S_l,S_m$ that define an edge
  $\edge{ijklm}$ in the 3D Apollonius diagram, and a sixth query site
  $S_q$, the predicate determines the portion of $\edge{ijklm}$
  that will disappear in the Apollonius diagram of the six sites due to
  the insertion of $S_q$.

  Our focus is on the algorithmic analysis of the predicate with the aim
  to minimize its algebraic degree. We decompose the main predicate into
  three sub-predicates, which are then evaluated with the aid of four
  additional primitive operations. We show that the maximum algebraic degree
  required to answer any of the sub-predicates and primitives, and, thus,
  our main predicate is 10.

  Among the tools we use is the 3D inversion transformation. In the scope of this paper and due to space limitations, only non-degenerate configurations are considered, \ie different Voronoi vertices are distinct 
  and the predicates never return a ``degenerate'' answer.
  Most of our analysis is carried out in the inverted space, which is 
  where our geometric observations and analysis is captured in 
  algebraic terms.
\end{abstract}

\maketitle




\section{Introduction}

  Voronoi diagrams have been among the most studied 
  constructions in computational geometry since their 
  inception \cite{green1978computing, brown1979voronoi, 
  aurenhammer2000voronoi, guibas1990randomized}, due to their numerous
  applications, including motion planning and collision detection,
  communication networks, graphics, and growth of microorganisms 
  in biology.

  Despite being a central topic in research for many 
  years, generalized Voronoi diagrams, and especially the Voronoi 
  diagram of spheres (also known as the 3D Apollonius diagram) 
  have not been explored sufficiently \cite{kim2006region}; this is also pointed out by Aurenhammer et al. \cite{aurenhammer2013voronoi}. 
  Moreover, due to 
  recent scientific discoveries in biology and chemistry, 3D Apollonius 
  diagrams are becoming increasingly important for representing 
  and analysing the molecular 3D structure and surface \cite{joonghyun2005computation} or 
  the structure of the protein \cite{kim2005protein}.

  The methods used to calculate the Apollonius diagram usually rely 
  on the construction of a different diagram altogether. 
  Some methods include the intersection of cones \cite{aurenhammer1987power} with the lifted power diagram and lower envelope calculations \cite{WillThesis, will1998fast, hanniel2008computing}. 
  Boissonnat et al. use the convex hull to describe its construction 
  \cite{Boissonnat2003, Boissonnat2005}. 
  Aurenhammer's lifting method has also been implemented for two dimensions
  \cite{anton2002exact}. 
  Karavelas and Yvinec \cite{karavelas2002dynamic} create the 2­d Apollonius diagram from its dual, using the predicates developed in \cite{Emiris2006Predicates}. In \cite{Emiris2006Predicates}, it is also reported  that the Apollonius diagram can be obtained as a 
  concrete case of the abstract Voronoi diagrams of Klein et al. 
  \cite{klein1992randomized}.

  Kim et al. made a major research contribution in the 
  domain of the Voronoi diagrams of spheres including one patent 
  \cite{kim2010calculating} for the computation of 3D Voronoi diagrams. Their work provides many new algorithms related to the Voronoi diagrams including the computation of three-dimensional Voronoi diagrams 
  \cite{kim2010calculating}, Euclidean Voronoi diagram of 3D balls and its
  computation via tracing edges \cite{kim2005euclidean} and the Euclidean Voronoi diagrams of 3D spheres and applications to protein structure analysis \cite{kim2005protein}.

  Hanniel and Elber \cite{hanniel2008computing} provide an 
  algorithm for computation of the Voronoi diagrams for planes, 
  spheres and cylinders in $\RR^3$. Their algorithm relies on 
  computing the lower envelope of the bisector surfaces similar 
  to the algorithm of Will \cite{will1998fast}. However, none of 
  the current research efforts provide the exact method 
  for computing the Apollonius diagram (or its dual Delaunay graph) 
  of spheres.


  In this paper, we are inspired by the the approach presented by Emiris and Karavelas in \cite{Emiris2006Predicates} for the evaluation of the 2D Apollonius diagram. 
  In order to extend their work towards an algorithm that would incrementally construct the Apollonius diagram for 3D spheres, we 
  develop equivalent predicates as the ones presented in their paper for 
  the 2D case. Our main goal is to implement the most degree demanding 
  predicate, called the \conflict predicate: 
  given five sites $S_i, S_j,S_k,S_l,S_m$ that define a finite edge
  $\edge{ijklm}$ in the 3D Apollonius diagram, and a sixth query site
  $S_q$, the predicate determines the portion of $\edge{ijklm}$
  that will disappear in the Apollonius diagram of the six sites due to
  the insertion of $S_q$.

  In order to accomplish this task, we developed various subpredicates and primitives. The creation of these tools was made taking into consideration the 
  modern shift of predicate design towards lower level algorithmic issues.
  Specifically, a critical 
  factor that influenced our designs was our goal to minimize the algebraic degree of the tested quantities (in terms of the input parameters) during a predicate evaluation. Such a minimization problem has become a main concern that 
  influences algorithm design especially in geometric predicates, where 
  zero tolerance in all intermediate computations is needed to obtain an 
  accurate result  \cite{devillers2002algebraic, geismann2001computing,berberich2002computational, wein2002high, wolpert2003jacobi} .

  Our main contribution in the research area is the development of a list 
  of subpredicates that where not implemented, either explicitly or implicitly, in the current bibliography and can be used within the scope of an incremental 
  algorithm that constructs the 3D Apollonius diagram of a set of spheres. 
  Our most outstanding result is the fact that all subpredicates 
  presented in this paper along with the 
  \conflict predicate require at most 10-fold degree demanding operation 
  (with respect to the input quantities). This is quite an unexpected 
  result since the equivalent \conflict predicate in the 2D Apollonius diagram required 16-fold operations \cite{Emiris2006Predicates,klein1992randomized}. Our approach of resolving the ``master'' 
  predicate and especially the observations made in the inverted plane, could also be implemented in the 2D case to yield lower algebraic degrees.

  This paper is organised as follows. In Section 2, we review the preliminaries of the Apollonius diagram of 3D spheres and the orientation of hyperbolic trisectors in such a diagram. An introduction
  to the inversion technique is also made along with useful remarks regarding the correlation between the original and the inverted space. 
  In Section 3, we present in detail the \conflict predicate along with 
  the assumptions made in the scope of this paper. An outline of the 
  subpredicates developed is then provided along with the geometric properties that derive from each one. Finally, we provide the main 
  algorithm that ultimately combines all the aforementioned tools 
  to answer the \conflict predicate. Section 4 is devoted to the implementation and algebraic analysis for each subpredicate. Finally, 
  in Section 5, we conclude the paper.

 \section{Preliminaries and Definitions} 
  \label{sub:definitions}
  
  Let $\mathcal{S}$ be a set of closed spheres $S_n$ (also referred as 
  \emph{sites}) 
  in $\EE^3$, with centers $C_n=(x_n,y_n,z_n)$ and radii $r_n$. 
  In this paper, we will assume that no one of these sites is
  contained inside another. Define the Euclidean distance
  $\delta(p,S)$ between a point $p\in\EE^3$ and a sphere $S=\{C,r\}$
  as $\delta(p,S)=\|p-C\|-r$, where $\|\cdot\|$ stands for the
  Euclidean norm. The \emph{Apollonius diagram}
  is then defined as the subdivision of the plane induced by 
  assigning each point $p\in\EE^3$ to its nearest neighbor with 
  respect to the distance function $\delta(\cdot,\cdot)$. 

  For each $i\neq j$, let 
  $H_{ij}=\{y\in\EE^3 :\delta(y,S_i)\leq \delta(y,S_j)\}$. Then 
  the (closed) \emph{Apollonius cell} $V_i$ of $S_i$ is defined to be 
  $V_i=\cap_{i\neq j} H_{ij}$. The set of points that belong to exactly 
  two Apollonius cells are called the \emph{Apollonius faces}, 
  whereas the connected set of points that 
  belong to exactly three Apollonius cells are called 
  \emph{Apollonius edges}. Points that belong to more than 
  three Apollonius cells are called \emph{Apollonius vertices}; 
  the\emph{ Apollonius diagram} $\mathcal{VD}(\mathcal{S})$ of 
  $\mathcal{S}$ is defined as the collection of the Apollonius cells, 
  faces, edges and vertices. 

  An Apollonius vertex $v$ is a point that belongs to 4 or more Apollonius
  cells. Without loss of generality we may assume that an Apollonius
  vertex is tangent to exactly 4 sites $S_i,S_j,S_k,S_l$, since
  otherwise we may apply a pertrubation scheme to resolve the degeneracy.
  The sphere centered at $v$ and tangent to $S_i,S_j,S_k,S_l$ is either
  externally tangent to them, or is contained inside all 4 four sites;
  in the former case it is called an \emph{external Apollonius sphere}, 
  while in the latter an \emph{internal Apollonius sphere}.
  Let $T_n$, $n=i,j,k,l$, be the point of tangency of the Apollonius
  sphere and $S_n$. The tetrahedron $T_iT_jT_kT_l$ can 
  either be positively or negatively oriented, or even flat
  \cite{devillers2012qualitative}.
  We assume below that $T_iT_jT_kT_l$ is not flat; otherwise we may
  employ the perturbation scheme described in
  \cite{devillers2012qualitative} and, thus,
  consider it as non-flat. The Apollonius vertex corresponding to a
  positively (resp., negatively) oriented tetrahedrom $T_iT_jT_kT_l$
  will be denoted $v_{ijkl}$ (resp., $v_{ikjl}$). Observe that a
  cyclic permutation of the indices does not affect our choice of
  Apollonius vertex. 

  The \emph{trisector} $\tri{ijk}$ of three different sites $S_i,S_j$
  and $S_k$ is the locus of points that are equidistant from the the
  three sites. In the absense of degeneracies its Hausdorff dimension
  is 1, and it is either (a branch of) a hyperbola, a line, an ellipse,
  a circle, or a parabola \cite{WillThesis}; in this paper, due to
  space limitations, we focus on the cases where the trisector is an
  open curve, and more specifically hyperbolic or linear.
  An \emph{Apollonius edge} $\edge{ijklm}$ is a connected subset of 
  the trisector
  $\tri{ijk}$ of three different sites and is defined by five sites
  $S_i,S_j,S_k,S_l$ and $S_m$. The first three sites define
  the supporting trisector $\tri{ijk}$ of the edge, whereas the last
  two define its endpoints $v_{ijkl}$ and $v_{ikjm}$.
  
 \subsection{Inversion} 
  \label{sub:inversion}
  The 3-dimensional \emph{inversion transformation} is a
  mapping from $\RR^3$ to $\RR^3$ that maps a point $z\in\RR^3$ to the
  point $W(z)=(z-z_0)/\|z-z_0\|^2$. The point $z_0$ is called the
  \emph{pole of inversion}.
  Inversion maps spheres that do not pass through the pole to
  spheres, and spheres that pass through the pole to planes.
  
  In the Apollonius diagram context we call \zspace the space where
  the sites live. Since the Apollonius diagram does not change when we
  add to the radii of all spheres the same quantity, we will, most of
  the times, reduce the radii of the spheres   $S_i,S_j,S_k,S_l,S_m$
  by the radius of one of them, the sphere $S_I$. The new spheres have
  obviously the same centers, whereas their radii become
  spheres become $r^\star_n=r_n-r_I$, $n\in\{i,j,k,l,m\}$. For
  convience, we call the image space of this radius-reducing
  transformation the $\mathcal{Z}^\star$-space. We may then apply
  inversion, with $C_I$ as the pole, to get a new set of spheres or
  planes; we call \wspace the space where the radius-reduced, inverted
  sites live.

  Since the sites $S_i,S_j,S_k,S_l,S_m$ are not contained inside each
  is tangent to another, the image of the sphere $S_n$ in \wspace
  is a sphere $\inv{S}_n$, centered at 
  $\inv{C}_n=(u_n,v_n,w_n)$ with radius $\rho_n$, where
  \begin{alignat*}{4}
  u_n &= \dfrac{\inv{x}_n}{\inv{p}_n}, &\quad 
  v_n &= \dfrac{\inv{y}_n}{\inv{p}_n}, &\quad
  w_n &= \dfrac{\inv{z}_n}{\inv{p}_n}, &\quad
  \rho_n &= \dfrac{\inv{r}_n}{\inv{p}_n},\\
  \inv{x}_n &= x_n-x_I, &\quad
  \inv{y}_n &= y_n-y_I, &\quad
  \inv{z}_n &= z_n-z_I, &\quad
  \inv{r}_n &= r_n-r_I,
  \end{alignat*}

  \noindent
  and $\inv{p}_n=(\inv{x}_n)^2+(\inv{y}_n)^2+(\inv{z}_n)^2-(\inv{r}_n)^2$.
  Note that $\inv{p}_n$ is positive due to the non-inclusion 
  assumption. We also define the quantities
  \begin{gather*}
  D^{\pi\theta}_{ijk} = 
  \begin{vmatrix}
  \pi_i & \theta_i & 1\\
  \pi_j & \theta_j & 1\\
  \pi_k & \theta_k & 1
  \end{vmatrix}, 
  \ \ \
  D^{\pi\theta\eta}_{ijk} = 
  \begin{vmatrix}
  \pi_i & \theta_i & \eta_i\\
  \pi_j & \theta_j & \eta_j\\
  \pi_k & \theta_k & \eta_k
  \end{vmatrix}, 
  \ \ \ 
  D^{\pi\theta\eta}_{ijkl} = 
  \begin{vmatrix}
  \pi_i & \theta_i & \eta_i & 1\\
  \pi_j & \theta_j & \eta_j & 1\\
  \pi_k & \theta_k & \eta_k & 1\\
  \pi_l & \theta_l & \eta_l & 1\\
  \end{vmatrix}, \\
  D^{\pi\theta\eta\zeta}_{ijkl} = 
  \begin{vmatrix}
  \pi_i & \theta_i & \eta_i & \zeta_i\\
  \pi_j & \theta_j & \eta_j & \zeta_j\\
  \pi_k & \theta_k & \eta_k & \zeta_k\\
  \pi_l & \theta_l & \eta_l & \zeta_l\\
  \end{vmatrix}, 
  E^{\pi\theta\eta}_{ijk} = 
  \begin{vmatrix}
  \inv{\pi_i} & \inv{\theta_i} & \inv{\eta_i}\\
  \inv{\pi_j} & \inv{\theta_j} & \inv{\eta_j}\\
  \inv{\pi_k} & \inv{\theta_k} & \inv{\eta_k}
  \end{vmatrix},
  E^{\pi\theta\eta\zeta}_{ijkl} = 
  \begin{vmatrix}
  \inv{\pi_i} & \inv{\theta_i} & \inv{\eta_i} & \inv{\zeta_i}\\
  \inv{\pi_j} & \inv{\theta_j} & \inv{\eta_j} & \inv{\zeta_j}\\
  \inv{\pi_k} & \inv{\theta_k} & \inv{\eta_k} & \inv{\zeta_k}\\
  \inv{\pi_l} & \inv{\theta_l} & \inv{\eta_l} & \inv{\zeta_l}\\
  \end{vmatrix}, 
  \end{gather*}

  \noindent
  for $\pi,\theta,\eta,\zeta \in\{ x,y,z,r,u,v,w,\rho\}$, and it holds that
  \begin{equation*}
  D^{\pi\theta}_{ijk} = E^{klp}_{ijk}(\inv{p_i}\inv{p_j}\inv{p_k})^{-1},\ \ 
  D^{\pi\theta\eta}_{ijk} = E^{klm}_{ijk}(\inv{p_i}\inv{p_j}
    \inv{p_k})^{-1}
  \end{equation*}
  for $\pi,\theta \in\{ u,v,w,\rho\}$ and $k,l,m\in\{ x,y,z,r\}$.

  


 \subsection{Orientation of a hyperbolic or linear trisector} 
  \label{sub:orientation_hyperbolic_trisector}
\begin{figure}[htbp]
   \centering
   \includegraphics[width=0.95\textwidth]{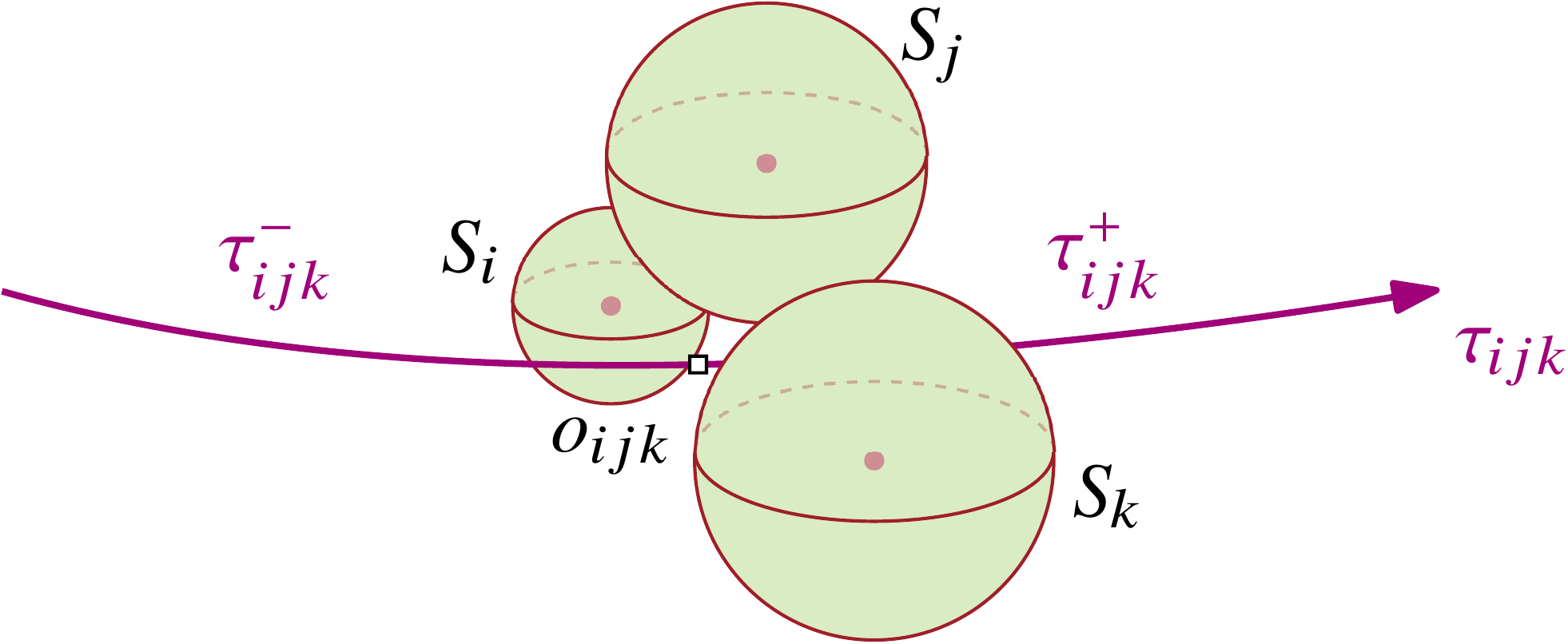}
   \caption{The case where the trisector $\tri{ijk}$ of the 
   spheres $S_i$, $S_j$ and $S_k$ is \emph{hyperbolic}. Notice the 
   orientation of $\tri{ijk}$ based on the ``right-hand rule''. 
   The point $o_{ijk}$ of the trisector, which is coplanar with the 
   centers $C_i$, $C_j$ and $C_k$, separates the $\tri{ijk}$ 
   into two semi-trisectors, $\tri{ijk}^-$ and $\tri{ijk}^+$.}
   \label{fig:hyperbolic_trisector}
  \end{figure}

  Under the assumption that the trisector $\tri{ijk}$ of the sites
  $S_i,S_j,S_k$ is a line or a hyperbola, the three centers
  $C_i,C_j,C_k$ cannot be collinear \cite{WillThesis}. 
  A natural way of orienting $\tri{ijk}$
  is accomplished via the well-known ``right-hand rule''; if we fold 
  our right hand to follow the centers $C_i,C_j$ and $C_k$ (in that order), 
  our thumb will be showing the positive ``end'' of $\tri{ijk}$ 
  (see Figure~\ref{fig:hyperbolic_trisector}).

  By orienting $\tri{ijk}$, we clearly define an ordering on the points of
  $\tri{ijk}$, which we denote by $\prec$.
  Let $\oo$ be the intersection of $\tri{ijk}$ and the plane 
  $\Pi_{ijk}$ going through the centers $C_i,C_j$ and $C_k$.
  We can now parametrize $\tri{ijk}$ as follows: if $\oo\prec p$
  then $\map{p}=\delta(p,S_i)-\delta(\oo,S_i)$; 
  otherwise $\map{p}=-(\delta(p,S_i)-\delta(\oo,S_i))$. 
  The function $\map{\cdot}$ is a 1-1 and onto mapping from
  $\tri{ijk}$ to $\RR$. Moreover, we define $\map{S}$, where 
  $S$ is an external tangent sphere to the sites $S_i,S_j$ and $S_k$,
  to be $\map{c}$, where $c\in\tri{ijk}$ is the center of $S$.

  We also use $\tri{ijk}^+$ (resp. $\tri{ijk}^-$) to denote 
  the positive (resp. negative) semi-trisector \ie, the set of 
  points $p\in\tri{ijk}$ such that $\oo\prec p$ (resp. $p\prec\oo$).

\newpage


\section{Non-Degenerate Case Analysis for Hyperbolic Trisectors} 
 \label{sec:non_degenerate_hyperbolic}

 \subsection{Voronoi Edges on Hyperbolic Trisectors} 
  \label{sub:voronoi_edges}

  In order to better understand our initial problem, more insight 
  regarding the properties of a Voronoi diagram is required. 
  Let us look closer at 
  an edge $e_{ijklm}$ (we drop the subscript for convenience) 
  of $\mathcal{V}(\mathcal{S})$, 
  where $\mathcal{S}$ is a set of given sites that includes 
  $S_n$ for $n\in\{i,j,k,l,m\}$ and does not include $S_q$. 
  This edge $e$ lies on the trisector $\tri{ijk}$,  
  the locus of points that are equidistant to the sites
  $Si,S_j$ and $S_k$. 
  In the scope of this paper, 
  we assume that $\tri{ijk}$ is of Hausdorff dimension 1 and 
  is either (a branch of) a hyperbola or a line. To ensure that the 
  spheres $Si,S_j$ and $S_k$ meet this criteria, 
  the predicate \tritype$(S_i,S_j,S_k)$, described in 
  Section~\ref{ssub:the_incone_and_tritype_predicates} 
  and analyzed in Section~\ref{sub:the_tritype_predicate_analysis}, 
  must return ``hyperbolic''.

  We now focus on the edge $e$, \ie the open continuous subset of 
  $\tri{ijk}$ whose closure is bounded by the Voronoi vertices 
  $v_{ijkl}$ and $v_{ikjm}$. If $\tts{t}$ denotes the external 
  Apollonius sphere of the sites $S_i,S_j$ and $S_k$ that is 
  centered at $t$, the most crucial
  property of a point $t\in e$ is that $\tts{t}$ does not intersect 
  with any other site of $\mathcal{S}$. We call this 
  \emph{the Empty Sphere Principle} since it is a property 
  that derives from the empty circle principle of a generic Voronoi 
  diagram and its basic properties.

  Using this property, we can show that the left and right endpoint
  of $e$ are the Apollonius vertices $v_{ijkl}$ and $v_{ikjm}$ 
  respectively. To prove that the left endpoint is indeed 
  $v_{ijkl}$ and not $v_{ikjl}$, consider a point $t\in\tri{ijk}$ 
  such that $t\equiv v_{ikjl}$ and move it infinitesimally on the 
  trisector towards its positive direction. The initial Apollonius
  sphere $\tts{v_{ikjl}}$ was tangent to $S_l$, and assuming $v_{ikjl}$ 
  was the left endpoint of $e$, the sphere $\tts{t}$ should not 
  longer be tangent nor intersect $S_l$ since $t\in e$ and the 
  Empty Sphere Principle must hold. 
  However, due to the negative orientation of the tetrahedron 
  $T_iT_jT_kT_l$, where $T_n$ is the tangency point of the sphere $S_n$ 
  and $\tts{v_{ikjl}}$ for $n=i,j,k,l$, and the orientation of $\tri{ijk}$, 
  the sphere $\tts{t}$ contains $T_l$ and therefore intersects $S_l$, 
  yielding a contradiction. Therefore, we have proven that the left 
  point of $e$ is necessarily $v_{ijkl}$ and we can prove that the 
  right endpoint is $v_{ikjm}$ (and not $v_{ijkm}$) in a similar way.




 \subsection{Problem Outline and Assertions} 
  \label{sub:problem_outline}
 
  For clarity reasons, we restate the \conflict predicate, highlighting 
  its input, output as well as the assertions we are making for the rest
  of this paper. 

  The \conflict predicate, one of the fundamental predicates required for
  the construction of the 3D Apollonius diagram (also known as the 3D
  Additively Weighted Voronoi diagram), takes as input five sites 
  $S_i, S_j,S_k,S_l$ and $S_m$ that define an edge $\edge{ijklm}$ in the 
  3D Apollonius diagram as well as a sixth query site $S_q$. The 
  predicate determines the portion of $\edge{ijklm}$ (we drop the subscript for convenience)
  that will disappear in the Apollonius diagram of the six sites due to
  the insertion of $S_q$ and therefore its output is one of the following  

  \begin{itemize}
  \item 
  \noconflict : no portion of $e$ is destroyed by the insertion of $S_q$ in the Apollonius diagram of the five sites.
  \item 
  \fullconflict : the entire edge $e$ is destroyed by the addition of $S_q$ in the Apollonius diagram of the five sites.
  \item 
  \leftvertex : a subsegment of $e$ adjacent to its origin vertex ($v_{ijkl}$) disappears in the Apollonius diagram of the six sites.
  \item 
  \rightvertex : is the symmetric case of the \leftvertex case; a subsegment of $e$ adjacent to the vertex $v_{ikjm}$ disappears in the Apollonius diagram of the six sites.
  \item 
  \verticesconflict : subsegments of $e$ adjacent to its two vertices disappear in the Apollonius diagram of the five sites.
  \item
  \middleconflict : a subsegment in the interior of $e$ disappears in the Apollonius diagram of the five sites.
  \end{itemize}

  In Section~\ref{sub:the_main_algorithm}, we prove that these are indeed
  the only possible answers to the studied predicate, under the assumption
  that no degeneracies occur. Specifically, all analysis presented in this paper is done under the following two major assertions: 
  \begin{itemize}
  \item The trisector $\tri{ijk}$ of the sites $S_i,S_j$ and $S_k$ is 
  ``hyperbolic'' \ie, it is either a branch of a hyperbola or a straight
  line. Therefore, the spheres must lie in \emph{convex position}; in other 
  words, there must exist two distinct planes commonly tangent to all three spheres.
  \item None of the subpredicates called during the algorithm presented 
  in Section~\ref{sub:the_main_algorithm} returns a degenerate answer. 
  Mainly, this is equivalent to the statement: \emph{All of the existing Apollonius vertices defined by the sites $S_i,S_j,S_k$ and $S_n$, for 
  $n\in\{l,m,q\}$, are distinct and the respective Apollonius spheres are all finite \ie, they are not centered at infinity}. Such assertion 
  dictates that the edge $e$ is finite as none of its bounding vertices 
  $v_{ijkl}$ and $v_{ikjm}$ can lie at infinity.
  \end{itemize}


 

 \subsection{SubPredicates and Primitives} 
  \label{sub:subpredicates}


  In this section, we describe the various 
  subpredicates used throughout the evaluation of the \conflict predicate 
  via the main algorithm presented in 
  Section~\ref{sub:the_main_algorithm}. For convenience, only the input, output and specific geometric observations is provided in this section, whereas a detailed implementation along with a algebraic degree analysis of each subpredicate is found in Section~\ref{sec:algebraic_analysis}.

  \subsubsection{The \insphere predicate} 
  \label{sub:the_insphere_predicate}

 The $\text{\insphere}(S_i,S_j,S_k,S_a,S_b)$ predicate 
 returns $-,+$ or $0$ if and only if the sphere $S_b$ 
 intersects, does not intersect or is tangent to the 
 external Apollonius sphere of the sites $S_i,S_j,S_k$
 and $S_b$, centered at $v_{ijka}$. It is assumed that 
 $v_{ijka}$ exists and none of the first four inputed sites 
 are contained inside one another. In \cite{Iordanov}, it is 
 shown that the evaluation of the \insphere predicate requires
 operations of maximum algebraic degree 10, whereas in \cite{anton2011exact}
 an implicit \insphere predicate could be evaluated via the Delaunay graph, 
 using 6-fold degree operations (although it is not clear if 
 we could easily distinguish if we are testing against the Apollonius 
 sphere centered at $v_{ijka}$ or $v_{ikja}$). 

 Since degenerate configurations are beyond the scope of this paper, 
 the \insphere tests evaluated during the main algorithm (see Section~\ref{sub:the_main_algorithm}) will always return $+$ or $-$. 
 We should also remark that, in bibliography, the \insphere predicate is also referred to 
 as the {\sc{}VertexConflict} predicate to reflect the fact that a negative 
 (resp. positive)
 outcome of $\text{\insphere}(S_i,S_j,S_k,S_a,S_b)$ amounts to the 
 Apollonius vertex $v_{ijka}$ in $\mathcal{VD}(\Sigma)$ vanishing (resp. remaining) in $\mathcal{VD}(\Sigma\cup\{S_b\})$, where $\Sigma$ contains 
 $S_i,S_j,S_k$ and $S_a$ but not $S_b$.

 \begin{lemma}
 The \insphere predicate can be evaluated by determining the sign
 of quantities of algebraic degree at most 10 (in the input 
 quantities).
 \end{lemma}


  \subsubsection{The \incone and {\sc{}Trisector} predicates} 
   \label{ssub:the_incone_and_tritype_predicates}

\begin{figure}[htbp]
    \centering
    \includegraphics[width=0.95\textwidth]{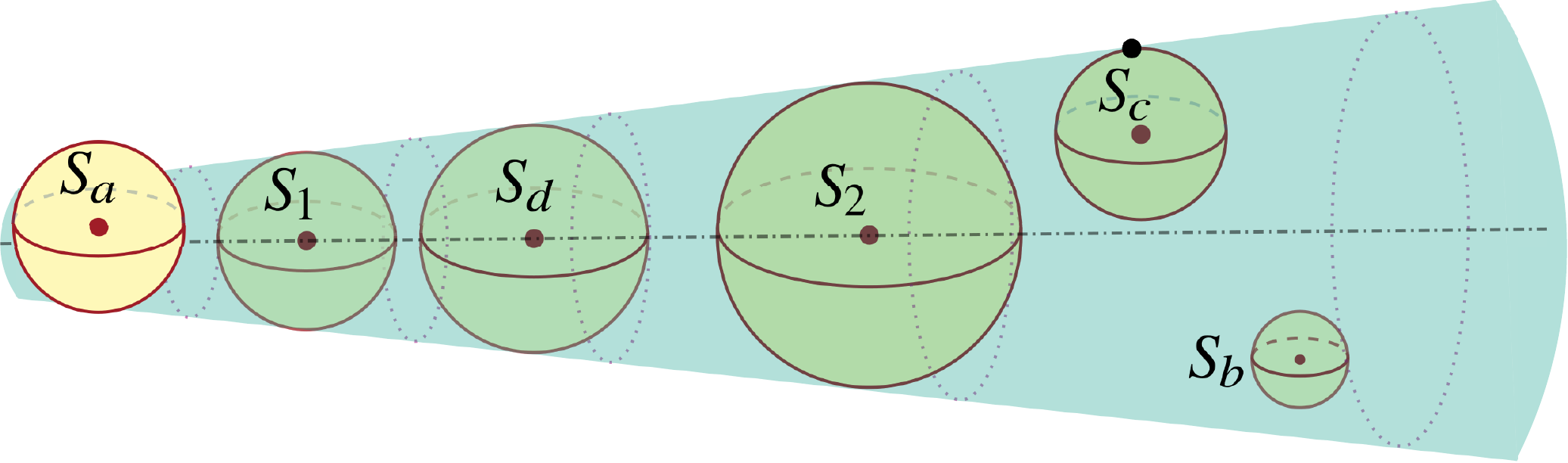}
    \caption{Some of the possible locations of a sphere $S_n$ against 
    the cone $\cone$ defined by $S_1$ and $S_2$. The 
    $\text{\incone}(S_1,S_2;S_n)$ returns \outside, \inside, 
    \ptouch\xspace and \ctouch\xspace for $n=a,b,c$ and $d$ respectively.}
    \label{fig:03}
   \end{figure}

   Given three spheres $S_a,S_b$ and $S_c$, such that $S_a$ and $S_b$
   are not contained one inside the other, we want to determine the
   relative geometric position of $S_c$ with
   respect to the uniquely defined closed semi-cone $\cone(S_a,S_bb)$ that is 
   tangent to both $S_a$ and $S_b$ and includes their centers 
   (see Figure~\ref{fig:03}). 
   We shall call this the $\text{\incone}(S_a,S_b;S_c)$ predicate. 

   In case the radii of $S_a$ and $S_b$ are equal, $\cone(a,b)$ 
   (we drop the parenthesis for convenience)
   degenerates into a cylinder without this having an impact to 
   the predicate. If $S^\circ_c$ is used to denote the open sphere 
   that corresponds to $S_c$, then all possible answers of the 
   predicate $\text{\incone}(S_a,S_b;S_c)$ are

   \begin{itemize}
   \item
   $\outside$ ,
      \text{if at least one point of $S_c$ is outside $\cone$}, 
   \item
   $\inside$,
     \text{if $S^\circ_c$ lies inside $\cone$ and 
     $S_c\cap \cone=\emptyset$}, 
   \item
   $\ptouch$,
     \text{if $S^\circ_c$ lies inside $\cone$ and
     $S_c\cap \cone$ is a point}, 
   \item
   $\ctouch$,
     \text{if $S^\circ_c$ lies inside $\cone$ and
     $S_c\cap \cone$ is a circle.}
   \end{itemize}


   The last two answers are considered ``degenerate'' and therefore, we  
   may consider that whenever \incone is called during the algorithm 
   presented in Section~\ref{sub:the_main_algorithm}, it will either return 
   $\outside$ or $\inside$. 

   This predicate is basic tool used in various other sub-predicates 
   such as the \tritype, which returns the trisector type of a set 
   of three spheres. It is known (\cite{WillThesis}) that if the trisector 
   $\tri{abc}$ of $S_a,S_b,S_c$ has Hausdorff dimension 1, it can 
   either be a branch of a``hyperbola'', a ``line'', an 
   ``ellipse'', a ``circle'' or a ``parabola''; these are the 
   possible answers of the $\text{\tritype}(S_a,S_b,S_c)$ predicate.
   However, since the ``line'' and the ``circle'' type are sub-case of 
   the ``hyperbolic'' and ``elliptic'' trisector types respectively, 
   we can characterize a trisector as either ``hyperbolic'', ``elliptic'' 
   or ``parabolic''. 

   During the execution of the main algorithm of Section~\ref{sub:the_main_algorithm}, the $\text{\tritype}\allowbreak(S_i,S_j,S_k)$ has to be evaluated. 
   Being able to distinguish the type of the trisector $\tri{ijk}$ is essential since all the analysis presented in this paper assumes that  
   $\tri{ijk}$ is hyperbolic.



   The analysis followed to determine the outcome of the \incone 
   or the \tritype predicate can be found in Sections~\ref{sub:the_incone_predicate_analysis} and \ref{sub:the_tritype_predicate_analysis} respectively, where the following lemma is proved.

  \begin{lemma}
  The \incone and \tritype predicates can be evaluated by determining 
  the sign of quantities of algebraic degree at most 4
  (in the input quantities).
  \end{lemma}

  \subsubsection{The {\sc{}Distance} predicate} 
   \label{ssub:the_distance_predicate}
   
   When the trisector $\tri{ijk}$ is a hyperbola or a line, there 
   exist two distinct planes, denoted by $\Pi^{-}_{ijk}$ and 
   $\Pi^{+}_{ijk}$, such that each one is commonly tangent 
   to the sites $S_i,S_j,S_k$ and leave their centers on the same 
   halfspace.

   Observe that $\Pi^{-}_{ijk}$ and
   $\Pi^{+}_{ijk}$ correspond to the two Apollonius spheres
   \emph{at infinity}, in the sense that they are centered at 
   \emph{infinity} and are cotangent to the spheres $S_i,S_j$ and 
   $S_k$. These planes are considered as oriented, and subdivide $\RR^3$
   into a positive and a negative halfspace, the positive being the
   halfspace containing the centers of the spheres. 

\begin{figure}[htbp]
    \centering
    \includegraphics[width=0.95\textwidth]{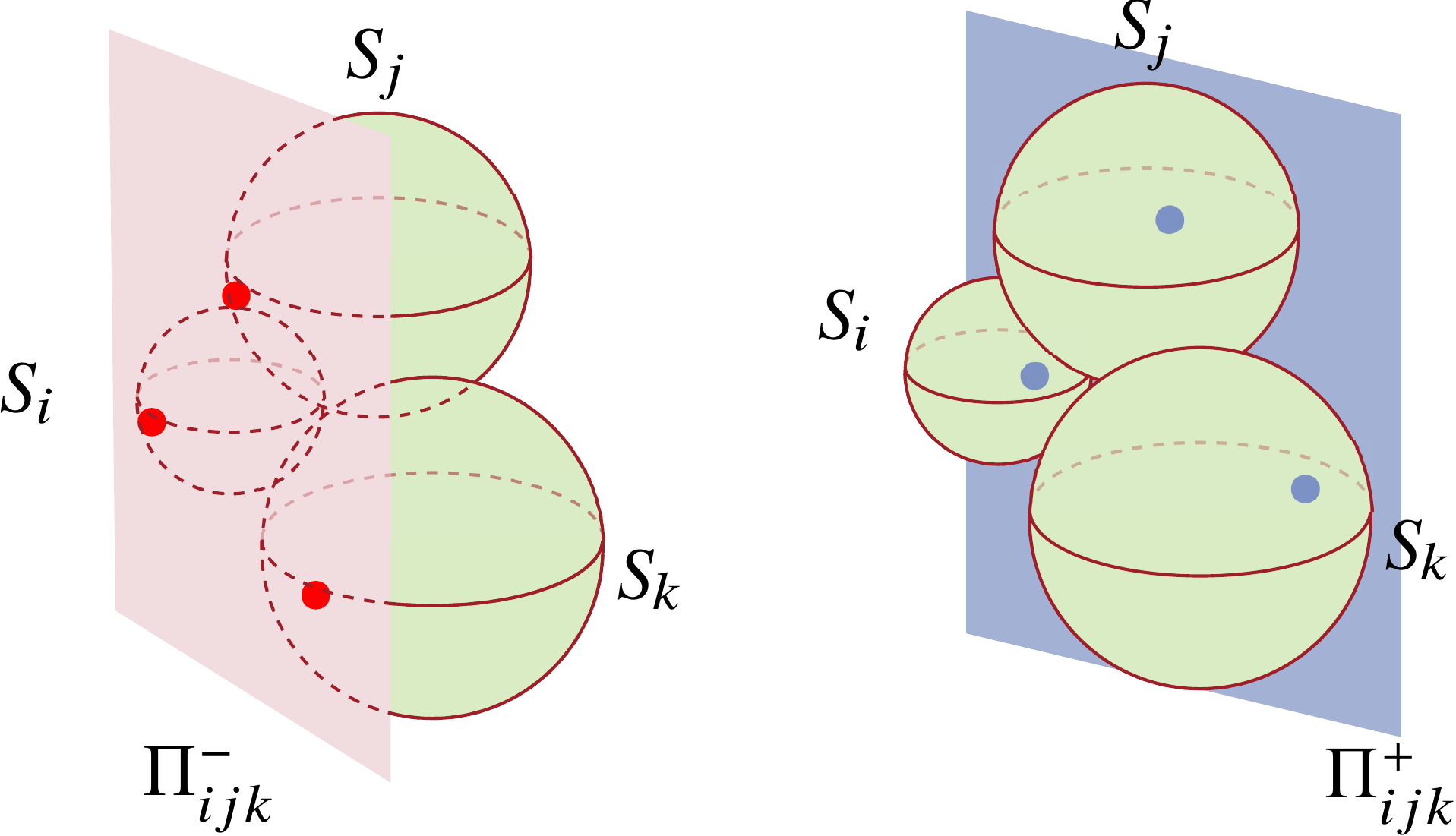}
    \caption{If the spheres $S_i, S_j$ and $S_k$ lie in a convex 
    position there exist two distinct planes, $\Pi_{ijk}^-$ and 
    $\Pi_{ijk}^+$, cotangent to all spheres. These planes 
    are considered as the Apollonius sphere of the sites 
    $S_i, S_j$ and $S_k$, centered at $p\in\tri{ijk}$, as $\map{p}$ 
    goes to $\pm\infty$ respectively.}
    \label{fig:04}
   \end{figure}

   Given a point $p$ on $\tri{ijk}$, we denote by $\tts{p}$ the
   tritangent Apollonius sphere of $S_i,S_j$ and $S_k$ centered at
   $p$. If we move $p$ on $\tri{ijk}$ such that 
   $\map{p}$ goes towards $-\infty$ or $+\infty$, the 
   sphere $\tts{p}$ becomes the corresponding Apollonius sphere at
   infinity \ie, the plane $\Pi^{-}_{ijk}$ or
   $\Pi^{+}_{ijk}$ (see Figure~\ref{fig:04}).

   Given the sites $S_i,S_j,S_k$ and $S_\alpha$, the
   \distance$(S_i,S_j,S_k,S_\alpha)$ predicate determines whether
   $S_\alpha$ intersects, is tangent to, or does not intersect the
   (closed) negative halfspaces delimited by the two planes $\Pi^{-}_{ijk}$
   and $\Pi^{+}_{ijk}$. The ``tangency'' case is considered as 
   degenerate and is beyond the scope of this paper.  
   This predicate is used in the evaluation of the \shadow predicate. 
   and is equal to
   $\text{\distance}(S_i,S_j,S_k,S_\alpha)=
    \big(\sgn(\delta(S_\alpha,\Pi^{-}_{ijk})),
    \sgn(\delta(S_\alpha,\Pi^{+}_{ijk}))
    \big)$,
   where $\delta(S,\Pi)=\delta(C,\Pi)-r$, and $\delta(C,\Pi)$
   denotes the signed Euclidean of $C$ from the plane $\Pi$ and $S$ is a sphere of radius $r$, centered at $C$. As for the
   \exist predicate, we reduce it to the computation
   of the signs of the two roots of a quadratic equation and prove the
   following lemma (see 
   Section~\ref{sub:the_distance_predicate_analysis}
   for this analysis).
   
   \begin{lemma}
   The \distance predicate can be evaluated by determining 
   the sign of quantities of algebraic degree at most 6 
   (in the input quantities).
   \end{lemma}

  \subsubsection{The {\sc{}Existence} predicate} 
   \label{ssub:the_exist_predicate}

   The next primitive operation we need for answering the 
   \conflict predicate is what we call the \exist predicate:
   given four sites $S_a,S_b,S_c$ and $S_n$, we would 
   like to determine the number of Apollonius spheres of the quadruple
   $S_a,S_b,S_c,S_n$.
   In general, given four sites there can be ``0'', ``1'', ``2'' or ``infinite'' Apollonius spheres 
   (cf. \cite{devillers2012qualitative}) including 
   the Apollonius sphere(s) at infinity.
   The $\text{\exist}(S_a,S_b,S_c,S_n)$ predicate only counts the 
   Apollonius spheres that are \emph{not} centered at infinity and since 
   degenerate configurations of the input sites are beyond the scope of this 
   paper, it is safe to assume that the outcome will always be ``0'',``1'' or ``2''. 
   It is also clear that in case of a 
   ``1'' outcome, the corresponding Apollonius center will either be 
   $v_{abcn}$ or $v_{acbn}$ but not both; the case where $v_{abcn}\equiv v_{acbn}$ is ruled out by our initial no-degeneracies assumption.

   


   The analysis of the \exist predicate can be found in 
   Section~\ref{sub:the_existence_predicate_analysis} where we 
   prove the following lemma.

   \begin{lemma}
   The \exist  predicate can be evaluated by determining 
   the sign of quantities of algebraic degree at most 8 
   (in the input quantities).
   \end{lemma}


  \subsubsection{The {\sc{}Shadow} predicate} 
   \label{ssub:the_shadow_predicate}

   We now come to the second major subpredicate used by the \conflict
   predicate: the \shadow predicate.

\begin{figure}[htbp]
   \centering
   \includegraphics[width=0.95\textwidth]{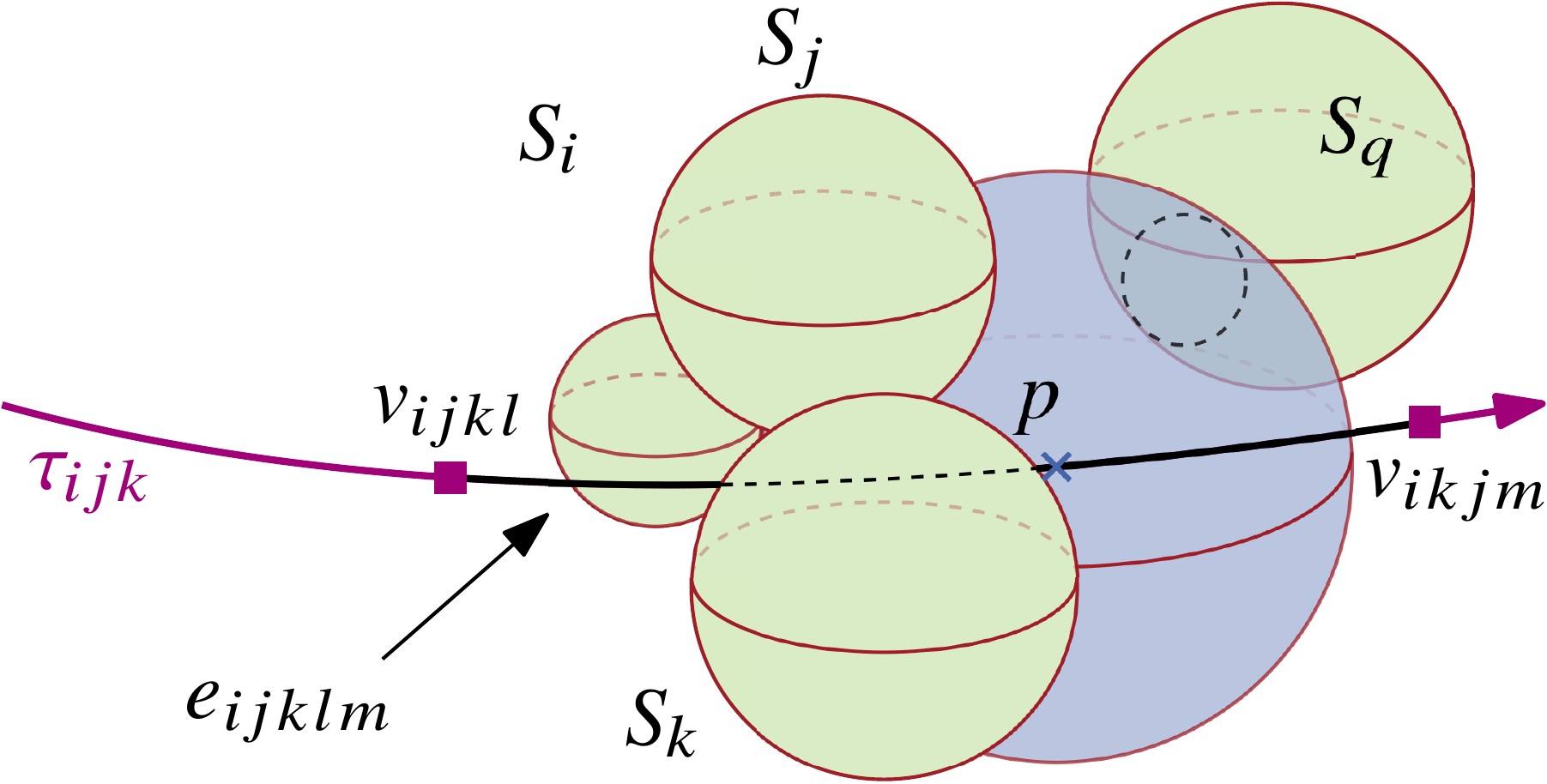}
   \caption{A finite edge $e_{ijklm}$ of the Voronoi diagram 
   $\mathcal{VD(S)}$ of the set 
   $\mathcal{S}=\{S_n : n = i,j,k,l,m\}$ 
   is the locus of points $p\in\tri{ijk}$
   such that $v_{ijkl}\prec p \prec v_{ikjm}$. Any sphere 
   $\tts{p}$, centered at $p$ and cotangent to $S_i,S_j$ and $S_k$ 
   does not intersect any sphere of $\mathcal{S}$ 
   (\emph{Empty Sphere Principle}). However, after inserting $S_q$ 
   in the existing Voronoi diagram, $\tts{p}$ may intersect it. In 
   $\mathcal{VD(S}\cup\{ S_q \})$, all points 
   $p\in\sh{S_q}$ will no longer exist on the ``updated'' 
   edge $e_{ijklm}'$.}
   \label{fig:02}
  \end{figure}

  Given three sites $S_i,S_j$ and $S_k$, we define the 
  \emph{shadow region} $\sh{S_\alpha}$ of a site $S_\alpha$, with 
  respect to the trisector $\tri{ijk}$, to be the locus of points $p$ 
  on $\tri{ijk}$ such that $\delta(C_\alpha,\tts{p})<r_\alpha$. The 
  shadow region of the sites $S_l, S_m$ and $S_q$ play an important 
  role when answering $\text{\conflict}(S_i,S_j,S_k,S_l,S_m,S_q)$ 
  (see Figure~\ref{fig:02} and Section~\ref{sub:the_main_algorithm}).

   The $\text{\shadow}(S_i,S_j,S_k,S_\alpha)$ predicate returns the type of
   $\sh{S_\alpha}$ seen as an interval, or union of intervals, in
   $\RR$. More precisely, the \shadow predicate returns the topological
   structure of the set $\map{\sh{S_\alpha}}=
   \map{\{p\in\tri{ijk}\mid{}\delta(C_\alpha,\tts{p})<r_\alpha\}}$, 
   which we denote by $SRT(S_\alpha)$. 
   
   Clearly, the boundary points of $\text{\shadow}(S_i,S_j,S_k,S_\alpha)$ 
   are the
   points $p$ on $\tri{ijk}$ for which
   $\delta(C_\alpha,\tts{p})=r_\alpha$. These points are nothing
   but the centers of the Apollonius spheres of the four sites
   $S_i,S_j,S_k$ and $S_\alpha$, and, as such, there can only be 0, 1 or
   2 (assuming no degeneracies). 
   This immediately suggests that $SRT(S_\alpha)$ can have one of the 
   following 6 types:
   $\emptyset$, $(-\infty,\infty)=\RR$, $(-\infty,\phi)$, $(\chi,+\infty)$, 
   $(\chi,\phi)$, or $(-\infty,\phi)\cup(\chi,+\infty)$,
   where $\phi,\chi\neq\pm\infty$. 

\begin{figure}[htbp]
    \centering
    \includegraphics[width=0.95\textwidth]{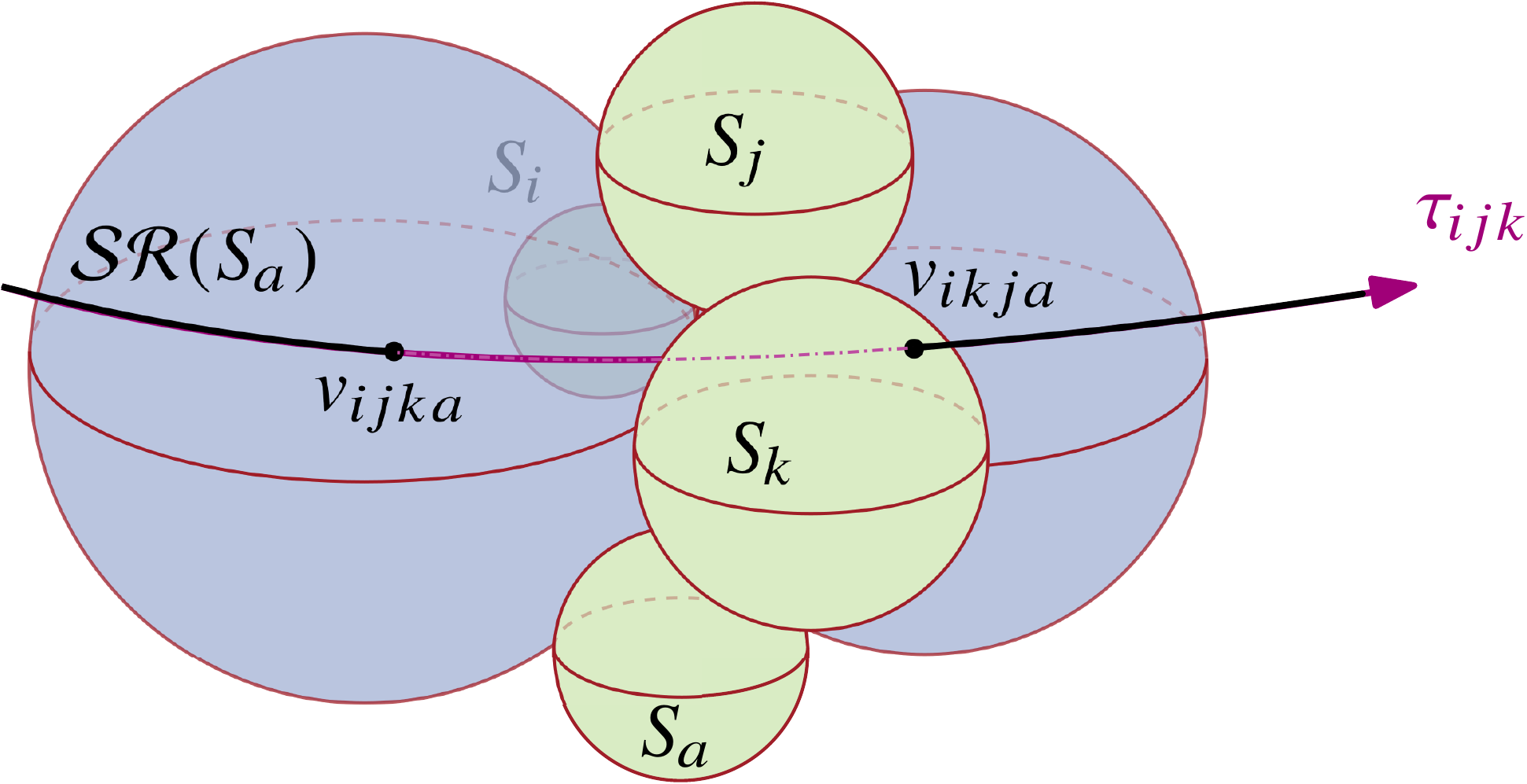}
    \caption{Since there are two Apollonius spheres of the 
    sites $S_n$, for $n\in\{i,j,k,a\}$, centered at $v_{ijka}$ and 
    $v_{ikja}$, the $\sh{S_a}$ on $\tri{ijk}$ must have two endpoints. 
    In this specific configuration notice that, for every 
    point $p$ on the segments of $\tri{ijk}$ painted black, the 
    sphere $\tts{p}$ will intersect $S_a$. Therefore, the black 
    segments are indeed the shadow region $\sh{S_a}$ of the sphere 
    $S_a$ on the trisector $\tri{ijk}$.}
    \label{fig:05}
   \end{figure}

   For convenience, we will use the 
   $\sh{S_a}$ notation instead of $SRT(S_\alpha)$; for example, 
   the statement ``$\sh{S_a} = (-\infty,\phi)$'' will be 
   often used instead of ``$\sh{S_a}$'s type is $(-\infty,\phi)$'' or 
   ``$SRT(S_\alpha) = (-\infty,\phi)$'' (see Figure~\ref{fig:05} 
   for an example). This notation change further highlights the fact 
   that we are only interested in the topological structure of $\sh{S_a}$
   rather than the actual set itself.

   In Section~\ref{sub:the_shadowregion_predicate_analysis}, we 
   prove that the evaluation of the \shadow predicate only 
   requires the call of the respective \distance and 
   \exist predicate, yielding the following lemma.
   
   \begin{lemma}
   The \shadow  predicate can be evaluated by determining 
   the sign of quantities of algebraic degree at most 8 
   (in the input quantities).
   \end{lemma}

  \subsubsection{The {\sc{}Order} predicate} 
   \label{ssub:the_order_predicate}

   The most important sub-predicate used to evaluate the \conflict predicate is what we call the \order predicate.
   When $\text{\order}(S_i,S_j,S_k,S_a,S_b)$ is called, it returns 
   the order of appearance of any of the existing Apollonius vertices
   $v_{ijka},v_{ikja},v_{ijkb}$ and $v_{ikjb}$ on the oriented trisector 
   $\tri{ijk}$. 

   This sub-predicate is called during the main algorithm 
   that answers the $\text{\conflict}(S_i,S_j,S_k,S_l,S_m,S_q)$, for $(a,b)\in\{(l,q),(m,q)\}$, only in the case that either $v_{ijkq}$, $v_{ikjq}$ or both exist. Let us also recall that, in this paper, the trisector 
   $\tri{ijk}$ is ``hyperbolic'' and that $\edge{ijklm}$ is a valid 
   finite Apollonius edge; the Apollonius vertices $v_{ijkl}$ and $v_{ikjm}$ both exist on (the oriented) $\tri{ijk}$ and $v_{ijkl} \prec v_{ikjm}$.


   In order to answer the \order predicate, we first call the 
   $\text{\shadow}(S_i,S_j,S_k,\allowbreak S_n)$ predicate, 
   for $n\in\{a,b\}$, to obtain the type of $\sh{S_a}$ and $\sh{S_b}$.  
   From the shadow region types, two pieces of
   information is easily obtained; firstly, we determine which of the Apollonius vertices 
   $v_{ijka},v_{ikja}$ and $v_{ijkb},v_{ikjb}$ actually exist and
   secondarily, if both $v_{ijkn}$ and $v_{ikjn}$ exist for 
   some $n\in\{a,b\}$, then their ordering on the oriented 
   trisector is also retrieved. Such deductions derive from the 
   study of the shadow region, as shown in Section~\ref{sub:The_classic_configuration} (see Lemma~\ref{lemma:phi_chi}).  
   For example, if $\sh{S_a}=(\chi,\phi)$, then both $v_{ijka}$ and 
   $v_{ikja}$ exist and appear on the oriented trisector in this order:
   $v_{ikja} \prec v_{ijka}$. 

   Now that the existence and partial ordering of the Apollonius vertices
   $v_{ijka}$ and $v_{ikja}$ ( resp. $v_{ijkb}$ and $v_{ikjb}$) is known, 
   we must provide a way of ``merging'' them into a complete 
   ordering. For this reason, we examine all 
   possible complete orderings of the Apollonius vertices  on the
   oriented trisector $\tri{ijk}$. The study of these orderings is seen in 
   the inverted plane \wspace. In Section~\ref{ssub:the_yspace_analysis}, 
   we present the strong geometric relationship that holds 
   between the spheres of the original \zspace and their images in the 
   inverted plane. The observations we make regarding the connection
   of the two spaces allow us to interpret geometric configurations on one 
   space to equivalent ones on the other. 
   A full analysis of how we tackle all possible configurations is 
   presented in  Section~\ref{sub:the_order_predicate_analysis}. 

   In our analysis, we prove that the \order predicate, in the worst case, 
   amounts to evaluate up to 4 \insphere predicates plus some auxiliary 
   tests of lesser algebraic cost. We have therefore proven the 
   following lemma. 

   \begin{lemma}
   The \order predicate can be evaluated by determining the sign 
   of quantities of algebraic degree at most 10 (in the input quantities). 
   \end{lemma}

 \subsection{The main algorithm} 
  \label{sub:the_main_algorithm}
  In this section, we describe in detail how the predicate    
  \conflict$(S_i,S_j,S_k,S_l,S_m,S_q)$ is resolved 
  with the use of the subpredicates \incone, \tritype, 
  \exist, \shadow and \order. 

 We begin by determining the type of the trisector 
 $\tri{ijk}$; this is done via the call of the \tritype$(S_i,S_j,S_k)$
 predicate. Recall that in the scope of this paper, it is assumed that
 the  $\tri{ijk}$ is a hyperbola (or a line) and that none of the 
 subpredicates called return a degenerate answer.

 To answer the \conflict predicate, one must determine 
 which ``part'' of the edge $\edge{ijklm}$ remains in the Voronoi
 diagram after the insertion of the site $S_q$. This is plausible 
 by identifying the set of points of $\edge{ijklm}$ that still remain 
 in the updated Voronoi Diagram; each of these points must satisfy the 
 ``empty-sphere property'': a sphere, centered at that point and 
 tangent to the spheres $S_i,S_j,S_k$, must not intersect any other sites 
 of the Voronoi Diagram. As an immediate result, 
 a point $p$ of the edge $\edge{ijklm}$ in 
 $\mathcal{VD}(\mathcal{S})$ remains in 
 $\mathcal{VD}(\mathcal{S}\cup\{S_q\})$ if and only if $\tts{p}$ does 
 not intersect $S_q$. Since the shadow region of the sphere $S_q$ 
 with respect to the trisector $\tri{ijk}$ consists of all points 
 $p$ such that $\tts{p}$ intersects $S_q$, it must hold that the 
 part of the edge $\edge{ijklm}$ that no longer remains in  
 $\mathcal{VD}(\mathcal{S}\cup\{S_q\})$ is actually 
 $\edge{ijklm}\cap\sh{S_q}$ (see Figure~\ref{fig:02}) . In conclusion, the result of 
 the \conflict predicate is exactly the set $\edge{ijklm}\cap\sh{S_q}$
 seen as an interval or union of intervals of $\RR$.

 To determine the intersection type of $\edge{ijklm}\cap\sh{S_q}$, 
 we first take into account that the finite edge $\edge{ijklm}$ consists 
 of all points $p$ on the oriented trisector $\tri{ijk}$ bounded by the 
 points $v_{ijkl}$ and $v_{ikjm}$ from left and right respectively 
 (see Section~\ref{sub:voronoi_edges}). Next, we consider the 
 type of $\sh{S_q}$ which can be evaluated as shown in 
 Section~\ref{sub:the_shadowregion_predicate_analysis} and  
 is one of the following: 
 $(-\infty,\phi)$, $(\chi,+\infty)$, $(\chi,\phi)$, 
 $(-\infty,\phi)\cup(\chi,+\infty)$, $\emptyset$ or $\RR$. 

 If the edge 
 $\edge{ijklm}$ is seen as the interval $(\lambda_1,\mu_2)$, 
 evidently the intersection type of $E'=\edge{ijklm}\cap\sh{S_q}$ 
 must be one of the following 6 types, each corresponding to a different 
 answers of the \conflict predicate. 

 \begin{itemize}
  \item 
  If $E'$ is of type $\emptyset$, the predicate returns \noconflict.
  \item 
  If $E'$ is of type $\edge{ijklm}$, the predicate returns \fullconflict.
  \item 
  If $E'$ is of type $(\lambda_1,\phi)$, the predicate returns \leftvertex.
  \item 
  If $E'$ is of type $(\chi,\mu_2)$, the predicate returns \rightvertex.
  \item 
  If $E'$ is of type $(\lambda_1,\phi)\cup(\chi,\mu_2)$, the 
  predicate returns \verticesconflict. 
  \item
  If $E'$ is of type $(\chi,\phi)$, the predicate 
  returns \middleconflict.
 \end{itemize}

 This observation suggests that, if we provide a way to identify the 
 type of $E'$, we can answer the \conflict predicate. Taking into
 consideration that
 \begin{itemize}
  \item 
  $\lambda_1$ and $\mu_2$ correspond to $v_{ijkl}$ and $v_{ikjm}$
  respectively as shown in Section~\ref{sub:voronoi_edges}, and 
  \item
  $\chi,\phi$ correspond to $v_{ikjq}$ and $v_{ijkq}$ respectively 
  as stated in Lemma~\ref{lemma:phi_chi} that we prove in 
  Section~\ref{sub:The_classic_configuration},
 \end{itemize}
 it becomes apparent that if we order all Apollonius vertices 
 $v_{ijkl}, v_{ikjm}$ and any of the existing among
 $v_{ijkq}, v_{ikjq}$, bearing in mind the type of $\sh{S_q}$, 
 we can deduce the type of $E'$. 

 For example, let us assume that
 $\sh{S_q}$ type is $(-\infty,\chi)\cup(\phi,+\infty)$. If
 $v_{ijkl} \prec v_{ikjq} \prec v_{ikjm} \prec v_{ijkq}$ on 
 the oriented trisector $\tri{ijk}$, or equivalently
 $\lambda_1<\chi<\mu_2<\phi$, 
 we can conclude that $E'$ is of type $(\lambda_1,\phi)$ and the
 \conflict predicate would return \leftvertex. 

 
 Therefore, it is essential that we are able to provide an ordering 
 of the Apollonius vertices $v_{ijkl}, v_{ikjm}$ and any of 
 the existing among $v_{ijkq}, v_{ikjq}$. Such a task is 
 accomplished via the call of the 
 \order$(S_i,S_j,S_k,S_a,S_b)$ predicate 
 $(a,b)=(l,q)$ and $(m,q)$. The outcomes of these predicates 
 consist of the orderings of all possible Apollonius vertices of 
 the sites $S_i,S_j,S_k,S_a$ and $S_i$, $S_j$, $S_k$, $S_b$ on the 
 trisector $\tri{ijk}$. 
 These partial orderings can then be merged into a complete ordering,
 which contains the desired one. 
 The results' combination principle is identical to the one 
 used when we have to order a set of numbers but we can only 
 compare two at a time. 

 A detailed algorithm that summarizes the analysis of this Section 
 and can be followed to answer the 
  $\text{\conflict}(S_i,S_j,S_k,S_l,S_m;S_q)$ is described in the following steps.
  \begin{description}
   \item[Step 1] The \tritype$(S_i,S_j,S_k)$ is called to 
   determine the type of the trisector $\tri{ijk}$. In this Section, 
   we assume that $\tri{ijk}$ return ``hyperbolic''.
   \item[Step 2] We evaluate $SRT(q)=\text{\shadow}(S_i,S_j,S_k,S_q)$. 
   If $SRT(q)=\emptyset$ or $\RR$, we return \noconflict or 
   \fullconflict respectively. Otherwise, we know that 
   $SRT(q)$ has one of the forms
   $(-\infty,\phi)$, $(\chi,+\infty)$, $(\chi,\phi)$ or $
   (-\infty,\phi)\cup(\chi,+\infty)$, where $\phi$ and 
   $\chi$ correspond to the Apollonius vertices $v_{ijkq}$ and 
   $v_{ikjq}$ respectively. 
   \item[Step 3] 
   We evaluate 
   \order$(S_i,S_j,S_k,S_l,S_q)$ and break down our analysis 
   depending on how many of the vertices $v_{ijkq}$ and $v_{ikjq}$ exist;
   if only one vertex exist (which is equivalent to $SRT(q)$ being $(-\infty,\phi)$ or $(\chi,+\infty)$ ), denote it by $v_q$ and go to Step 3a.
   Otherwise, if both exist, go to Step 3b.
   \begin{description}
    \item[Step 3a] 
    If only $v_{ikjq}$ or $v_{ijkq}$ exist, 
    then $SRT(q)$, which was evaluated in Step 2, is of the form 
    $(-\infty,\phi)$ or $(\chi,+\infty)$ 
    respectively. From the outcome of \order$(S_i,S_j,S_k,S_l,S_q)$, 
    we know whether $v_q\prec v_{ikjl}$ or $v_{ikjl} \prec v_q$. 
    If $v_q \prec v_{ijkl}$, and therefore 
    $v_q\prec v_{ijkl}\prec v_{ikkm}$, the \conflict 
    predicate returns \noconflict if $SRT(q)=(-\infty,\phi)$ or 
    \fullconflict if $SRT(q)=(\chi,+\infty)$.

    If $v_{ijkl}\prec v_q$, we evaluate 
    $\text{\order}(S_i,S_j,S_k,S_m,S_q)$ and determine whether
    $v_q\prec v_{ikjm}$ or $v_{ikjm}\prec v_q$. In the first case, 
    we conclude that  $v_{ijkl}\prec v_q\prec v_{ikjm}$ and 
    the \conflict predicate returns \leftvertex if 
    $SRT(q)=(-\infty,\phi)$ or 
    \rightvertex if $SRT(q)=(\chi,+\infty)$. In the second case, 
    we get that $v_{ijkl}\prec v_{ikjm}\prec v_q$ and the predicate returns 
    \fullconflict if $SRT(q)=(-\infty,\phi)$ or 
    \noconflict if $SRT(q)=(\chi,+\infty)$. 
    
    \item [Step 3b] 
    If both $v_{ijkq}$ and $v_{ikjq}$ exist, 
    then $SRT(q)$ is of the form $(\chi,\phi)$ or 
    $(-\infty,\phi)\cup(\chi,+\infty)$,
    where $\phi$ and $\chi$ correspond 
    to $v_{ijkq}$ and $v_{ikjq}$ respectively. We now call the  
    $\text{\order}(S_i,S_j,S_k,S_l,S_q)$ predicate. If it returns that 
    $v_{ijkq},v_{ikjq}\prec v_{ijkl}$, the \conflict predicate 
    immediately returns \noconflict if $SRT(q)=(\chi,\phi)$, 
    otherwise it returns \fullconflict.

    If this is not the case, we have to call the 
    $\text{\order}(S_i,S_j,S_k,S_m,S_q)$ predicate to acquire
    the ordering of $v_{ijkq},v_{ikjq}$ and $ v_{ijkm}$. 
    If $v_{ijkm}\allowbreak \prec \allowbreak v_{ijkq},v_{ikjq}$, the \conflict predicate immediately 
    returns \noconflict if $SRT(q)=(\chi,\phi)$, otherwise it returns
    \fullconflict. 
    
    In any other case, we must combine the information from the 
    two \order predicates with the fact that  $v_{ikjl}\prec v_{ijkm}$, 
    to obtain the complete ordering of $v_{ikjl}$, $v_{ijkm}$, $v_{ijkq}$ 
    and $v_{ikjq}$. The list of possible \conflict predicate 
    answers for the cases that have not been already handled is 
    found in this list. 
    \begin{description}
     \item[\noconflict] 
     If $SRT(q)=(-\infty,\phi)\cup(\chi,+\infty)$, 

     $v_{ikjq}\prec v_{ijkl}\prec v_{ikjm}$ and 
     $v_{ijkl}\prec v_{ikjm}\prec v_{ijkq}$.
     \item[\fullconflict] 
     If $SRT(q)=(\chi,\phi)$, 
     $v_{ijkl}\prec v_{ikjm}\prec v_{ikjq}$ and

     $v_{ijkq}\prec v_{ijkl}\prec v_{ikjm}$.
     \item[\leftvertex] 
     If $SRT(q)=(\chi,\phi)$, 
     $v_{ijkl}\prec v_{ikjq} \prec v_{ikjm}$ 

     and $v_{ijkq} \prec v_{ijkl}\prec v_{ikjm}$,
     or $SRT(q)=(-\infty,\phi)\cup(\chi,+\infty)$,

     $ v_{ijkl}\prec v_{ikjq}\prec v_{ikjm}$ and
     $v_{ijkl}\prec v_{ikjm}\prec v_{ijkq}$.
     \item[\rightvertex] 
     If $SRT(q)=(\chi,\phi)$
     $v_{ijkl}\prec v_{ikjm}\prec v_{ikjq}$ and 
     $v_{ijkq}\prec v_{ijkl}\prec v_{ikjm}$, 
     or $SRT(q)=(-\infty,\phi)\cup(\chi,+\infty)$,
     $v_{ikjq}\prec v_{ijkl}\prec v_{ikjm}$ and
     $v_{ijkl}\prec v_{ijkq}\prec v_{ikjm}$.
     \item[\verticesconflict] 
     If $SRT(q)=(-\infty,\phi)\cup(\chi,+\infty)$ and

     $v_{ijkl}\prec v_{ijkq},v_{ikjq}\prec v_{ikjm}$ 
     \item[\middleconflict]
     If $SRT(q)=(\chi,\phi)$ and 
     $v_{ijkl}\prec v_{ijkq},v_{ikjq}\prec v_{ikjm}$.
    \end{description}
   \end{description}
  \end{description}

\begin{figure}[tbp]
   \centering
   \includegraphics[width=0.85\textwidth]{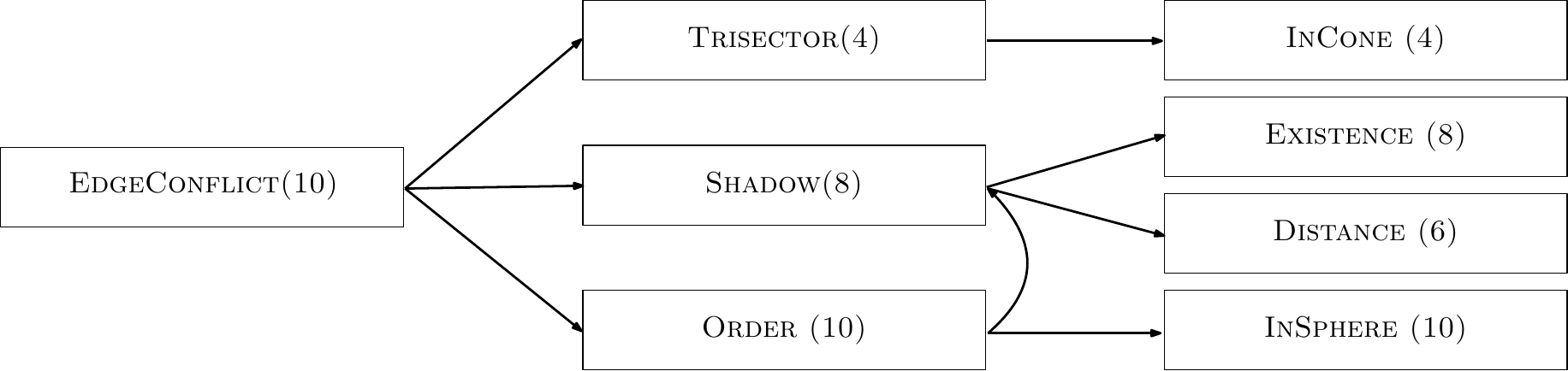}
   \caption{The layout of predicates and their subpredicates used to 
     answer the \conflict predicate. The number next to each predicate corresponds to its algebraic cost. It is assumed that every
     subpredicate returns a non-degenerate answer.}
   \label{fig:predicates}
  \end{figure}

  A sketch of the subpredicates used when answering the 
  \conflict predicate is shown in Figure \ref{fig:predicates}. 
  Since the highest algebraic degree needed in the evaluation of the 
  subpredicates used is 10, we have proven the following theorem.

  \begin{theorem}
  The \conflict predicate can be evaluated by determining the sign of
  quantities of algebraic degree at most 10 (in the input quantities). 
  \end{theorem}

\section{Implementation and Analysis of SubPredicates} 
 \label{sec:algebraic_analysis}


 In this Section, we provide a detailed description on how to answer every 
 subpredicate involved in the algorithm presented in 
 Section~\ref{sub:the_main_algorithm}. For each primitive, beside 
 analyzing how we derive the outcome, we also compute its algebraic 
 degree \ie, the maximum algebraic degree of all quantities that have 
 to be evaluated to obtain the subpredicate's result.
 



\subsection{The \incone predicate} 
  \label{sub:the_incone_predicate_analysis}

  To answer the \incone$(S_a,S_b,S_c)$ predicate 
  we first determine the number of possible tangent planes 
  to the sites $S_a, S_b,$ and $S_c$ that leave them all 
  on the same side; there can be either 0, 1, 2 or 
  $\infty$ such planes. Bear in mind that the \incone predicate can
  be called only if no one of the spheres $S_a$ and $S_b$ are contained 
  inside another.

  If $S_a$ and $S_b$ have different radii, then 
  $\cone$ will denote the cone that contains and is tangent to these spheres, whereas $\cone^-$ will symbolize the symmetric cone with the same 
  axis and apex. Let us no consider each of the four possible cases 
  regarding the number of cotagent planes, since it is indicative of the 
  relative position of the three spheres.

  \begin{enumerate}
  \item 
  If no such plane exists, there are three cases to consider; 
  \begin{itemize}
   \item 
   $S_c$ lies strictly inside the cone $\cone$;  
   the predicate returns $\inside$.
    \item 
   $S_c$ lies strictly inside the cone $\cone^-$;
   the predicate returns  $\outside$.
   \item 
   $S_c$ fully intersects the cone $\cone$ in the sense that 
   there is a circle on $S_c$ that is outside $\cone$. In this case, 
   the predicate returns  $\outside$.
   \item There is a circle on $S_c$ that is outside $\cone$; $S_c$ could lie strictly inside the cone $\cone^-$ or fully intersect the cone $\cone$, 
   $\cone^-$ or both. In all these cases, the predicate returns $\outside$.
  \end{itemize}

  \item 
  If there is only one such plane, then ${S_c}$ touches ${\cone}$ in 
  a single point. There are two cases to consider; 
  \begin{itemize}
   \item $S_c$ lies strictly inside the cone $\cone$; the predicate 
   returns \ptouch.
   \item $S_c$ fully intersects the cone $\cone$ (there is a circle in 
   $S_c$ that is outside $\cone$). In this case, the predicate returns  
   $\outside$.
  \end{itemize}

  \item 
  If there are two such planes, the spheres must lie in convex position,
  hence the predicate returns  $\outside$.
  \item 
  If there are infinite such planes, the spheres $S_a,S_b$ and $S_c$
  have collinear centers and the points of tangency of each sphere 
  with the cone is a single circle. The predicate returns $\ctouch$ 
  in this scenario. 
  \end{enumerate}

  In the case no cotagent plane to all sites $S_a,S_b$ and $S_c$ exist, 
  we must be able to tell if $S_c$ lies inside the cone $\cone^-$. 
  However, this check is only needed in the case $r_a\neq r_b$; if 
  $r_a=r_b$, the cone $C_{ij}$ degenerates into a cylinder and 
  $\cone^-$ does not exist. 

  Let us consider the case $r_a\neq r_b$ in detail. First, observe that 
  we can assume without loss of generality that $r_a<r_b$ as this follows 
  from the definition of the \incone predicate. Indeed, since 
  \incone$(S_a,S_b,S_c)$ and \incone$(S_b,S_a,S_c)$ represent the same 
  geometric inquiry, we can exchange the notation of the spheres $S_a$ 
  and $S_b$ in case $r_a>r_b$. 

  Taking this into consideration, we denote 
  $K$ to be the apex of the cone $\cone$ and $\Pi_c$ to be the plane that 
  goes through $K$ and perpendicular to axis $\ell$ of the cone. We also 
  denote by $\Pi_c^+$ the half-plane defined by the plane $\Pi_c$ and 
  the centers $C_a$ and $C_b$, whereas its compliment half-plane is 
  denoted by $\Pi_c^-$.

  It is obvious that if the center of the sphere $S_c$ does not lie 
  in $\Pi_c^+$ the predicate must return $\outside$ since $S_c$ has 
  at least one point outside the cone $\cone$. Moreover, if the center 
  $C_c$ lies in $\Pi_c^+$ then $S_c$ cannot lie inside $\cone^-$ and this 
  case is ruled out. To check if $C_c$ lies on $\Pi_c^+$, we first 
  observe that $C_a$ and $C_b$ define the line $\ell$ hence for 
  every point $P$ of $\ell$ stands that $\ov{OP}=\ov{OC_a}+t\ov{C_aC_b}$ 
  for some $t\in\RR$.
  We now claim that for a sphere with center $P(t)$ to be tangent 
  to the cone it must have radius $r(t)$ that 
  is linearly dependent with $t$, i.e. $r(t) = k_1t+k_0$. To evaluate 
  $k_1$ and $k_0$, we observe that for $t=0$, $P(0)\equiv C_a$ hence 
  $r(0)=r_a$ and respectively for $t=1$, $P(1)\equiv C_b$ hence 
  $r(1)=r_b$. We conclude that 
  $r(t)=t\cdot (r_b-r_a)+r_a, \ t\in\RR$.
  The cone apex lies on $\ell$ so
  $\ov{OK} = \ov{OC_a}+t_c\ov{C_aC_b}$
  for $t_c\in\RR$ such that
  $r(t_c)=0 $ or equivalently $ t_c = r_a/(r_a-r_b)$. 
  In this way we have evaluated the cone apex coordinates which derive 
  from the relation
  $\ov{OK} = \ov{OC_a}+\ov{C_bC_a}\cdot{r_a}/(r_b-r_a)$. 
  Since $\Pi_c$ is perpendicular to $\ell$ and therefore to $\ov{C_aC_b}$, 
  and $\ov{C_aC_b}$ points towards the positive side of $\Pi_c$, 
  the point $C_c$ lies on the positive half plane $\Pi_c^+$ iff the quantity
  $M = \ov{C_aC_b}\cdot\ov{KC_c}$ is strictly positive.

  To evaluate the sign of $M$ we have 
  \begin{align*} \sgn(M) &= 
  \sgn(\ov{C_aC_b}\cdot\ov{KC_c})
  = \sgn(\ov{C_aC_b}\cdot\ov{OC_c}-\ov{C_aC_b}\cdot\ov{OK})\\
  &= 
  \sgn(\ov{C_aC_b}\cdot(\ov{OC_c}-\ov{OC_a}-\dfrac{r_a}{r_b-r_a}\ov{C_bC_a})) \\
  &= \sgn(\ov{C_aC_b}\cdot((r_b-r_a)\ov{C_aC_c}+r_a\ov{C_aC_b}))\sgn(r_b-r_a)\\
  &= 
  \sgn((r_b-r_a)\ov{C_aC_c}\cdot\ov{C_aC_b}+r_a\ov{C_aC_b}\cdot\ov{C_aC_b})
  \end{align*}
  \noindent
  so determining $\sgn(M)$ requires operations of degree 3.

  If $M$ is non-positive the predicate returns $\outside$, however more 
  analysis is need if this is not the case. For the rest of this section, 
  we assume that $M$ is positive and
  break the analysis depending on the collinearity of the centers 
  of the spheres $S_a,S_b$ and $S_c$. Remember that  
  $C_a,C_b$ and $C_c$ are collinear if and only if the 
  cross product $\ov{C_aC_b}\times\ov{C_aC_c}$ is the zero vector, which 
  is a 2-degree demanding operation in the input quantities.

  \subsubsection{The Centers \texorpdfstring{$C_a,C_b,C_c$}{Ci,Cj,Ck} 
    are Collinear} 
  \label{ssub:incone_collinear_centers}

  If $C_a,C_b$ and $C_c$ are collinear, they all lie on the line $\ell$, 
  and therefore  $\ov{OC_c} = \ov{OC_a}+t_o\ov{C_aC_b}$
  for some $t_o\in\RR$. Equivalently, we get that
  $t_o\ov{C_aC_b} = \ov{C_aC_c}$ and since $C_a$ and $C_b$ cannot be 
  identical we can evaluate $t_o=X/Y$ where 
  $(X,Y)=(x_c-x_a,x_b-x_a)$ or $(y_c-y_a,y_b-y_a)$ or $(z_c-z_a,z_b-z_a)$,
  if $x_b-x_a\neq 0$ or $y_b-y_a\neq 0$ or $y_b-y_a\neq 0$ respectively. 

  Denote $r(t)$ as before, we evaluate the sign $S$ of $r_c-r(t_o)$, 
  \begin{equation*}
  S = \sgn(r_c-r(t_o)) = \sgn( (r_a-r_c) Y +(r_b-r_a) X )
  \end{equation*}
  \noindent
  which requires operations of degree 2.

  We can now answer the predicate because if $r_c<r(t_o)$, ie. $S$
  is negative, then $S_c$ lies 
  strictly on the cone; otherwise, if $r_c>r(t_o)$, ie. $S$ is 
  positive, then $S_c$ intersects the cone. 
  If $r_c=r(t_c)$, ie. $S$ is zero, then $S_c$ touches $\cone$ in a circle.
  In conclusion, we get that 
  \begin{equation*}
  \text{\incone}(S_a,S_b,S_c) = 
  \begin{cases}
  \inside, &  \text{if } S<0,\\
  \ctouch, &  \text{if } S=0,\\
  \outside, & \text{if } S>0,
  \end{cases}
  \end{equation*}

  \subsubsection{Non-Collinear Centers} 
  \label{ssub:incone_non_collinear_centers}
  If $C_a,C_b$ and $C_c$ are not 
  collinear and must examine the number of possible tritangent 
  planes to the sites $S_n$ for $n\in\{i,j,k\}$.  
  Denote $\Pi: ax+by+cz+d=0$ a plane tangent to 
  $S_a,S_b$ and $S_c$ that leaves the spheres on the same half-plane,
  and assume without loss of generality that $a^2+b^2+c^2=1$. Since the sphere $S_n$ for 
  $n\in\{i,j,k\}$ touches the plane $\Pi$, we get that
  $\delta(S_n,\Pi)=ax_n+by_n+cz_n+d=r_n$. 
  %
  %
  %
  %
  %
  We examine the resulting system of equations
   \begin{align*}
  ax_a+by_a+cz_a &= r_a-d\\
  ax_b+by_b+cz_b &= r_b-d\\
  ax_c+by_c+cz_c &= r_c-d\\
  a^2+b^2+c^2 &= 1
  \end{align*}

   and distinguish 
  the following cases 
  \begin{itemize}
  \item 
  if $D^{xyz}_{abc}\neq 0$, we can express $a,b$ and $c$ linearly 
  in terms of $d$.
  Substituting these expressions in the last equation, we 
  get a quadratic equation that vanishes at $d$; 
  the sign of the discriminant $\Delta'$
  of this quadratic reflects the number
  of possible $d's$ and therefore tangent planes to the spheres 
  $S_a,S_b$ and $S_c$. 
  \item 
  if $D^{xyz}_{abc}=0$ and since the centers of the spheres $S_a,S_b$ 
  and $S_c$ are not collinear, one of the quantities 
  $D^{xy}_{abc},D^{xz}_{abc}$ or $D^{yz}_{abc}$
  is non-zero, without loss of generality assume $D^{xy}_{abc}\neq 0$.
  In this case, we can express $a$ and $b$ linearly in terms of $c$, 
  whereas
  $d = D^{xyr}_{abc}/D^{xy}_{abc}$. From the last equation, we get 
  a quadratic equation that vanishes at $c$, 
  and the sign of the discriminant $\Delta''$ again reflects 
  the number of possible $c's$ and therefore tangent planes to the 
  spheres $S_a,S_b$ and $S_c$. 
  \end{itemize}

  \noindent
  Writing down the expressions of the discriminants, we finally evaluate 
  that $\Delta' = 4(D^{xyz}_{abc})^2\Delta$ 
  and $ \Delta''=4(D^{xy}_{abc})^2\Delta$,  where 
  \begin{equation*}
  \Delta = (D^{xy}_{abc})^2+(D^{xz}_{abc})^2+(D^{yz}_{abc})^2
   -(D^{xr}_{abc})^2-(D^{yr}_{abc})^2-(D^{zr}_{abc})^2.
  \end{equation*}
  \noindent
  Since the signs of the discriminants $\Delta_1,\Delta_2$ and $\Delta$
  are identical, we evaluate $\sgn(\Delta)$ and proceed as follows:
  \begin{enumerate}
  \item 
  If $\Delta>0$, there are two planes tangent to all three spheres 
  $S_a, S_b$ and $S_c$; the predicate returns $\outside$.
  \item 
  If $\Delta=0$, there is a single plane tangent to the spheres 
  $S_a,S_b$ and $S_c$ and we have to distinguish between the two possible cases; $S_c$ lies strictly inside $\cone$ or $S_c$ intersects $\cone$.

  Assume we are in the latter case, observe that for a proper
  $\epsilon<0$, the deflated sphere $\tilde{S_c}$, with 
  radius $\tilde{r_c}=r_c+\epsilon$, would point-touch the cone $\cone$.
  (Note that the tangency points of the cone $\cone$ and the
  spheres $S_c$ and $\tilde{S_c}$  are not the same!) Therefore, if we 
  consider the analysis of the predicate \incone$(S_a,S_b,\tilde{S_c})$, 
  we would get that the ``perturbed'' discriminant 
  $\tilde{\Delta}$ would vanish for 
  this $\epsilon<0$!

  For the evaluation of $\tilde{\Delta}$, we simply substitute 
  $r_c$ in $\Delta$ with $\tilde{r_c}=r_c+\epsilon$
  and rewrite $\tilde{\Delta}=\tilde{\Delta}(\epsilon)$ as a polynomial 
  in terms of $\epsilon$:
  $\tilde{\Delta}(\epsilon)=\Delta_2\epsilon^2+\Delta_1\epsilon+\Delta_0$,
  where $\Delta_0 = \Delta=0$ and 
  \begin{align*}
  \Delta_2 &= -[(x_b-x_a)^2+(y_b-y_a)^2+(z_b-z_a)^2] ( <0 ),\\
  \Delta_1 &= -2\big((x_b-x_a)D^{xr}_{abc}+(y_b-y_a)D^{yr}_{abc}
    +(z_b-z_a)D^{zr}_{abc}\big).
  \end{align*}

  Since $\epsilon=0$  is a root of the quadratic (in terms of $\epsilon$) 
  $\tilde{\Delta}(\epsilon)$, a simple use of Vieta's formula shows that 
  $\tilde{\Delta}(\epsilon)$ has a negative root, if and only if 
  $\sgn(\Delta_1)$ is strictly negative. In conjuction with our previous 
  remarks, the predicate should return $\outside$ 
  if $\Delta_1<0$; otherwise it must return $\ptouch$. Indead, if $S_c$ 
  point touches the cone $\cone$ from the ``inside'', one can inflate $S_c$
  to $\tilde{S_c}$ so that $\tilde{S_c}$ point-touches $\cone$!
  \item 
  If $\Delta<0$, there is no plane tangent to the 
  spheres $S_a,S_b$ and $S_c$. Since $S_c$ lying inside the cone $\cone'$ is 
  ruled out, we have to distinguish between the two possible cases; $S_c$ lies strictly inside $\cone$ or $S_c$ intersects $\cone$.

  It follows that, either $S_c$ lies strictly within the cone and the
  predicate must return $\inside$, or $S_c$ fully intersects the cone 
  and the predicate must return $\outside$. 
  Using the same analysis as in case $\Delta=0$, we observe that
  if we inflate (or deflate) $S_c$, the perturbed sphere 
  $\tilde{S}_c$ with radius $\tilde{r_c}=r_c+\epsilon$ will
  touch the cone $\cone$ for two different values $\epsilon_1$ and 
  $\epsilon_2$. 
  The predicate must return $\inside$ if we must inflate $S_c$ 
  to touch $\cone$, ie. if $\epsilon_1,\epsilon_2>0$ whereas 
  the predicate must return $\outside$ if we must deflate $S_c$, 
  to point-touch $\cone$, ie. if $\epsilon_1,\epsilon_2<0$. 
  As shown in the case $\Delta=0$, the perturbed discriminant that will 
  appear during the evaluation of \incone$(S_a,S_b,\tilde{S_c})$ is
  $\tilde{\Delta}(\epsilon)=\Delta_2\epsilon^2+\Delta_1\epsilon+\Delta_0$
  where
  \begin{align*}
  \Delta_2 &= -[(x_b-x_a)^2+(y_b-y_a)^2+(z_b-z_a)^2] ( <0 ),\\
  \Delta_1 &= -2\big((x_b-x_a)D^{xr}_{abc}+(y_b-y_a)D^{yr}_{abc}
    +(z_b-z_a)D^{zr}_{abc}\big),\\
  \Delta_1 &= \Delta < 0.
  \end{align*}
  
  Since the sphere $\tilde{S_c})$ point-touches the cone $\cone$ for 
  $\epsilon=\epsilon_1,\epsilon_2$, the discriminant 
  $\tilde{\Delta}(\epsilon)$ must vanish for these epsilons, as mentioned 
  in the previous case. Therefore, $\epsilon_1$ and $\epsilon_2$ 
  are the roots of the quadratic (in terms of $\epsilon$)
  $\Delta_2\epsilon^2+\Delta_1\epsilon+\Delta_0$ and we know, using 
  Vieta's rule and the fact that $\Delta_0<0$, that 
  $\epsilon_1,\epsilon_2$ are both negative (resp. positive) if 
  and only if $\Delta_1$ is negative (resp. positive).
  \end{enumerate}

  Note that, since evaluating the sign of $\Delta$ and $\Delta_1$ 
  require operations of degree 4 and 3 (in the input quantities) 
  respectively, we have shown that the maximum algebraic cost of 
  the \incone predicate is 4, yielding the following lemma.

  \begin{lemma}
  The \incone predicate can be evaluated by determining 
  the sign of quantities of algebraic degree at most 4 
  (in the input quantities).
  \end{lemma}

 \subsection{The \tritype predicate} 
  \label{sub:the_tritype_predicate_analysis}
  Assuming the trisector $\tri{ijk}$ exists and has Hausdorff dimension 1,
  observe that the trisector's type and the relative position of the 
  spheres are closely related.
  Specifically, 
  \begin{itemize}
  \item 
  If the spheres are in \emph{convex position}, \ie there exist two 
  distinct commonly tangent planes that leave them on the same side, 
  then their trisector can either be a hyperbola or a line, in the special 
  case $r_i=r_j=r_k$. The predicate returns ``hyperbolic'' in this scenario.
  \item 
  If the spheres are in in \emph{strictly non-convex position}, \ie 
  one of them lies strictly inside the cone defined by the other two, 
  then their trisector can either be an ellipse or a circle, 
  in the special case where $C_i,C_j$ and $C_k$ are collinear,
  The predicate returns ``elliptic'' in this scenario.
  \item 
  If the spheres are in \emph{degenerate non-convex position}, \ie 
  they are in non-convex position and the closure of all three touch 
  their convex hull, then their trisector is a parabola and the 
  predicate returns ``parabolic''.
  \end{itemize} 

\begin{figure}[htbp]
   \centering
   \includegraphics[width=0.4\textwidth]{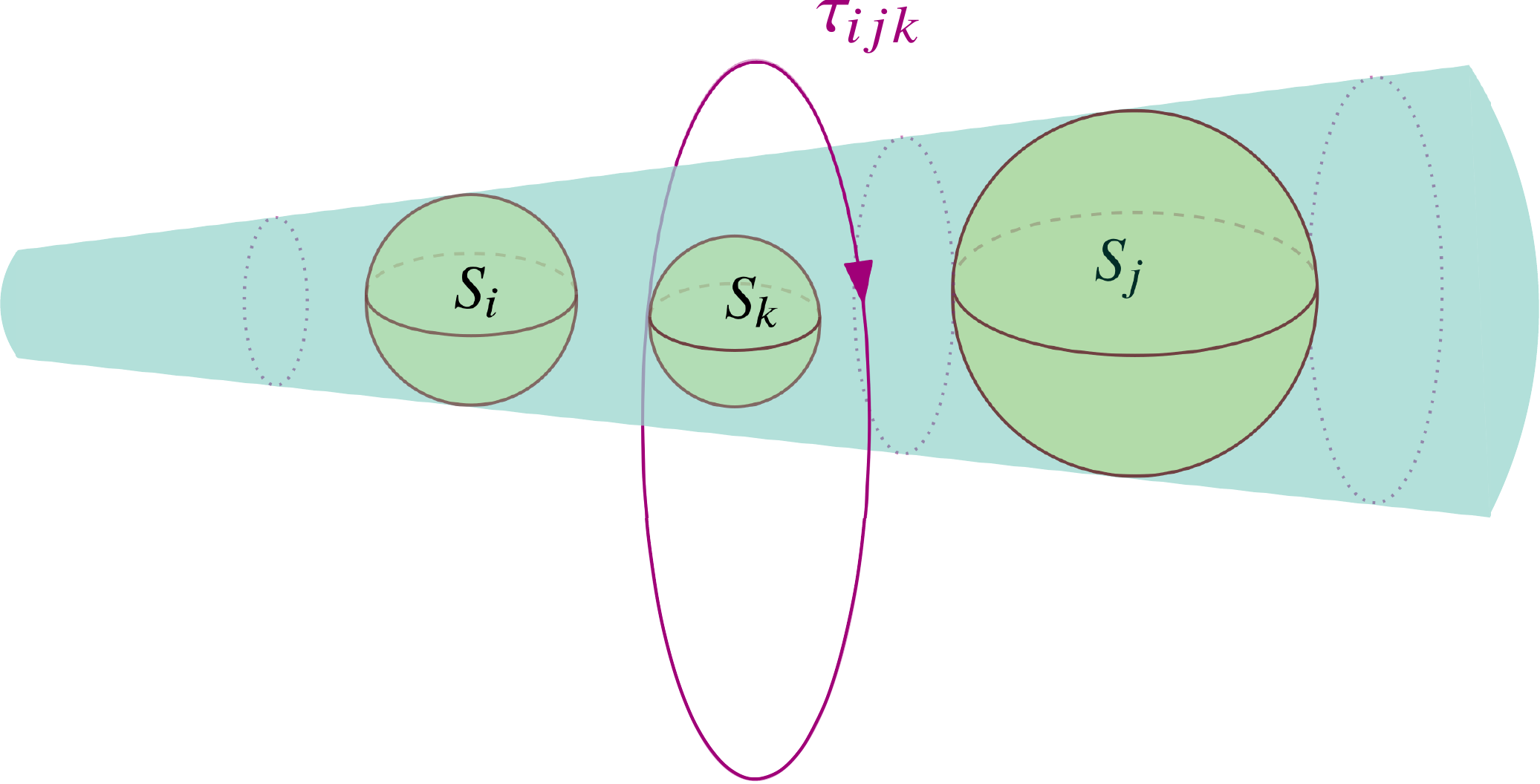} \hspace{1cm}
   \includegraphics[width=0.4\textwidth]{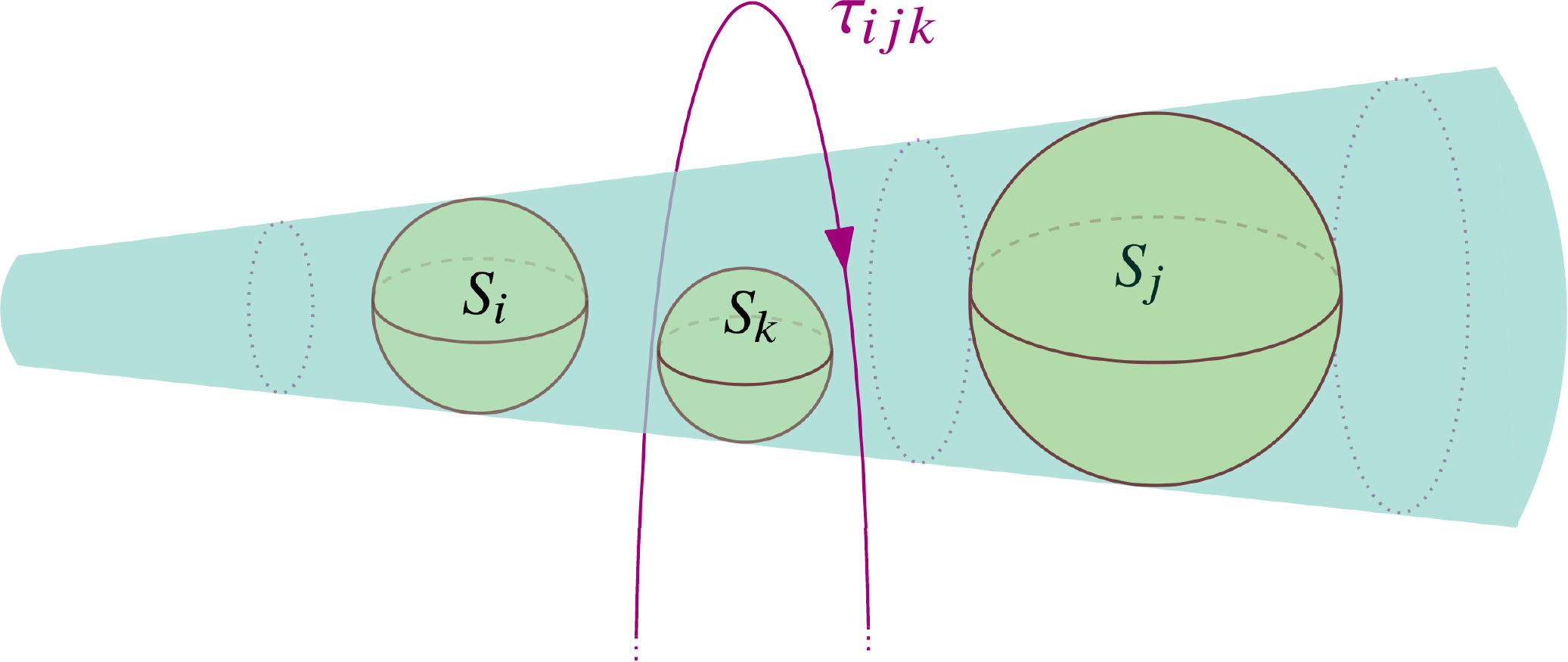}
   \caption{The case of an \emph{elliptic} (Left) and a 
   \emph{parabolic} trisector (Right). In either case, we can orient 
   these trisectors using the ``right-hand rule''; if our thumb points
   from $C_i$ towards $C_j$ the direction pointed by our hand if 
   it ``wraps'' $\tri{ijk}$ is the positive direction. }
   \label{fig:07}
  \end{figure}

  Therefore, one can answer the \tritype predicate if the relative position of the spheres three input spheres is identified. We accomplish such 
  task by combining the  outcomes of the three \incone predicates with inputs
  $(S_i,S_j,S_k)$, $(S_i,S_k,S_j)$ and $(S_j,S_k,S_i)$. Since the 
  trisector $\tri{ijk}$ is assumed to have Hausdorff 
  dimension 1, we know that no one of the spheres $S_i, S_j$ and 
  $S_k$ are contained inside another and therefore we may call the 
  \incone predicate on these inputs.
  \begin{itemize}
    \item If at least one outcome is $\inside$ or $\ctouch$, 
  then the spheres are in \emph{strictly non-convex position} and the 
  trisector's type is ``elliptic''.
    \item If at least one outcome is 
  $\ptouch$, then the spheres are in \emph{degenerate non-convex 
  position} and the \tritype predicate returns ``parabolic''.
    \item Finally, if all three outcomes are $\outside$, then the sites are
  in \emph{convex position} and the trisector's type is ``hyperbolic''.(see 
  Figure~\ref{fig:07}).
  \end{itemize}

  We have proven that the \tritype predicate can be resolved by calling 
  the \incone predicate at most three times (for example, if the first 
  \incone returns $\inside$ the trisector must be ``elliptic''). Since
  the \incone predicate is a 4-degree demanding operation in the input 
  quantities, we have proven the following lemma. 

  \begin{lemma}
  The \tritype predicate can be evaluated by determining 
  the sign of quantities of algebraic degree at most 4 
  (in the input quantities).
  \end{lemma}

 \subsection{The \exist predicate} 
  \label{sub:the_existence_predicate_analysis}
  

  To answer the \exist$(S_i,S_j,S_k,S_a)$ predicate, we 
  first break up our analysis depending on whether the radii of 
  all sites $S_n$, for $n\in\{i,j,k,a\}$, are equal; if $r$ is 
  indeed their common radius, we deflate them by $r$. 
  The existence of a sphere tangent to the original spheres amounts 
  to the existence of a sphere tangent to the centers of the 
  sites where the latter cotangent sphere is the former inflated by $r$.
  Regarding the existence of the Apollonius sphere of the centers
  $C_n$ for $n\in\{i,j,k,a\}$, we consider the tetrahedron
  $C_iC_jC_kC_a$ formed and check it for flatness; this is a 
  3-degree demanding operation. 
  If it is not flat, there are is a 
  single sphere $S$ passing through all the centers, hence there are 
  two spheres tangent to the sites, one internal and one external 
  (inflate or deflate $S$ by $r$) and the \exist predicate 
  returns 1 (since we only count external Apollonius spheres). 
  If the tetrahedron is flat, i.e. the centers lie on the same 
  plane $\Pi_{ijka}$, we check them for co-circularity
  (this is a 5-degree demanding operation); 
  if co-circular, there are infinite number of spheres tangent to 
  the sites (either internal or external), hence the \exist
  predicate returns $\infty$ (degenerate answer). If they are not 
  co-circular, there can not be a sphere tangent to all sites and the \exist
  predicate returns 0. Note that if the \exist
  predicate would return $\infty$ we apply a QSP scheme to resolve the 
  degeneracy.

  Let us now consider the case where the radii of the sites 
  $S_n$ for $ n\in\{i,j,k,a\}$ are not all equal.
  Since a name exchange of the sites does not affect the \exist
  predicate, we name exchange them such that 
  $r_a=\min\{r_i,r_j,r_k,r_a\}$. 
  We now deflate all spheres by $r_a$ and then invert all sites 
  with respect to the point $C_a$; we call this ``inversion 
  through the sphere $S_a$''. 

  In the inverted \wspace, a plane tritangent to the inverted 
  spheres $\inv{S_i},\allowbreak\inv{S_j},\allowbreak\inv{S_k}$  
  amounts to a sphere 
  tangent to all sites $S_i, S_j, S_k $ and $S_l$ in the original 
  \zspace.
  For the corresponding sphere to be an external Apollonius sphere 
  of the sites,  the following conditions must stand:
  \begin{enumerate}
  \item 
  The plane must leave all inverted spheres on one side, called
  the positive side of the plane, and 
  \item 
  The origin $\OO=(0,0,0)$ of the \wspace, that corresponds to 
  the ``point at infinity'' in the \zspace, must also lie on the 
  positive side of the plane.
  \end{enumerate}

  Considering the Condition 1, we denote $\inv{\Pi}_{ijk}: au+bv+cw+d=0$ 
  a plane tangent to 
  $\inv{S_i},\inv{S_j}$ and $\inv{S_k}$ in the inverted space 
  and assume without loss of generality that $a^2+b^2+c^2=1$. 
  Since the signed Euclidean distance of a point $P(u_p,v_p,w_p)$ 
  from the plane $\inv{\Pi}_{ijk}$ is 
  $\delta(P,\Pi_{ijk})=au_p+bv_p+wz_p+d$ and the sphere 
  $\inv{S_n}$, for $ n\in\{i,j,k\}$, touches $\inv{\Pi}_{ijk}$ and lies on 
  its positive side, we get that 
  $\delta(\inv{C_n},\Pi_{ijk})=au_n+bv_n+wz_n+d=\rho_n$. 

  A tuple $(a,b,c,d)$ that satisfies the resulting 
  system of equations, 
  amounts to a tangent plane in \wspace and an Apollonius 
  sphere in \zspace. 
  In order for the condition 2 to be valid, 
  the point $\OO=(0,0,0)$ of the \wspace must lie on the positive
  side of $\Pi_{ijk}$ i.e. the signed distance of $\OO$ from the plane 
  $\Pi_{ijk}$ must be positive. Equivalently, we want 
  $\delta(\mathcal{O},\Pi_{ijk})= d $ to be
  positive, hence we are only interested in 
  the solutions $(a,b,c,d)$ that satisfy $d>0$. 

  The rest of this section is devoted to the algebraic analysis of the 
  aforementioned system of equations to determine the number of such 
  solutions with the minimum algebraic cost. The conclusion of our analysis
  is that such a task is possible by evaluating expressions of algebraic
  degree at most 8 (in the input quantities) yielding the lemma at the 
  end of this section. 

  Our main tool is Crammer's rule and therefore two major cases rise 
  during the analysis of the system
  \begin{align*}
  au_i+bv_i+cw_i &= \rho_i-d,\\
  au_j+bv_j+cw_j &= \rho_j-d,\\
  au_k+bv_k+cw_k &= \rho_k-d,\\
  a^2+b^2+c^2 &= 1.
  \end{align*}

  If $D^{uvw}_{ijk}\neq 0$, we can express $a,b$ and $c$ in
  terms of $d$ as follows
  \begin{equation*}
  a = \dfrac{D^{vw\rho}_{ijk}-d D^{vw}_{ijk}}{D^{uvw}_{ijk}}, \quad
  b = -\dfrac{D^{uw\rho}_{ijk}-d D^{uw}_{ijk}}{D^{uvw}_{ijk}}, \quad
  c = \dfrac{D^{uv\rho}_{ijk}-d D^{uv}_{ijk}}{D^{uvw}_{ijk}}.
  \end{equation*} 

  We will then substitute the expressions of $a,b$ and $c$ to the 
  equation $a^2+b^2+c^2=1$ and conclude that $d$ is a root
  of $M(d)=M_2d^2+M_1d+M_0$, where 
  \begin{align*}
  M_2 &= (D^{uv}_{ijk})^2+(D^{uw}_{ijk})^2+(D^{vw}_{ijk})^2, \\
  M_1 &= D^{vw\rho}_{ijk}D^{vw}_{ijk}+D^{uw\rho}_{ijk}D^{uw}_{ijk}
      +D^{uv\rho}_{ijk}D^{uv}_{ijk},\\
  M_0 &= (D^{vw\rho}_{ijk})^2+(D^{uw\rho}_{ijk})^2+(D^{uv\rho}_{ijk})^2
      -(D^{uvw}_{ijk})^2.
  \end{align*}

  The signs of $M_1$ and $M_0$ are determined using the  
  the following equalities
  \begin{align*}
  \sgn(M_1) 
  &= -\sgn(D^{vw\rho}_{ijk}D^{vw}_{ijk}+D^{uw\rho}_{ijk}D^{uw}_{ijk}
    +D^{uv\rho}_{ijk}D^{uv}_{ijk})\\
  &= -\sgn(E^{yzr}_{ijk}E^{yzp}_{ijk}+E^{xzr}_{ijk}E^{xzp}_{ijk}
    +E^{xyr}_{ijk}E^{xyp}_{ijk}),\\
  \sgn(M_0) 
  &= \sgn((D^{vw\rho}_{ijk})^2+(D^{uw\rho}_{ijk})^2+(D^{uv\rho}_{ijk})^2
    -(D^{uvw}_{ijk})^2)\\
  &= \sgn((E^{yzr}_{ijk})^2+(E^{xzr}_{ijk})^2+(E^{xyr}_{ijk})^2
    -(E^{xyz}_{ijk})^2),
  \end{align*}
  \noindent
  whereas $M_2$ always positive unless $E^{xyr}_{ijk}, E^{xzr}_{ijk}$ 
  and $E^{yzr}_{ijk}$ are all zero; in this case $M_2=0$.
  The expressions that appear in the evaluation of $M_2, M_1$ and 
  $M_0$ have maximum algebraic degree 7 in the input quantities.


  First, we shall consider the case $M_2\neq 0$; 
  this is geometrically equivalent to the non-collinearity of the 
  inverted centers $\inv{C_n}$ for $ n\in\{i,j,k\}$. Since $M_2$ 
  is assumed to be strictly positive, the quadratic $M(d)$ has 
  0, 1 or 2 real roots depending on whether the sign of the discriminant
  $\Delta_M = M_1^2-4M_2M_0$ is negative, zero or positive 
  respectively. We evaluate the discriminant $\Delta_M$ of $M(d)$ to be
  \begin{equation*}
  \Delta_M = 4 (D^{uvw}_{ijk})^2 \big((D^{uv}_{ijk})^2+(D^{uw}_{ijk})^2
   +(D^{vw}_{ijk})^2-(D^{u\rho}_{ijk})^2-(D^{v\rho}_{ijk})^2-(D^{w\rho}_{ijk})^2
   \big),
  \end{equation*}
  hence determining $\sgn(\Delta_M)$ requires 8-fold algebraic operations
  since
  \begin{equation*}
  \sgn(\Delta_M) = 
  \sgn \left( (E^{xyp}_{ijk})^2+(E^{xzp}_{ijk})^2+(E^{yzp}_{ijk})^2
   -(E^{xrp}_{ijk})^2-(E^{yrp}_{ijk})^2-(E^{zrp}_{ijk})^2 \right).
  \end{equation*}

  If $\Delta_M<0$, the predicate returns ``0'', whereas if $\Delta_M = 0$
  we get that $M(d)$ has a double root $d = -M_1/(2M_2)$; 
  the predicate returns ``1 double'' if $\sgn(d) = -\sgn(M_1)$ is positive, 
  otherwise it returns ``0''. Finally, if $\Delta_M>0$, $M(d)$ 
  has two distinct roots $d_1<d_2$, whose sign we
  check for positiveness. Using Vieta's rules and 
  since $\sgn(M_1) = -\sgn(d_1+d_2)$ and
  $\sgn(M_0) = \sgn(d_1)\sgn(d_2)$ we conclude that, 
  if $M_0$ is negative the predicate returns ``1''. 
  Otherwise, the predicate's outcome is ``0'' if $\sgn(M_1)$
  is positive or ``2'' if $\sgn(M_1)$ is negative. 

  The signs of $M_1$ and $M_0$ are determined using the  
  the following equalities
  \begin{align*}
  \sgn(M_1) 
  &= -\sgn(D^{vw\rho}_{ijk}D^{vw}_{ijk}+D^{uw\rho}_{ijk}D^{uw}_{ijk}
    +D^{uv\rho}_{ijk}D^{uv}_{ijk})\\
  &= -\sgn(E^{yzr}_{ijk}E^{yzp}_{ijk}+E^{xzr}_{ijk}E^{xzp}_{ijk}
    +E^{xyr}_{ijk}E^{xyp}_{ijk}),\\
  \sgn(M_0) 
  &= \sgn((D^{vw\rho}_{ijk})^2+(D^{uw\rho}_{ijk})^2+(D^{uv\rho}_{ijk})^2
    -(D^{uvw}_{ijk})^2)\\
  &= \sgn((E^{yzr}_{ijk})^2+(E^{xzr}_{ijk})^2+(E^{xyr}_{ijk})^2
    -(E^{xyz}_{ijk})^2).
  \end{align*}
  \noindent
  The expressions that appear in the evaluation of $M_1$ and 
  $M_0$ have maximum algebraic degree 7 in the input quantities.

  We now analyse the case where $M_2=0$. Since the quantities 
  $D^{uv}_{ijk}, D^{uw}_{ijk}$ and $D^{vw}_{ijk}$ are all zero, we 
  get that $a = D^{vw\rho}_{ijk}/D^{uvw}_{ijk}$
  $b = D^{uw\rho}_{ijk}/D^{uvw}_{ijk}$ and 
  $c = D^{uv\rho}_{ijk}/D^{uvw}_{ijk}$. Therefore, since 
  $d = \rho_i-(au_i+bv_icw_i)$ we can evaluate the sign of $d$ as
  \begin{align*}
   \sgn(d) &=\sgn(D^{uvw}_{ijk})
   \sgn(D^{uvw}_{ijk}\rho_i-D^{vw\rho}_{ijk}u_i-D^{uw\rho}_{ijk}v_i-D^{uv\rho}_{ijk}w_i)\\
   &= \sgn(E^{xyz}_{ijk})\sgn(E^{xyz}_{ijk}r^{\star}_i-E^{yzr}_{ijk}x^{\star}_i-E^{xzr}_{ijk}y^{\star}_i-E^{xyr}_{ijk}z^{\star}_i)
  \end{align*}
 
  The evaluation of the sign of $d$ therefore demands operations of algebraic degree 4 (in the input quantities). If $d>0$ the predicate 
  returns ``1'' otherwise it returns ``0''.

  Last, we consider the case $D^{uvw}_{ijk}=0$. We can safely assume that 
  at least one of the quantities $D^{uv}_{ijk},D^{uw}_{ijk}$ and $D^{vw}_{ijk}$ since otherwise the centers $\inv{C_n}$ for $ n\in\{i,j,k\}$ would be 
  collinear
  \footnote{If $D^{uv}_{ijk}=D^{uw}_{ijk}=D^{vw}_{ijk}=0$, 
  the projections of the points $\inv{C_n}$, $n\in\{i,j,k\}$, on all
  three planes $w=1$, $v=1$ and $u=1$ would form a flat triangle. 
  For each projection, this is equivalent to either 
  some of the projection points coinciding or all three being collinear. 
  Since the original centers $\inv{C_n}$ are distinct points for 
  $n\in\{i,j,k\}$, they must be collinear for such 
  a geometric property to hold.}
  ,yielding a contradiction.
  Assume without loss of generality that $D^{uv}\neq 0$, we can solve the 
  system of equations in terms of $c$ and we get that
  \begin{equation*}
  a = \dfrac{-D^{v\rho}_{ijk}+cD^{vw}_{ijk}}{D^{uv}_{ijk}},\quad
  b = \dfrac{D^{u\rho}_{ijk}-cD^{uw}_{ijk}}{D^{uv}_{ijk}},\quad
  d = \dfrac{D^{uv\rho}_{ijk}}{D^{uv}_{ijk}}.
  \end{equation*}

  If we substitute $a$ and $b$ in the equation $a^2+b^2+c^2=1$, 
  we get that $c$ is a root of a quadratic polynomial
  $L(c)=L_2c^2+L_1c+L_0$, where 
  \begin{align*}
  L_2 &= (D^{uv}_{ijk})^2+(D^{uw}_{ijk})^2+(D^{vw}_{ijk})^2,\\
  L_1 &= -2(D^{v\rho}_{ijk}D^{vw}_{ijk}+D^{u\rho}_{ijk}D^{uw}_{ijk}),\\
  L_0 &= (D^{u\rho}_{ijk})^2+(D^{v\rho}_{ijk})^2-(D^{uv}_{ijk})^2.
  \end{align*}
  We evaluate the discriminant of $L(c)$ to be
  \begin{equation*}
  \Delta_L = 4 (D^{xy}_{ijk})^2 \left[(D^{xy}_{ijk})^2+(D^{xz}_{ijk})^2+(D^{yz}_{ijk})^2
   -(D^{xr}_{ijk})^2-(D^{yr}_{ijk})^2-(D^{zy}_{ijk})^2 \right],
  \end{equation*}
  and therefore $\sgn(\Delta_L)=\sgn(\Delta_M)$. The evaluation of  
  $\sgn(\Delta_M)$ is known to require 8-fold algebraic operations 
  as shown in a previous case.

  If $\Delta_M<0$, the predicate returns ``0''; there is no tangent plane
  in the inverted space. Otherwise, we determine the sign of $d$ to be
  \begin{equation*}
  \sgn(d)=\sgn\big(D^{uv\rho}_{ijk}/D^{uv}_{ijk}\big)
         =\sgn\big(E^{xyr}_{ijk}/E^{xyp}_{ijk}\big)
         = \sgn(E^{xyr}_{ijk})\sgn (E^{xyp}_{ijk}) 
  \end{equation*}
  and the following cases arise
  \begin{enumerate}
  \item 
  If $\Delta_M=0$, the predicate returns ``0'' if $d<0$ or ``1'' if $d>0$.
  \item 
  If $\Delta_M>0$, the predicate returns ``0'' if $d<0$ or ``2'' if $d>0$.
  \end{enumerate}

  Note that in the last case, the algebraic degrees of $\Delta_M$ and 
  $d$ are 8 and 4 respectively and that, in every possible scenario, 8 
  was the maximum algebraic degree of any quantity we had to evaluate. 
  
  \begin{lemma}
  The \exist predicate can be evaluated by determining 
  the sign of quantities of algebraic degree at most 8 
  (in the input quantities).
  \end{lemma}


 \subsection{The \distance predicate} 
  \label{sub:the_distance_predicate_analysis}

 In this section, we provide a detailed analysis regarding the evaluation of 
 the $\text{\distance}(S_i,S_j,S_k,S_a)$ predicate, as this was defined 
 in Section~\ref{ssub:the_distance_predicate}. As stated there, the outcome 
 of this primitive is the tuple 
 $(\sgn{(\delta(S_a,\Pi_{ijk}^{-}))},\sgn{(\delta(S_a,\Pi_{ijk}^{+}))})$, 
 where the planes $\Pi_{ijk}^{-}$ and $\Pi_{ijk}^{+}$ are commonly tangent 
 to the sites $S_i,S_j$ and $S_k$. The existence of these planes is 
 guaranteed since the trisector $\tri{ijk}$ is assumed to be ``hyperbolic'' 
 (see Section~\ref{ssub:the_incone_and_tritype_predicates}). Also take into
 consideration that in the scope of this paper, this subpredicate never
 returns a degenerate answer \ie, neither $(\sgn{(\delta(S_a,\Pi_{ijk}^{-}))}$ nor $\sgn{(\delta(S_a,\Pi_{ijk}^{+}))})$ can equal zero.

 We shall now consider such a plane $\Pi: ax+by+cz+d=0$ tangent to all
 sites $S_i,S_j$ and $S_k$, that leaves them all on the same side 
 (this site is denoted as the \emph{positive} side). 
 If we assume without loss of generality
 that $a^2+b^2+c^2=1$, it must stand that
 $\delta(C_n,\Pi)=ax_n+by_n+cz_n+d=r_n$ for $ n\in\{i,j,k\}$, where 
 $\delta(C_n,\Pi)$ denotes the signed Euclidean of the center 
 $C_n$ from the plane $\Pi$. If we consider the distance  
 $\epsilon=\delta(S_a,\Pi)=\delta(C_a,\Pi)-r_a$ of the sphere $S_a$ from 
 this plane, then the following system of equations must hold.
 
 \begin{align*}
 ax_i+by_i+cz_i +d &= r_i, \\
 ax_j+by_j+cz_j +d &= r_j, \\
 ax_k+by_k+cz_k +d &= r_k, \\
 ax_a+by_a+cz_a +d &= r_a+\epsilon, \\
 a^2+b^2+c^2=1
 \end{align*}

 Due to the initial assumption of a hyperbolic trisector, 
 only two such planes, $\Pi^{-}_{ijk}$ and $\Pi^{+}_{ijk}$, are cotangent 
 to the spheres $S_i,S_j$ and $S_k$ and therefore
 algebraicly satisfy the system of equations above.
 In other words, there exist only two distinct algebraic solutions
 $(a_\nu,b_\nu,c_\nu,d_\nu,\epsilon_\nu)$ for $\nu\in\{1,2\}$, 
 and apparently $\{\epsilon_1,\epsilon_2\}= \{\delta_a^+,\delta_a^-\}$, where $\delta_a^+=\delta(S_a,\Pi^{+}_{ijk})$ and $\delta_a^-=\delta(S_a,\Pi^{-}_{ijk})$.

 Bearing in mind that the answer to the \distance predicate is 
 actually the tuple $(\delta_a^+,\delta_a^-)$, 
 we want to determine the signs of $\epsilon_1$ and $\epsilon_2$ and 
 correspond them to $\delta_a^+$ and $\delta_a^-$. 
 Regarding the signs of $\{\epsilon_1,\epsilon_2\}$ 
 we algebraicly study the system of equations above. 

 First, we consider the case $D^{xyz}_{ijka}\neq 0$ in detail. Under this assumption and with the use of Crammer's rule, we may express 
 $a, b$ and $c$ with respect to $\epsilon$ as
 \begin{align*}
 a = \dfrac{D^{yzr}_{ijka}-\epsilon D^{yz}_{ijk}}{D^{xyz}_{ijka}}, \quad
 b = \dfrac{-D^{xzr}_{ijka}+\epsilon D^{xz}_{ijk}}{D^{xyz}_{ijka}},\quad
 c = \dfrac{D^{xyr}_{ijka}+\epsilon D^{xy}_{ijk}}{D^{xyz}_{ijka}}. \quad
 \end{align*}

  


 After substituting these expressions in the equation $a^2+b^2+c^2=1$,
 we obtain that 
 $\Lambda(\epsilon)=\Lambda_2\epsilon^2+\Lambda_1\epsilon+\Lambda_0=0$, 
 \begin{align*}
 \Lambda_2 &= (D^{xy}_{ijk})^2+(D^{yz}_{ijk})^2+(D^{xz}_{ijk})^2, \\
 \Lambda_1 &= -2(D^{yzr}_{ijka}D^{yz}_{ijk}+D^{xzr}_{ijka}D^{xz}_{ijk}
              -D^{xyr}_{ijka}D^{xy}_{ijk} ), \\
 \Lambda_0 &= (D^{xyr}_{ijka})^2+(D^{xzr}_{ijka})^2+(D^{yzr}_{ijka})^2
              -(D^{xyz}_{ijka})^2.
 \end{align*}

 Take into consideration that $\Lambda_2$ cannot be zero since otherwise
 the centers $C_i,C_j$ and $C_k$ would be collinear, yielding 
 a contradiction; we have assumed that $\tri{ijk}$ is 
 hyperbolic. Since $\Lambda(\epsilon)$, which is definitely a 
 quadratic in terms of $\epsilon$, has the aforementioned 
 $\epsilon_1$ and $\epsilon_1$ as roots, we may use Vieta's formula 
 to determine the signs of $\epsilon_1,\epsilon_2$. All we need is
 the signs of $\Lambda_1$ and $\Lambda_0$, quantities of algebraic degree 
 5 and 6 respectively (we already proved $\Lambda_2$ is 
  positive).

 If both roots are positive, negative or zero, the predicate returns 
 $(+,+)$, $(-,-)$ or $(0,0)$ respectively. If the sign
 of the roots differ, we must consider a way of distinguishing which 
 of them corresponds to $\delta_a^+$ and $\delta_a^-$. Since the 
 $\epsilon_1$ and $\epsilon_2$ have different signs, the set 
 $\{\epsilon_1,\epsilon_2\}=\{\delta_a^+,\delta_a^+\}$ is one of the 
 following: $\{+,-\},\{0,-\}$ or $\{0,+\}$. 

 The cases we now consider are products of geometric observations 
 based on the three possible configurations of the centers $C_i, C_j, C_k$
 and $C_a$:
 \begin{itemize}
  \item  If $D^{xyz}_{ijka}$ is positive, then $C_a$ lies 
 on the positive (resp. negative) side of the plane $\Pi_{ijk}$. 
 In this case, only the geometric configurations
 $(\delta_a^-,\delta_a^+)= (+,-),(0,-)$ 
 or $(+,0)$ are possible. 
 \item  If $D^{xyz}_{ijka}$ is negative, then $C_a$ lies
 on the negative side of the plane $\Pi_{ijk}$. 
 In this case, only the geometric configurations
 $(\delta_a^-,\delta_a^+)= (-,+),(-,0)$ 
 or $(0,+)$ are possible. 
 \end{itemize}

 
 For example, if the roots of $\Lambda(\epsilon)$ turn out to be 
 one positive and one negative, the predicate will return $(+,-)$ if 
 $D^{xyz}_{ijka}$ is positive or $(-,+)$ if 
 $D^{xyz}_{ijka}$ is negative. 
 
 Lastly, we consider the case $D^{xyz}_{ijka}=0$, where the 
 centers $C_n$, for 
 $n\in\{\allowbreak i,\allowbreak j,\allowbreak k,\allowbreak a\}$, 
 are coplanar. In this context, 
 it is both algebraicly and geometricaly apparent that 
 $\epsilon_1=\epsilon_2=\epsilon$; the sphere $S_a$ either intersects, is tangent or does not intersect both planes $\Pi_{ijk}^-$ and $\Pi_{ijk}^+$.
 Specifically, if $D^{xyz}_{ijk}\neq 0$, then 
 $\epsilon= -{D^{xyzr}_{ijka}}/{D^{xyz}_{ijk}}$ and 
 we immediately evaluate 
 $\sgn(\epsilon)=-\sgn(D^{xyzr}_{ijka})\cdot\allowbreak\sgn(D^{xyz}_{ijk})$.
 If $D^{xyz}_{ijk}=0$ then at least one of the quantities 
 $D^{xy}_{ijk},D^{xz}_{ijk},D^{yz}_{ijk}$ does not equal zero, 
 since the centers $C_n$ for $n\in\{i,j,k\}$ are not collinear. 
 Assume without loss of generality that $D^{xy}_{ijk}\neq 0$, then
 $\epsilon={D^{xyr}_{ijka}}/{D^{xy}_{ijk}}$ and 
 we evaluate $\epsilon=\sgn(D^{xyr}_{ijka})\sgn(D^{xy}_{ijk})$. 
 In every case the predicate returns $\sgn(\epsilon),\sgn(\epsilon)$. 
 
 Taking into consideration that the evaluation of $\Lambda_1$ is the most 
 degree-demanding operation to resolve the \distance predicate, we 
 have proven the following lemma.
 
 \begin{lemma}
 The \distance predicate can be evaluated by determining 
 the sign of quantities of algebraic degree at most 6 
 (in the input quantities).
 \end{lemma}

 \subsection{The \shadow Predicate} 
 \label{sub:the_shadowregion_predicate_analysis}
 In this section, we provide a way of resolving the 
 \shadow$(S_i,S_j,S_k,S_\alpha)$ predicate as this was described in 
 Section~\ref{ssub:the_shadow_predicate}. Assuming that  the trisector
 $\tri{ijk}$ is ``hyperbolic'' and no degeneracies occur, we have shown 
 that the outcome, which is the topological structure of $\sh{S_a}$ on 
 $\tri{ijk}$, is one of the following: $\emptyset$, $(-\infty,\infty)=\RR$, 
 $(-\infty,\phi)$, $(\chi,+\infty)$, $(\chi,\phi)$, or 
 $(-\infty,\phi)\cup(\chi,+\infty)$, where $\phi,\chi\neq\pm\infty$.

 Initially, the predicates $\text{\exist}(S_i,S_j,S_k,S_\alpha)$ and
 $\text{\distance}(S_i,S_j,S_k,S_\alpha)$ are called; let us 
 denote by $E$ and $(\sigma_1,\sigma_2)$ their respective outcomes. 
 The geometric interpretation of these two quantities leads to the 
 resolution of the shadow region $\sh{S_\alpha}$ with respect to 
 $\tri{ijk}$. 

 Regarding the meaning of the signs $\sigma_1$ and $\sigma_2$, 
 assume that $\sigma_1=-$. In this case, the sphere $S_\alpha$ does not 
 intersect the plane $\Pi_{ijk}^-$ which is in fact the Apollonius sphere
 $\tts{\invmap{-\infty}}$. In other words, $\invmap{-\infty}$ 
 does not belong to $\sh{S_\alpha}$ \ie, $-\infty$ does not ``shows up'' in the predicate's outcome. 
 Using a similar argument, if $\sigma_2=-$ then $+\infty$ 
 shows up in the outcome whereas, if $\sigma_1$ or
 $\sigma_2$ is positive then $-\infty$ or 
 $+\infty$ shows up respectively. Note that $\sigma_1,\sigma_2\in\{+,-\}$ 
 under the assumption of no degeneracies.

 Regarding the geometric interpretation of $E$, we have mentioned that 
 the boundary points of $\sh{S_\alpha}$ correspond to centers of 
 Apollonius spheres of the sites $S_n$, for $n\in\{i,j,k,\alpha\}$
 that are not centered at infinity. 
 We have shown in Section~\ref{sub:the_existence_predicate_analysis} that the
 cardinality of these Apollonius spheres is in fact $E$ and assuming no 
 degeneracies there are either 0, 1 or 2. 

 The combined information of $\sigma_1, \sigma_2$ and $E$ is used to 
 determine the type of $\sh{S_\alpha}$, as follows:
 \begin{itemize}
 \item 
 If $E=0$, the shadow region $\sh{S_\alpha}$ has no boundary 
 points, hence it's type is either $(-\infty,+\infty)$ 
 or $\emptyset$. If $\sigma_1=-$ (or $\sigma_2=-$), the predicate 
 returns $(-\infty,+\infty)$ otherwise, if $\sigma_1=+$ 
 (or $\sigma_2=+$), it returns $\emptyset$.
 \item 
 If $E=1$, then $\sh{S_\alpha}$ has one boundary 
 points, hence it's type is either $(-\infty,\phi)$ 
 or $(\chi,+\infty)$. If $\sigma_1=-$ (or $\sigma_2=+$), the
 predicate returns $(-\infty,\phi)$ otherwise, 
 if $\sigma_1=+$ (or $\sigma_2=-$) it returns $(\chi,+\infty)$).
 \item 
 Finally, if $E=2$ then $\sh{S_\alpha}$ has two boundary 
 points, hence it's type is either $(\chi,\phi)$ 
 or $(-\infty,\phi)\cup(\chi,+\infty)$. If 
 $\sigma_1=+ $ (or $\sigma_2=+$) the predicate returns 
 $(\chi,\phi)$ otherwise, if $\sigma_1=-$ (or $\sigma_2=-$), it 
 returns $(-\infty,\phi)\cup(\chi,+\infty)$.
 \end{itemize}

 Since the evaluation of the \shadow predicate only requires the call 
 of the \distance and the \exist predicates, which demand operations 
 of maximum algebraic degree 6 and 8 respectively, we proved the 
 following lemma.
 
 \begin{lemma}
 The \shadow predicate can be evaluated by determining 
 the sign of quantities of algebraic degree at most 8 
 (in the input quantities).
 \end{lemma}

 \subsection{The \order predicate} 
 \label{sub:the_order_predicate_analysis}

  The major subpredicate called during the evaluation of the 
  \conflict predicate via the algorithm described in 
  Section~\ref{sub:the_main_algorithm} is the so called \order predicate.
  As already described in Section~\ref{ssub:the_order_predicate}, 
  the \order$(S_i,S_j,S_k,S_a,S_b)$ returns the order of appearance 
  on the oriented trisector $\tri{ijk}$ of any Apollonius vertices 
  defined by the sites $S_i,S_j,S_k$ and $S_a$ or $S_b$. This primitive
  needs to be evaluated only if both $\sh{S_a}$ and $\sh{S_b}$ are 
  not of the form $\emptyset$ or $\RR$, with $\{a,b\}\in\{l,m,q\}$. 
  Indeed, for $a\in\{l,m\}$ we know that $v_{ijkl}$ and $v_{ikjm}$ exist and therefore $\sh{S_l}$ and $\sh{S_m}$ can not be of that type. Moreover, 
  $\sh{S_q}$ can not be of this form since otherwise, the \order predicate 
  is not called at all.
  To conclude, the predicates \shadow$(S_i,S_j,S_k,S_n)$ for $n\in\{a,b\}$ are called in advance and their outcomes are considered to be known for the rest of the analysis. As already mentioned, $\tri{ijk}$ is also assumed to be a ``hyperbolic'' trisector.

  The analysis of the \order predicate consists most of our breakthrough 
  and contribution in this research area. We demonstrate the usefulness 
  of the inversion technique by proving the strong connection between the original and the inverted space. By exploiting this relation, we are able 
  to create useful tools which can also be used in both 2D and 3D Apollonius
  diagrams to improve results of the current bibliography. 

  The following section is organised as follows. In Section~\ref{ssub:the_w_space} we make an introduction to the inverted space 
  and some initial observations on how it is connected with the  
  original space. Afterwards, in Section~\ref{ssub:the_yspace_analysis}, 
  we define  a 2-dimensional sub-space of the inverted space, 
  to make useful geometric observations effortless. In the last 
  three sections, we break up our analysis of the \order predicate 
  according to the shadow region types $\sh{S_a}$ and $\sh{S_b}$. Ultimately, we prove the following lemma.

  \begin{lemma}
  The \order predicate can be 
  evaluated by determining the sign of quantities of algebraic 
  degree at most 10 (in the input quantities).
  \end{lemma}

  \subsubsection{The \texorpdfstring{\wspace}{W-space}} 
  \label{ssub:the_w_space}

  The original space where the sites $S_i,S_j,S_k,S_a$ and 
  $S_b$ lie is called the \zspace. However, most of our analysis
  is carried in the inverted \wspace, as defined in 
  Section~\ref{sub:inversion} with a slight modification.

  Specifically, observe that the definition of the \wspace 
  depends on the choice of the sphere $S_I$. Since a cyclic 
  permutation of the sites $S_i,S_j$ and $S_k$ does not alter the 
  outcome of the \order predicate, we assume that 
  $r_k=\min\{r_i,r_j,r_k\}$; otherwise we name
  exchange the sites so it does. We now select to invert \zspace
  ``through the sphere $S_k$'' \ie, we reduce the radii of 
  all initial sites by $r_k$ (and obtain $\mathcal{Z}^\star$-space)
  and then invert all points with $C_k$ as the pole. 

  In the resulting space, called the \wspace, is where most of our 
  geometric observations and analysis takes place. Notice that 
  when we reduce the sites $S_a$ or $S_b$ by 
  $r_k$ we may end up with a sphere of negative radius if 
  $r_a<r_k$ or $r_b<r_k$. Although the existence (or not) of 
  spheres with negative radius in \wspace makes the geometric 
  configurations quite different to handle, the algebraic methods 
  we present here can handle both cases without modifications. 
  For this reason, we shall assume for the rest of the Section that 
  all sites lying in \wspace have positive radii.

  The analysis that follows is based on the strong relation that 
  holds between the geometric configuration of the sites 
  $S_i,S_j,S_k,S_a$ and $S_b$ in \zspace and the
  corresponding configuration of the inverted sites 
  $\inv{S_i},\inv{S_j},\inv{S_a}$ and $\inv{S_b}$ in \wspace. 

  In \wspace, $\OO$ denotes the point $(0,0,0)$ which is the  
  image of the ``point at infinity'' of \zspace.  
  Given a point $p$ that lies on $\tri{ijk}$, $\tts{p}$ 
  denotes the external Apollonius sphere tangent to the
  sites  $S_i,S_j$ and $S_k$, centered at $p$.  
  Such a sphere $\tts{p}$ of \zspace corresponds in \wspace 
  to a plane, denoted $\itp{p}$, that is tangent to the 
  inverted sites $\inv{S_i}$ and $\inv{S_j}$ and therefore 
  tangent to the cone defined by them. 
  Notice that, since $\inv{S_i}$ and $\inv{S_j}$ are distinct 
  spheres of \wspace due to their pre-images $S_i$ and $S_j$ also 
  being distinct in \zspace, the cone $\cone(\inv{S_i},\inv{S_j})$ 
  is well defined. For the rest of this Section, we define $\wcone$
  to be the semi cone (or cylinder if $\rho_i=\rho_j$) 
  that contains $\inv{S_i}$ and $\inv{S_j}$.

  Let us observe what happens in \wspace when we consider this point 
  $p$ moving on $\tri{ijk}$ such that $\map{p}$ goes 
  from $-\infty$ towards $+\infty$. The corresponding plane
  $\itp{p}$ rotates remaining tangent to $\wcone$, with starting 
  and ending positions the planes denoted by $\itp{-\infty}$ 
  and $\itp{+\infty}$ respectively.

  It is obvious that these two planes correspond to the two 
  Apollonius spheres of $S_i,S_j$ and $S_k$ ``at infinity'' 
  \ie, the planes $\{\Pi_{ijk}^-,\Pi_{ijk}^+\}$.
  These planes of \zspace must be distinct in the case of a 
  ``hyperbolic'' trisector $\tri{ijk}$ as 
  shown in Sections~\ref{sub:the_incone_predicate_analysis} and 
  \ref{sub:the_distance_predicate_analysis}, and, as a result, 
  their images in \wspace must also be distinct. 

  Moreover, each of $\itp{-\infty}$ 
  and $\itp{+\infty}$ must go through $\OO$ because the their 
  pre-images are planes that go through the ``point at infinity'' 
  in \zspace. Combining these last two remarks, we conclude that 
  the points $\inv{C_i},\inv{C_j}$ and $\OO$ are not collinear
  and  $\OO$ lies strictly outside the semi cone $\wcone$. 
  It is of great importance to understand that the last fact holds only 
  because $\tri{ijk}$ is ``hyperbolic''; if we were studying the 
  ``elliptic'' trisector type, $\OO$ 
  would lie strictly inside the semi cone $\wcone$ and 
  in the degenerate case of a ``parabolic'' trisector, 
  $\OO$ would lie on the boundary of $\wcone$.

  For every point $p\in\tri{ijk}\backslash\{\pm\infty\}$, the sphere 
  $\tts{p}$ is an external Apollonius sphere and therefore 
  does not contain the ``point at infinity'' in \zspace. Correspondingly,
  its image in \wspace \ie, the plane $\itp{p}$,  must leave 
  the point $\OO$ and the centers of the spheres $\inv{S_i}$ and 
  $\inv{S_j}$ on the same side. This side of the plane$\itp{p}$ is  
  called \emph{positive} whereas the other is referred 
  to as \emph{negative}.

  Let us now consider the plane $\inv{\Pi}$ that goes through 
  the points $\inv{S_i},\inv{S_j}$ and the point $\OO$ of \wspace.
  The well-definition of $\inv{\Pi}$ follows from the  
  non-collinearity of the three points we proved earlier.
  This plane turns out to be the inversion image of the plane $\Pi_{ijk}$
  that goes through the centers $C_i, C_j$ and $C_k$ 
  (and apparently the point at infinity) in \zspace. 

  The latter plane separates \zspace into two half-spaces, 
  $\HH_+$ and $\HH_-$, where $\HH_+$ (resp. $\HH_-$) denotes 
  the set of points $N$ such that \orient$(N,C_i,C_j,C_k)$ 
  is positive (resp. negative). The plane $\inv{\Pi}$ also separates
  \wspace into two half-spaces, $\inv{\HH}_+$ and $\inv{\HH}_-$, 
  where $\inv{\HH}_+$ (resp. $\inv{\HH}_-$) denotes 
  the set of points $M$ such that \orient$(M,\inv{C_i},\inv{C_j},\OO)$
  is positive (resp. negative).

  If we now consider a point $C_n$ of \zspace and its inversion image
  $\inv{C_n}$ in \zspace, we can easily prove that
  \begin{align*}
  \text{\orient}(\inv{C_n},\inv{C_i},\inv{C_j},\OO)&=
  \sgn(\inv{p_i}\inv{p_j}\inv{p_n}D^{uvw}_{nij})=
  \sgn(D^{xyz}_{nijk})\\
  &= \text{\orient}(C_n,C_i,C_j,C_k).
  \end{align*}
  %
  This result indicates that $\inv{\HH_+}$ and $\inv{\HH_-}$ are 
  in fact the inversion images of the open semi-spaces 
  $\HH_+$ and $\HH_-$. 

  This simple correspondence of semi-spaces yields a remarkable
  result. If we consider a point $p\in\tri{ijk}^+$ (resp. $\tri{ijk}^-$),
  then the tangency points of the 
  Apollonius sphere $\tts{p}$ with the spheres $S_i$ and $S_j$ 
  must lie on  $\HH_+$ (resp. $\HH_-$).
  If this fact is considered through inversion, the tangency points 
  of the plane $\itp{p}$ with
  the semi cone $\wcone$ must lie on  $\inv{\HH}_+$ 
  (resp. $\inv{\HH}_-$). Furthermore, if we let $p$ move on 
  $\tri{ijk}^+$ (resp. $\tri{ijk}^-$) such that $\map{p}$ goes 
  to $+\infty$ (resp. $-\infty$), we can deduce that the plane
  $\itp{+\infty}$ (resp. $\itp{-\infty}$) 
  is the one that goes through the point $\OO$ and its tangency 
  points with the spheres $\inv{S_i}$ and $\inv{S_j}$ lie on 
  $\inv{\HH}_+$ (resp. $\inv{\HH}_-$).

  \subsubsection{The \texorpdfstring{\yspace}{Y-space}} 
  \label{ssub:the_yspace_analysis}

\begin{figure}[htbp]
   \centering
   \includegraphics[width=0.95\textwidth]{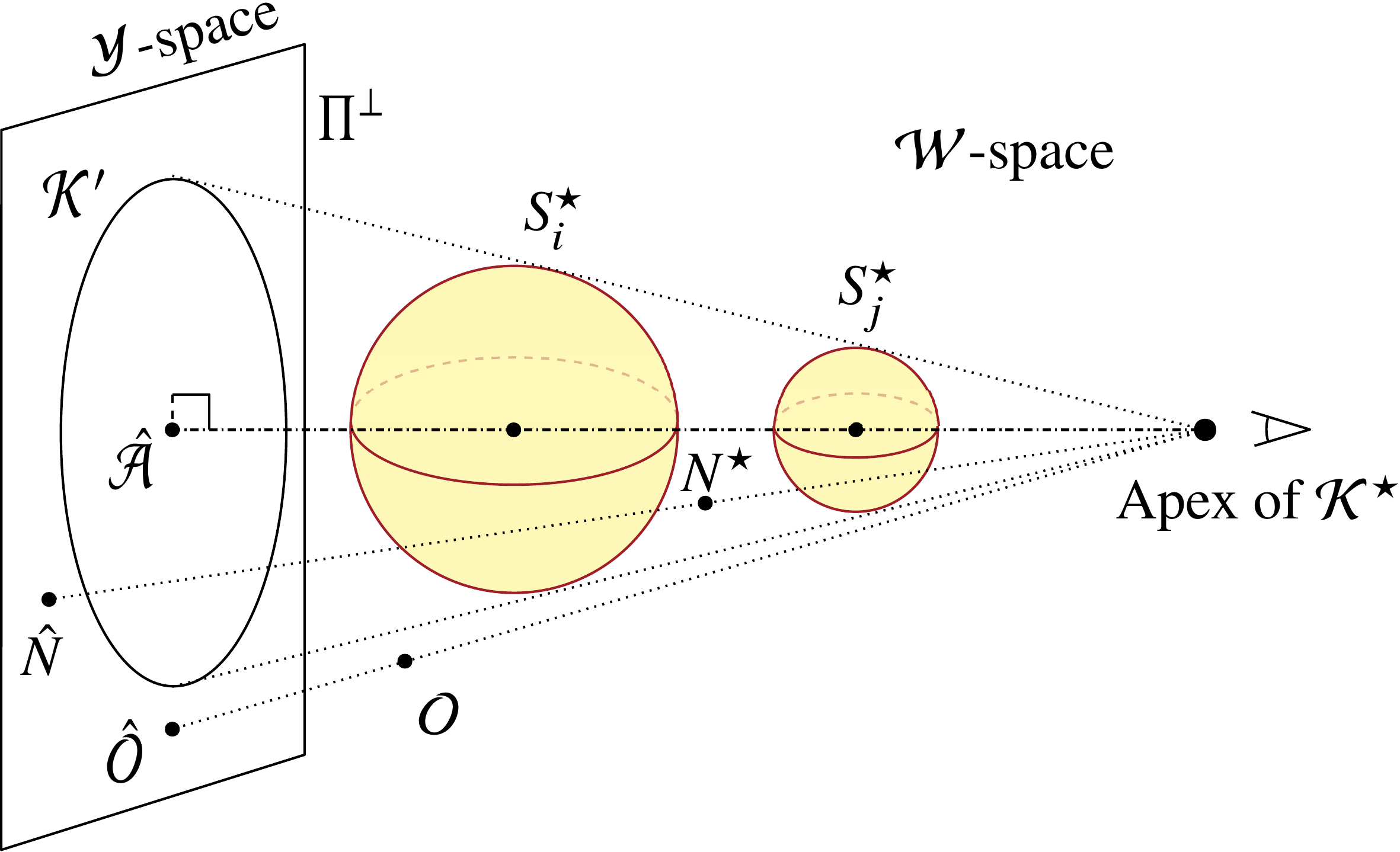}
   \caption{The \yspace, where most of the analysis of the 
   \order predicate is carried on, is in essence a projection of 
   \wspace to a plane $\Pi^\bot$ via the apex of the cone $\wcone$.  
   Since \yspace is a 2-dimensional space, an observation 
   regarding a \wspace geometric configuration is made easier 
   if we consider the corresponding configuration in \yspace. }
   \label{fig:08}
  \end{figure}

  All these observations are indicative of the strong connection 
  of the original and and the inverted space. However, since a 
  three-dimensional space such as \wspace makes observations and 
  case breakdowns too complex, we will now consider a sub-space to 
  carry our analysis. For this reason, we consider a (random) plane 
  $\Pi^\bot$ in \wspace that is perpendicular to the axis of the 
  semi-cone $\wcone$ at point $\yinv{\mathcal{A}}$ and 
  intersects it at a full circle $\ycone$;
  the intersection of \wspace and $\Pi^\bot$ is called the \yspace. 
  Notice that in every figure representing the \yspace, we always 
  depict the $\Pi^\bot$ plane such that the vector 
  $\ov{\inv{C_i}\inv{C_j}}$ points ``towards'' the reader 
  (see Figure~\ref{fig:08}). 

  In \yspace, we will use the following notation:
  \begin{itemize}
  \item  
  $\ill{p},\ill{\pm\infty},\yinv{\ell}$ 
  and $\ill{\oo}$  denote the intersection of the plane $\Pi^\bot$ with the planes 
  $\itp{p},\inv{\Pi}(\pm\infty),\inv{\Pi}$ and $\inv{\Pi}(\oo)$ 
  respectively.
  \item 
  $\yinv{\eta},\yinv{\theta},\yinv{o}$ and $\yinv{p}$ denote 
  the points of tangency of $\ycone$ and the lines 
  $\ill{-\infty}$, $\ill{+\infty}$, $\ill{\oo}$ and $\ill{p}$
  respectively (for $p\in\tri{ijk}$).
  \item 
  The intersection of $\inv{\HH}_+$ 
  (resp. $\inv{\HH}_-$) with the $\Pi^\bot$ plane is called 
  the positive (resp. negative) half-plane $\yinv{\HH}_+$
  (resp. $\yinv{\HH}_-$).
  \item
  The positive (resp. negative) side of the line
  $\ill{p}$ for a point $p\in\tri{ijk}$ to be the side that 
  contains (resp. does not contain) the point $\yinv{\mathcal{A}}$.  
  \item
  $\yinv{\OO}$ denotes the point of intersection of 
  the lines $\ill{-\infty}$ and $\ill{+\infty}$. 
  \end{itemize}

  We shall now prove an equivalency relation between the trisector 
  $\tri{ijk}$ and an arc of $\ycone$, which is the biggest idea 
  upon which the rest of our analysis is based.
  If a point $p$ moves on $\tri{ijk}$ such 
  that $\map{p}$ goes from $-\infty$ to $+\infty$ then, in \yspace, 
  the corresponding point $\yinv{p}$ moves on $\ycone$ from the 
  point $\yinv{\eta}$ to the point $\yinv{\theta}$, going
  through the point $\yinv{o}$. Observe that there is a 
  1-1 correspondence between the oriented trisector $\tri{ijk}$ and 
  the oriented arc $(\yinv{\eta} \yinv{o} \yinv{\theta})$. We denote this 
  1-1 and onto mapping from $\tri{ijk}$ to
  the arc $(\yinv{\eta} \yinv{o} \yinv{\theta})$ by $\arc{\cdot}$, such that $\arc{p}=\yinv{p}$. 
  
\begin{figure}[tbp]
  \centering
  \includegraphics[width=0.95\textwidth]{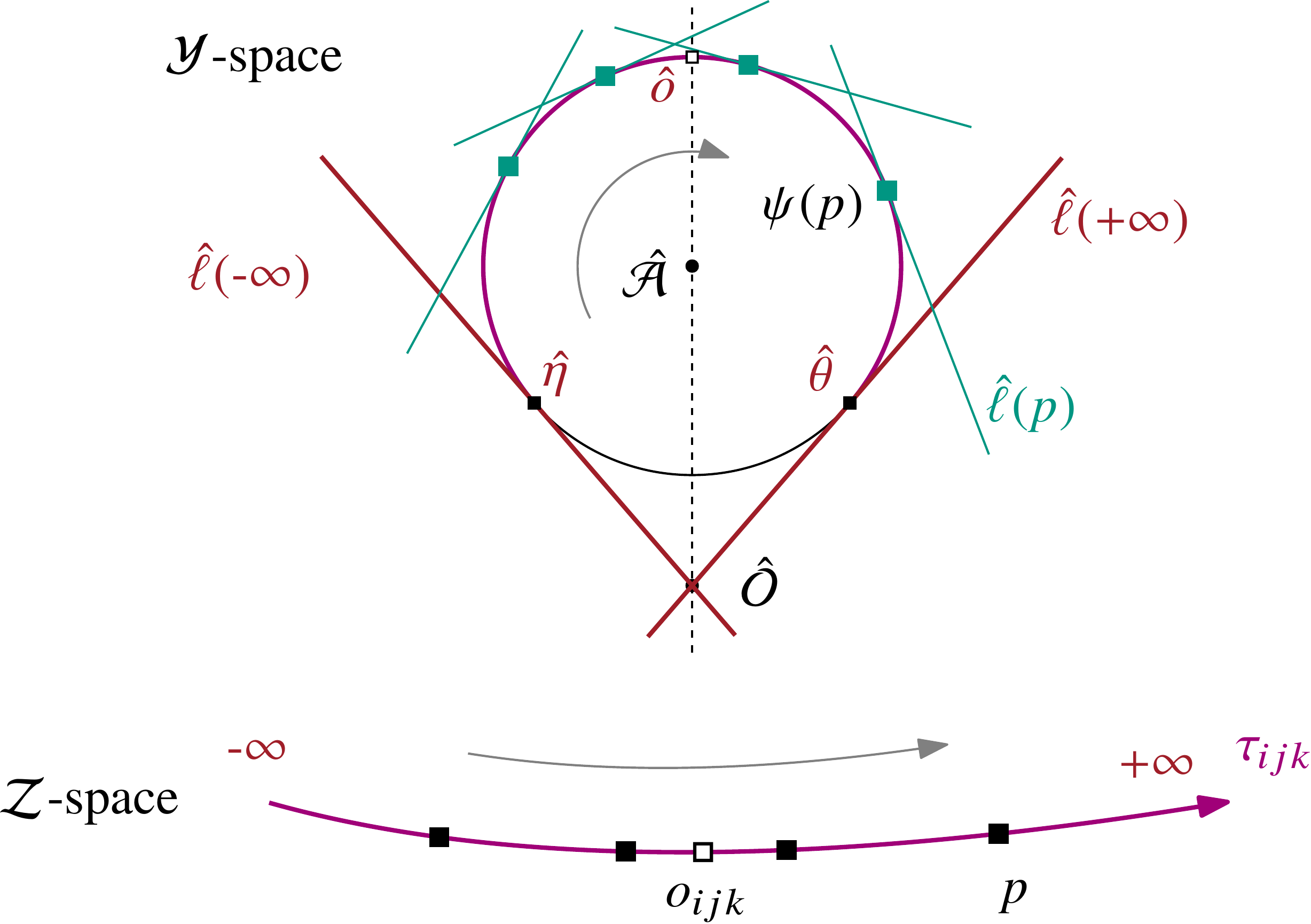}
  \caption{There is a strong relation between the \zspace 
  and the \yspace. When the point $p$ of \zspace traverses 
  the trisector $\tri{ijk}$ towards its positive direction, the corresponding point $\arc{p}$ of 
  \zspace traverses the arc $(\arc{\eta}, \arc{o}, \arc{\theta})$. 
  At the same time, $\ill{p}$ rotates in \yspace, remaining tangent 
  to $\cone'$, with starting and ending positions the lines 
  $\ill{-\infty}$ and $\ill{+\infty}$ respectively.}
  \label{fig:zspace_yspace_equivalency}
  \end{figure}

  What naturally follows is that the order of appearance of the 
  vertices $v_{ijka}$, $v_{ikja}$, $v_{ijkb}$ and $v_{ikjb}$ on the
  oriented trisector amounts to the order of appearance of the points 
  $\arc{v_{ijka}}$, $\arc{v_{ikja}}$, $\arc{v_{ijkb}}$ and $\arc{v_{ikjb}}$ 
  on the oriented arc $(\arc{\eta},\arc{o},\arc{\theta})$
  (see Figure~\ref{fig:zspace_yspace_equivalency}). 
  Consider however that we only need to order the Apollonius vertices that 
  actually exist among $v_{ijka},v_{ikja},v_{ijkb}$ and $v_{ikjb}$. 

  \begin{lemma}
  \label{lemma:ordering_equivalence}
  There is a 1-1 correspondence between the 
  order of appearance of the existing vertices among
  $v_{ijka}, v_{ikja}, v_{ijkb}$ and $v_{ikjb}$ on the oriented 
  trisector $\tri{ijk}$ and the order of appearance of the existing 
  points $\arc{v_{ijka}}$, $\arc{v_{ikja}}$, $\arc{v_{ijkb}}$ 
  and $\arc{v_{ikjb}}$ on the oriented arc 
  $(\arc{\eta},\arc{o},\arc{\theta})$.
  \end{lemma}

  The lemma suggests that the outcome of the \order predicate could 
  return the order of appearance of the images of the aforementioned Apollonius vertices on the arc $(\arc{\eta},\arc{o},\arc{\theta})$ 
  instead of the order of the original vertices on the trisector 
  $\tri{ijk}$. 


  Towards our goal of obtaining the ordering of the inverted 
  Apollonius vertices, we denote 
  the circle $\yinv{S_n}\in\text{\yspace}$ 
  for $n\in\{a,b\}$ which will be considered as the image 
  of $\inv{S_n}$ of \wspace. We need to define the image of these 
  spheres in a proper way such that they carry their geometric 
  properties from \wspace to \yspace. For this reason, we consider 
  the center $\yinv{C_n}$ of $\yinv{S_n}$ to be the 
  intersection of $\Pi^{\bot}$ with the line that goes through 
  the apex of the cone $\wcone$ and the center $\yinv{C_n}$. 
  Observe that such a line is well defined since the latter
  two points cannot coincide; if they did then $\sh{S_n}$ would be 
  $\emptyset$ or $\RR$ yielding a contradiction
  \footnote{In such a geometric configuration, there would not 
  exist a plane in \wspace co-tangent to all spheres 
  $\inv{S_i}$,$\inv{S_j}$ and $\inv{S_n}$. Equivalently, 
  in \zspace there would not exist an Apollonius sphere of 
  the sites $S_i$,$S_j$,$S_k$ and $S_n$ hence $\sh{S_n}$ would be 
  either $\emptyset$ or $\RR$ based on the analysis of 
  Section~\ref{sub:the_shadowregion_predicate_analysis}. }. Finally, 
  the radius of $\yinv{C_n}$ is such that $\yinv{C_n}$ is tangent 
  to each of the existing lines $\ill{v_{ijkn}}$ and $\ill{v_{ikjn}}$ 
  (at least one of them exists due to $\sh{S_n}$ not being $\emptyset$ 
  or $\RR$).

  Another crucial property we want to point out derives from the 
  inversion mapping we used to go from $\mathcal{Z}^\star$-space to 
  \wspace. The mapping $W(z)$ we used is known to be 
  \emph{inclusion preserving} \ie, the relative position of 
  two spheres in the original space is preserved in the inverted one. 
  For example, consider a sphere $S_\mu$ that 
  intersects the (existing) sphere Apollonius sphere
  $\tts{v_{ikjn}}$ (resp. $\tts{v_{ijkn}}$), for $n\in\{a,b\}$ in 
  \zspace. After reducing both spheres by $r_k$, their images 
  in $\mathcal{Z}^\star$ retain the same relative position \ie, they 
  intersect. 
  Applying the inversion mapping, it must stand that 
  the sphere $\inv{S_\mu}$ must intersect the negative side of
  $\inv{\Pi}(v_{ikjn})$ (resp. $\inv{\Pi}(v_{ijkn})$) since this 
  half space is precisely the inversion image of the interior of 
  $\tts{v_{ikjn}}$ (resp. $\tts{v_{ijkn}}$). Finally, 
  if we consider this configuration in \yspace, we deduce that
  $\yinv{S_\mu}$ must intersect the negative side of $\ill{v_{ikjn}}$
  (resp. $\ill{v_{ijkn}}$). 

\begin{figure}[htbp]
   \centering
   \includegraphics[width=0.7\textwidth]{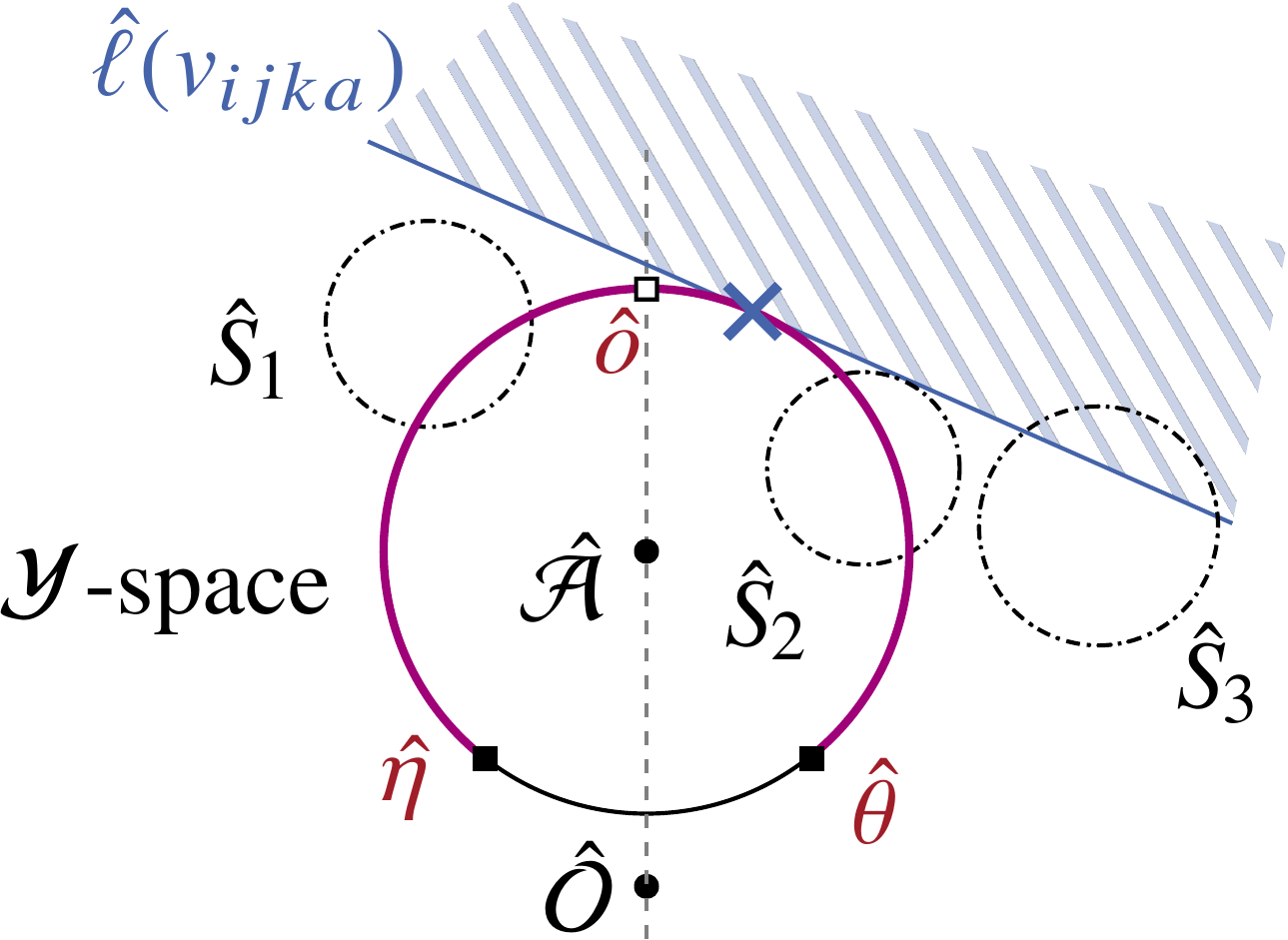}
   \caption{As Lemma~\ref{lemma:yinv_insphere_equiv} suggests, 
   the \insphere predicate with input $(S_i,S_j,S_k,S_a,S_n)$ 
   is positive, zero or negative for $n=0,1$ or $2$ respectively.}
   \label{fig:09b}
  \end{figure}

  In a similar way we can show that if $S_\mu$ 
  is tangent or does not intersect the sphere Apollonius sphere
  $\tts{v_{ikjn}}$ (resp. $\tts{v_{ijkn}}$) then, in \yspace, 
  $\yinv{S_\mu}$ is tangent to $\ill{v_{ikjn}}$
  (resp. $\ill{v_{ijkn}}$) or does not intersect its negative side (see Figure~\ref{fig:09b}).
  A fact tightly connected with these observations is that 
  the relative position of 
  $S_\mu$ and $\tts{v_{ikjn}}$ (resp. $\tts{v_{ijkn}}$) is provided
  by the \insphere predicate. Specifically,
  \begin{itemize}
   \item if \insphere $(S_i,S_k,S_j,S_n,S_m)$ is $-$, $0$ or $+$ then
   $S_m$ intersects, is tanget to or does not intersect the Apollonius
   sphere $\tts{v_{ikjn}}$ and,
   \item if \insphere $(S_i,S_j,S_k,S_n,S_m)$ is $-$, $0$ or $+$ then
   $S_m$ intersects, is tanget to or does not intersect the Apollonius
   sphere $\tts{v_{ijkn}}$.
  \end{itemize}
  Since these \insphere predicates can be evaluated as shown in 
  Section~\ref{sub:the_insphere_predicate}, 
  the relative position of $\yinv{S_\mu}$ with respect to
  any of the existing lines $\ill{v_{ikjn}}$ and $\ill{v_{ijkn}}$ can 
  be determined in \yspace, for $n\in\{a,b\}$.

  \begin{lemma}
  \label{lemma:yinv_insphere_equiv}
  The circle $\yinv{S_m}$ intersects, is tangent to or does 
  not intersect the negative side of $\ill{v_{ikjn}}$ 
  (resp. $\ill{v_{ijkn}}$) if and only if 
  the \insphere predicate with input $(S_i$,$S_k$,$S_j$,$S_n$,$S_m)$
   (resp. $(S_i$,$S_j$,$S_k$,$S_n$, $S_m)$ ) is negative, zero or 
  positive respectively. 
  \end{lemma}

  \subsubsection{The classic configuration} 
  \label{sub:The_classic_configuration}

  When the $\text{\order}(S_i, S_j, S_k, S_a, S_b)$ predicate is 
  called, we initially determine the shadow region types of 
  $\sh{S_a}$ and $\sh{S_b}$ via the appropriate \shadow predicates. 
  If the type of each shadow region is $(\chi,\phi)$ or $(\chi,+\infty)$ 
  (not necessary the same), we say that  we are in 
  \emph{classic configuration}. In such a setup, we can distinguish 
  simpler cases regarding the ordering the images of the Apollonius vertices 
  on the oriented arc $(\yinv{\eta} \yinv{o} \yinv{\theta})$.

  We therefore break up the analysis the \order predicate depending on 
  whe\-ther we are in a classic 
  (Section~\ref{sub:The_classic_configuration}) or 
  non-classic configuration (Section~\ref{sub:ordering_in_a_non_classic_configuration}), the latter 
  being reduced to the former using various observations.
  Let us now study in more detail what kind of information derives 
  from the fact that the sites $S_a$ and $S_b$ satisfy the conditions of a
  classic configuration. 

  Suppose $S_n$, for $n=a$ or $b$, is $(\chi,\phi)$ and therefore, 
  the endpoints $\{\chi,\phi\}$ must correspond to the two 
  Apollonius vertices $\{\map{v_{ijkn}},\map{v_{ikjn}})\}$ on
  the trisector $\tri{ijk}$ based on the remarks of 
  Section~\ref{ssub:the_shadow_predicate}. Let us consider
  $\chi$ and $\phi$ as $\map{v_n}$ and $\map{v_n^\prime}$ respectively, 
  with $\{v_n,v_n^\prime\}=\{v_{ijkn},v_{ikjn}\}$. Then, for every 
  $p\in\tri{ijk}$ such that $v_n\prec p\prec v_n^\prime$,
  the sphere $\tts{p}$ must intersect $S_n$ as this follows from the 
  definition of $\sh{S_n}$. 

  If we consider this point $p$ moving on $\tri{ijk}$, initialy starting 
  from the left-endpoint position $v_n$, then we observe that the 
  Apollonius sphere $\tts{v_n}$ intersects $S_n$ if we move its center
  infinitesimally towards the positive direction of $\tri{ijk}$. 
  Taking a closer look at the tangency points $T_i,T_j,T_k$ and $T_n$ 
  of $\tts{v_n}$ with the spheres $S_i,S_j,S_k$ and $S_n$ respectively, 
  and since the orientation of $\tri{ijk}$ is based in such a way 
  on the orientation of $C_i,C_j$ and $C_k$, it must hold that 
  $T_n$ must lie with respect to the plane formed by $T_i$, $T_j$ and 
  $T_k$ such that $T_iT_jT_kT_n$ be negative oriented. For that reason, 
  $v_n$ is in fact $v_{ikjn}$ and subsequently $v_n^\prime$ is $v_{ijkn}$.

  The same argument can be used to prove that if $\sh{S_n}$ is 
  $(\chi,+\infty)$ then $\chi$ corresponds to $\map{v_{ikjn}}$. 
  If we apply a similar analysis in all shadow region types that 
  contain a finite endpoint, it will lead to the following lemma.

  \begin{lemma}
  \label{lemma:phi_chi}
  If the type of the shadow region $\sh{S_n}$ of a sphere $S_n$ 
  on a hyperbolic trisector is one of the following: 
  $(-\infty,\phi)$, $(\chi,+\infty)$, $(\chi,\phi)$, or 
  $(-\infty,\phi)\cup(\chi,+\infty)$ where 
  $\phi,\chi\neq\pm\infty$, then $\chi\equiv\map{v_{ikjn}}$ 
  and $\phi\equiv\map{v_{ijkn}}$.
  \end{lemma}

  As the lemma suggest, in a classic configuration, 
  for $n\in\{a,b\}$,
  \begin{itemize}
   \item if $\sh{S_n}=(\chi,\phi)$, then 
   both $v_{ijkn}$ and $v_{ikjn}$ exist and $v_{ikjn}\prec v_{ijkn}$, 
   whereas
   \item if $\sh{S_n}=(\chi,+\infty)$, then 
   $v_{ikjn}$ exists while $v_{ijkn}$ does not. 
  \end{itemize}

\begin{figure}[htbp]
   \centering
   \includegraphics[width=0.7\textwidth]{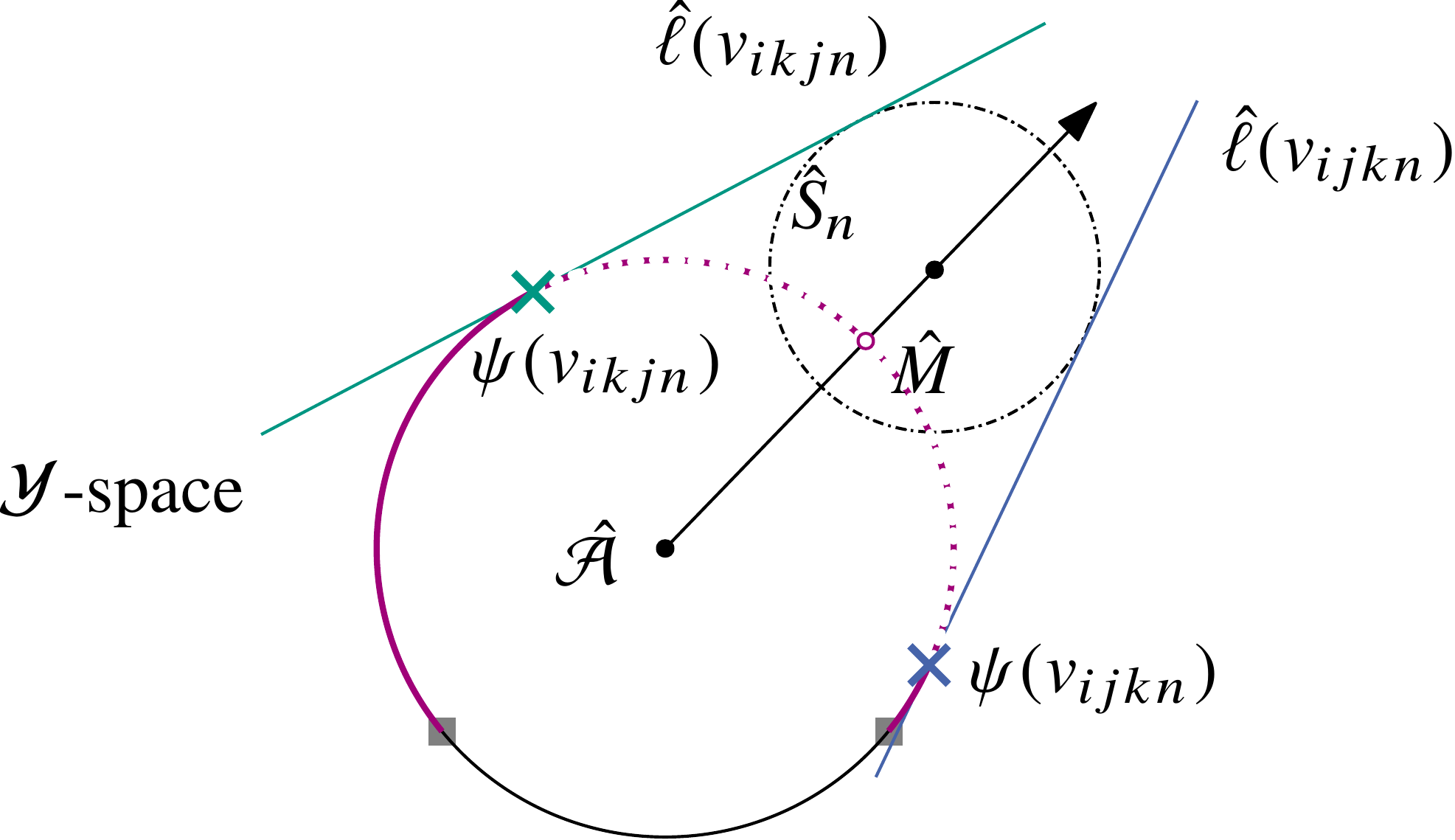}
   \caption{If both $v_{ikjn},v_{ijkn}$ exist and $\sh{S_a}$ 
   is $(\chi,\phi)$ with respect to the trisector $\tri{ijk}$, 
   $\yinv{C_n}$ must lie on the ray $(\yinv{\mathcal{A}},\yinv{M})$. 
   The dotted arc represent the image of $\sh{S_n}$ in 
   \yspace; it is obvious that, for any $\arc{p}$ in this arc, 
   $\yinv{S_n}$ intersects the negative side of the line 
   $\ill{p}$. As of Lemma~\ref{lemma:yinv_insphere_equiv},
   $S_n$ must intersect the sphere $\tts{p}$ which is equivalent 
   to $p\in\sh{S_n}$.}
   \label{fig:10a}
  \end{figure}

  An equally important result arises when pondering of the possible positions 
  of the circle $\yinv{S_n}$, for $n\in\{a,b\}$ with any of the 
  existing lines $\ill{v_{ikjn}}$ and $\ill{v_{ijkn}}$. Firstly, let us consider the scenario where $\sh{S_n}=(\chi,\phi)$
  and in consequence, both lines exist. In this case, both points 
  $\arc{v_{ikjn}}$ and $\arc{v_{ijkn}}$ exist on the oriented arc 
  such that $\arc{v_{ikjn}}\prec\arc{v_{ijkn}}$, as this follows from 
  all previous remarks. From the definition of  
  $\sh{S_n}=(\chi,\phi)$, it derives as a result that for a point 
  $p$ on the trisector $\tri{ijk}$ such that 
  $v_{ikjn}\prec p v_{ijkn}$, the sphere $\tts{p}$ intersects 
  with $S_n$. Using the ``inclusion preserving'' argument, it must 
  stand that, in \yspace, $\yinv{S_n}$ intersects with the negative side 
  of $\ill{p}$. Therefore, if $\yinv{M}$ is the midpoint $\arc{v_{ikjn}}$ 
  and $\arc{v_{ijkn}}$ on the arc $(\eta o \theta)$
  and $\mathbb{V}$ denotes the open ray from $\yinv{\mathcal{A}}$ towards
  $\yinv{M}$, then the circle $\yinv{S_n}$ must be centered at a point
  on $\mathbb{V}$ \ie, $\yinv{C_n}\in\mathbb{V}$ 
  (see Figure~\ref{fig:10a}).

\begin{figure}[htbp]
   \centering
   \includegraphics[width=0.7\textwidth]{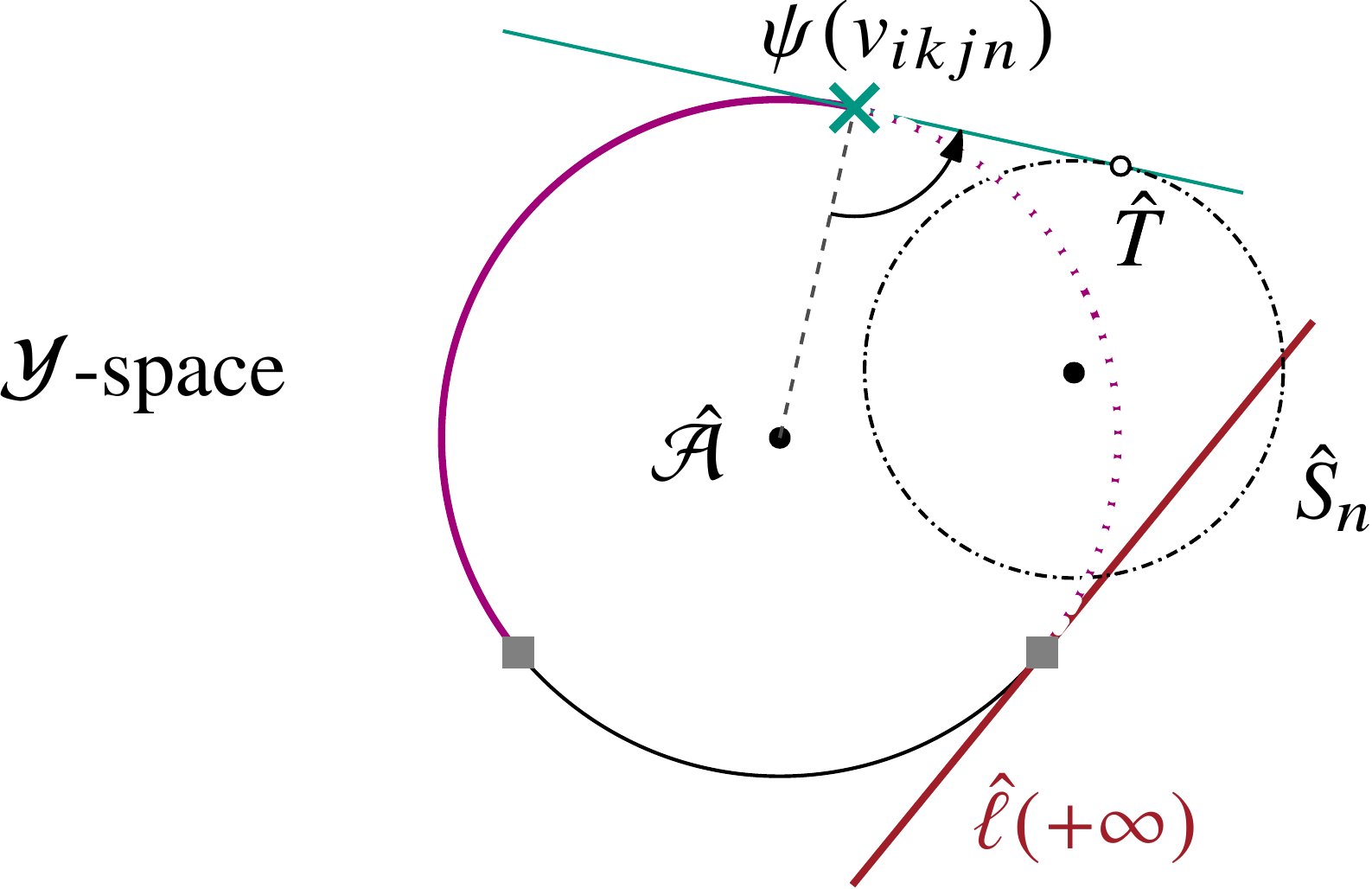}
   \caption{If $v_{ikjn}$ exists and $v_{ijkn}$ does not, 
   the shadow region of $S_n$ is known to be $(\chi,+\infty)$.
   This means that the tangency point $\yinv{T}$ of $\yinv{S_n}$
   with the line $\ill{v_{ikjn}}$ is on the side of the line 
   that forms a $90^\circ$ angle with the vector 
   $(\arc{v_{ikjn}},\yinv{\mathcal{A}})$. The image of 
   $\sh{S_n}$ in \yspace in this case is the dotted arc which is 
   indeed of the form $(\chi,+\infty).$ }
   \label{fig:10b}
  \end{figure}

  Lastly, let us examine the case where $\sh{S_n}=(\chi,+\infty)$. We begin 
  by observing that $S_n$ must intersect with $\Pi_{ijk}^+$ due to 
  the definition of the shadow region and this amounts, 
  in \yspace, to $\yinv{S_n}$ intersecting the negative 
  side of the line $\ill{+infty}$. Moreover, following a similar 
  analysis as in the case of $\sh{S_n}=(\chi,\phi)$, we come to 
  the conclusion that $\yinv{S_n}$ must be tangent to 
  $\ill{v_{ikjn}}$ at a point $\yinv{T}$ such that the (counterclowise)
  angle $(\yinv{\mathcal{A}},\arc{v_{ikjn}},\yinv{T})$ is $90^\circ$ 
  (not $270^\circ$, see Figure~\ref{fig:10b}). 
  This fact must hold for the line 
  $\ill{p}$ to intersect $\yinv{S_n}$, for every point 
  $p\in\tri{ijk}$ with 
  $\arc{v_{ikjn}}\prec \arc{p} \prec \yinv{\theta}$.

 \subsubsection{Ordering the Apollonius vertices in a classic configuration} 
  \label{sub:ordering_in_a_classic_configuration}
  
  In a classic configuration, we take for granted that, for 
  $n\in\{a,b\}$, either $\sh{S_n}=(\chi,\phi)$ and therefore 
  $v_{ikjn}\prec v_{ijkn}$ on the trisector $\tri{ijk}$ 
  or $\sh{S_n}=(\chi,+\infty)$ and 
  only $v_{ikjn}$ exists on $\tri{ijk}$. To order all of these 
  existing Apollonius vertices, we break down our analysis into 
  four sub-configurations.
  \begin{description}
   \item[Case A.] All vertices $v_{ikja},v_{ijka},v_{ikjb}$ and 
   $v_{ijkb}$ exist \ie, both $\sh{S_a}$ and $\sh{S_b}$ are 
   of type $(\chi,\phi)$.
   \item[Case B.] Only the vertices $v_{ikja}$ and 
   $v_{ikjb}$ exist \ie, both $\sh{S_a}$ and $\sh{S_b}$ are 
   of type $(\chi,+\infty)$.
   \item[Case C.] Only the vertices $v_{ikja},v_{ijka}$ and 
   $v_{ikjb}$ exist \ie, the type of $\sh{S_a}$ and $\sh{S_b}$ are 
   $(\chi,\phi)$ and $(\chi,+\infty)$ respectively.
   \item[Case D.] Only the vertices $v_{ikjb},v_{ijkb}$ and 
   $v_{ikja}$ exist \ie, the type of $\sh{S_a}$ and $\sh{S_b}$ 
   are $(\chi,+\infty)$ and $(\chi,\phi)$ respectively.
  \end{description}

  The last Case D is identical with the Case C if we name exchange the 
  spheres $S_a$ and $S_b$. Therefore, if Case D arises, we 
  evaluate \order$(S_i$, $S_j$, $S_k$, $S_b$, $S_a)$ instead, 
  which falls in Case C,
  and return the resulting ordering of $v_{ikjb},v_{ijkb}$ and $v_{ikja}$.
  Consequently, we only need to consider the Cases 
  A, B and C; the analysis of each configuration is deployed 
  separately in the following sections. 

  \paragraph*{\textbf{Analysis of Case A}}
   Given that all Apollonius vertices $v_{ijka},v_{ikja},v_{ijkb},v_{ikjb}$
   exist on $\tri{ijk}$ and  $v_{ikja}\prec v_{ijka}$ as well as 
   $v_{ikjb}\prec v_{ijkb}$ , the list of all possible orderings (and thus ouctomes of the \order predicate) is the following
   \begin{description}
   \item[\answer 1.] $v_{ikja}\prec v_{ijka}\prec v_{ikjb}\prec v_{ijkb}$,
   \item[\answer 2.] $v_{ikja}\prec v_{ikjb}\prec v_{ijka}\prec v_{ijkb}$,
   \item[\answer 3.] $v_{ikjb}\prec v_{ikja}\prec v_{ijka}\prec v_{ijkb}$,
   \item[\answer 4.] $v_{ikjb}\prec v_{ikja}\prec v_{ijka}\prec v_{ijka}$,
   \item[\answer 5.] $v_{ikjb}\prec v_{ijkb}\prec v_{ikja}\prec v_{ijka}$,
   \item[\answer 6.] $v_{ikja}\prec v_{ikjb}\prec v_{ijkb}\prec v_{ijka}$.
   \end{description}


  \begin{figure}[tbp]
   \centering
   \includegraphics[width=0.35\textwidth]{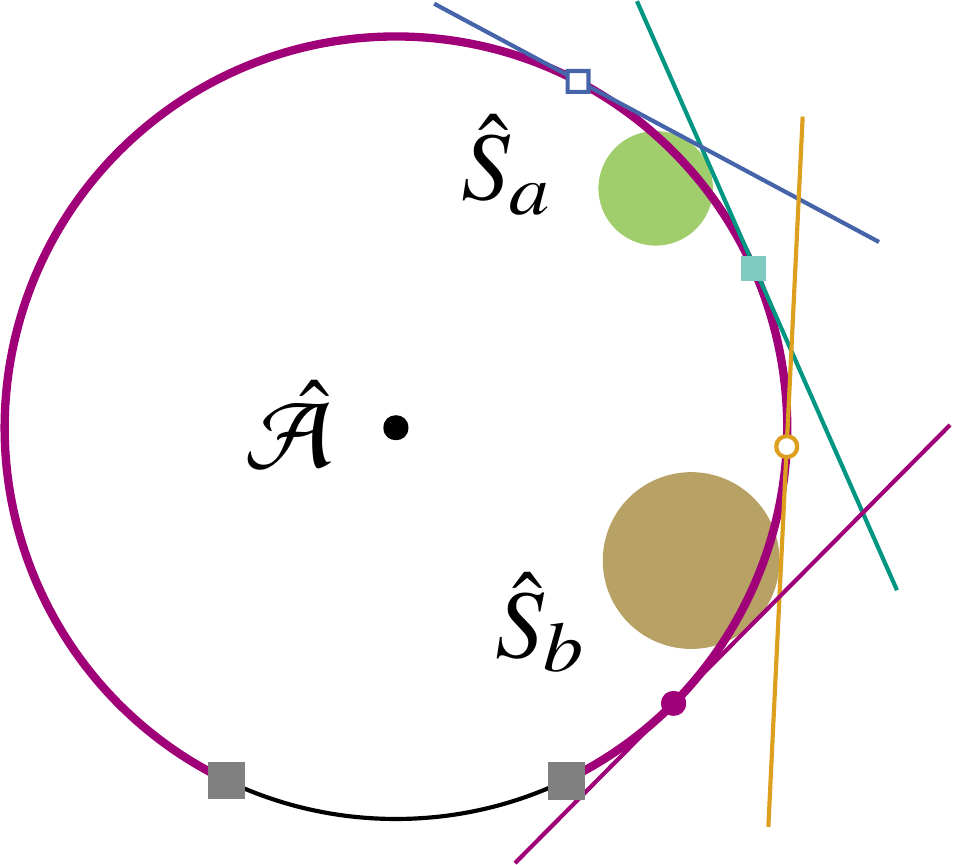}\hspace{1 cm}
   \includegraphics[width=0.35\textwidth]{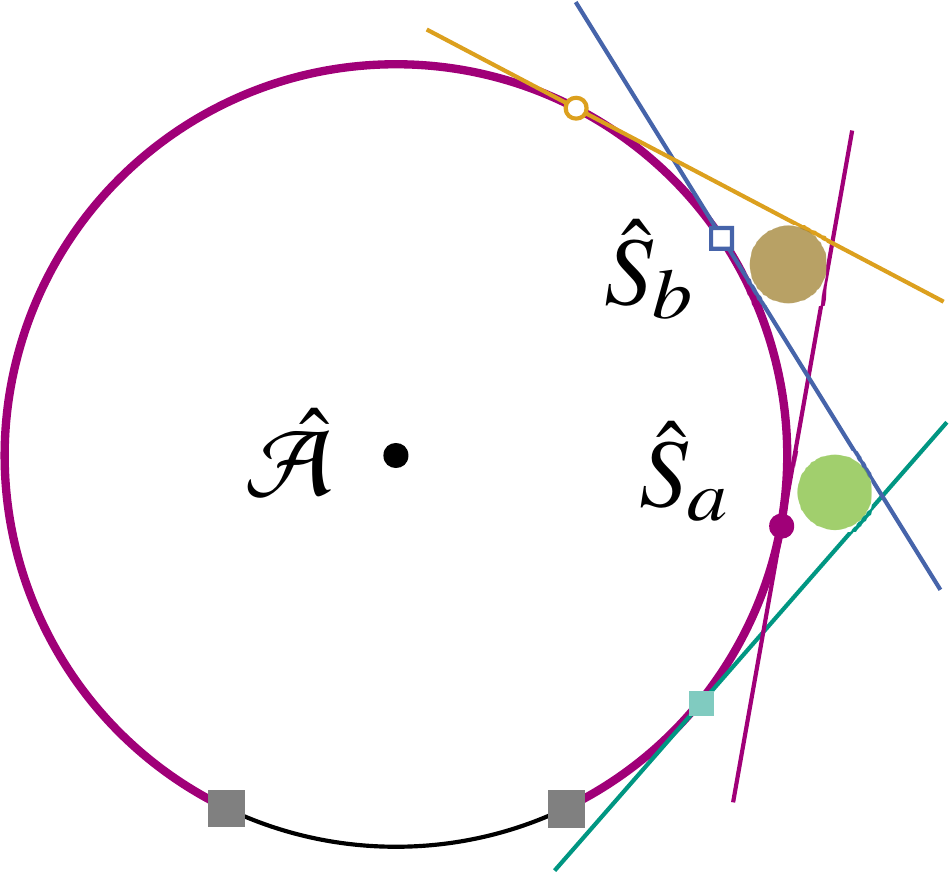}\\
   \includegraphics[width=0.35\textwidth]{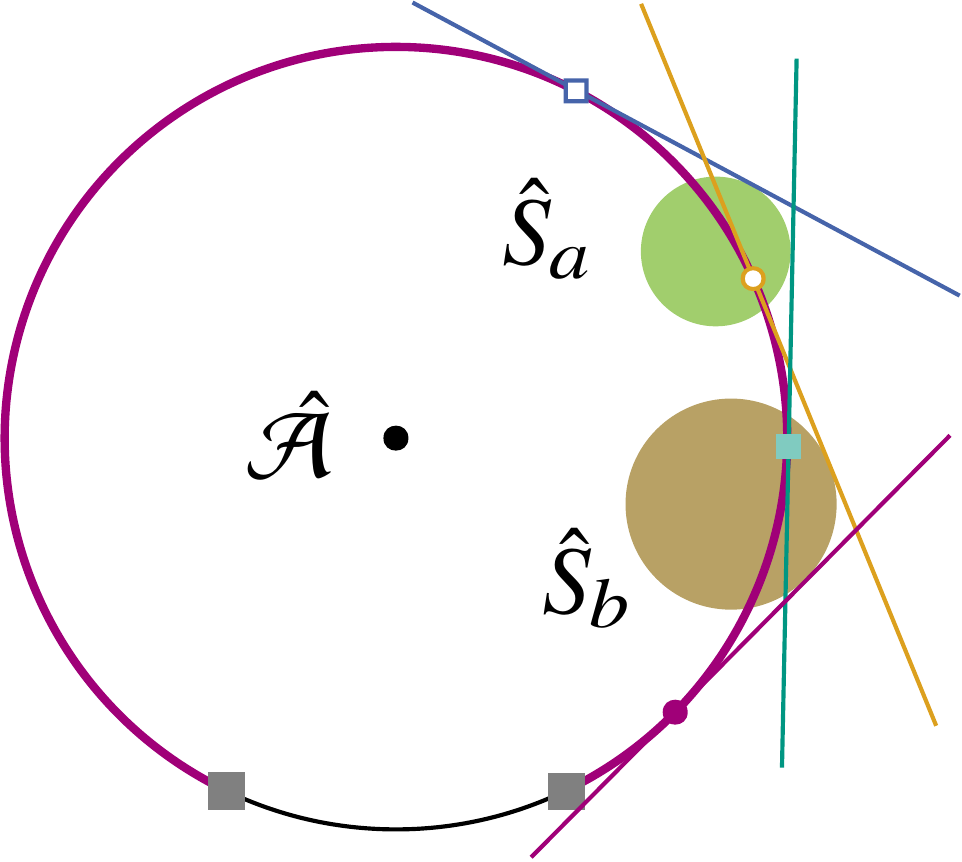}\hspace{1 cm}
   \includegraphics[width=0.35\textwidth]{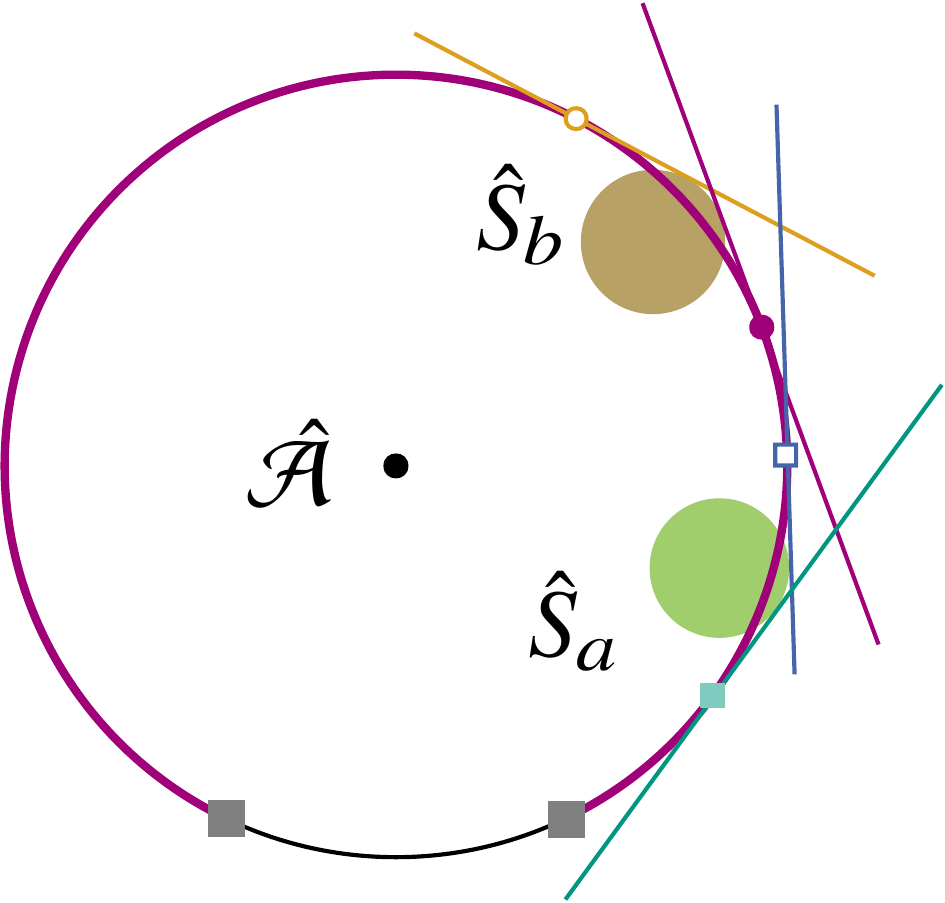}\\
   \includegraphics[width=0.35\textwidth]{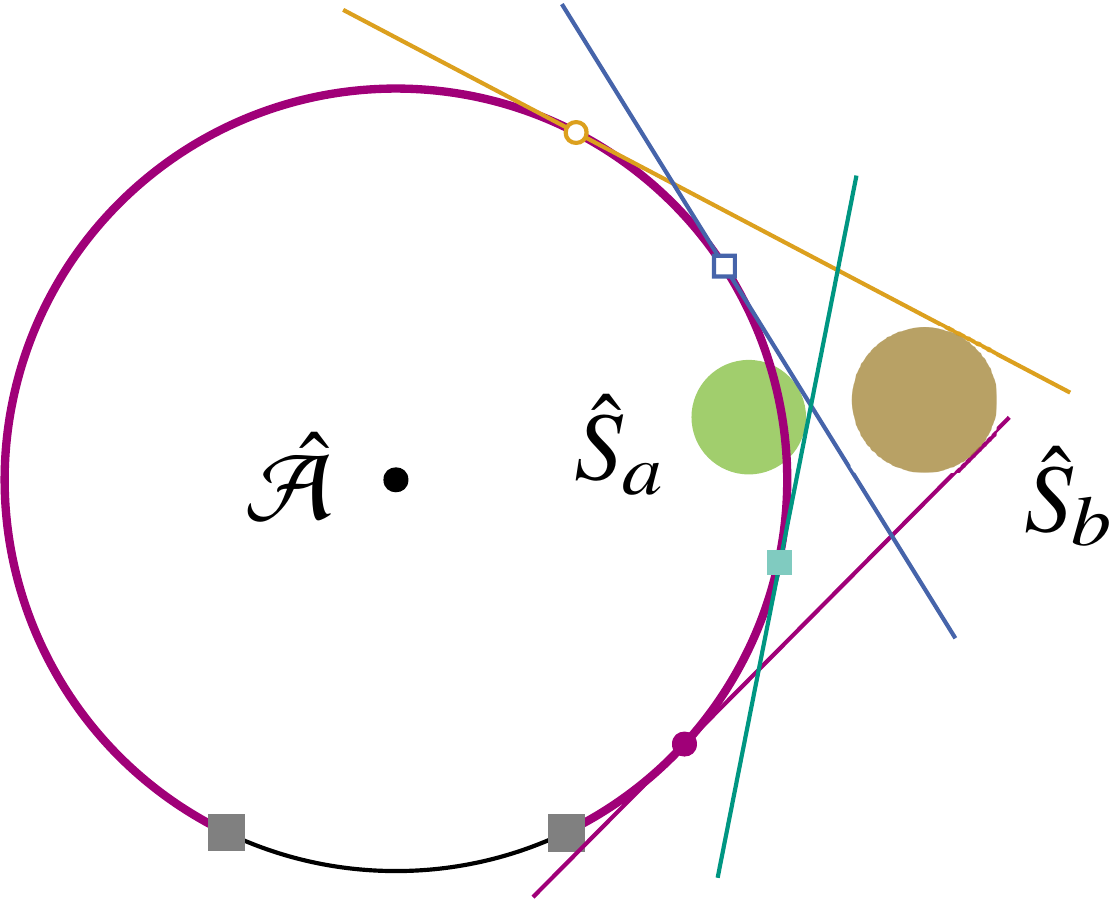}\hspace{1 cm}
   \includegraphics[width=0.35\textwidth]{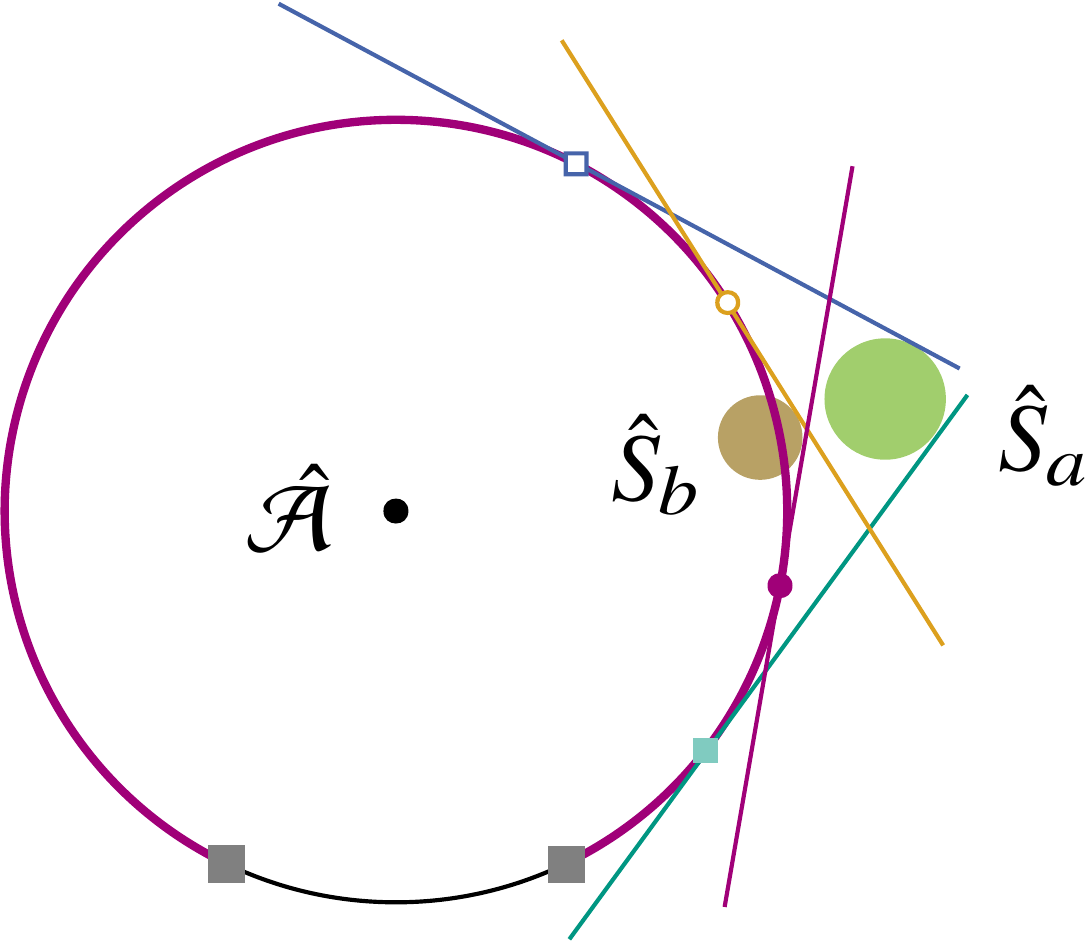}\\ \vspace{10pt}
   \includegraphics[width=0.8\textwidth]{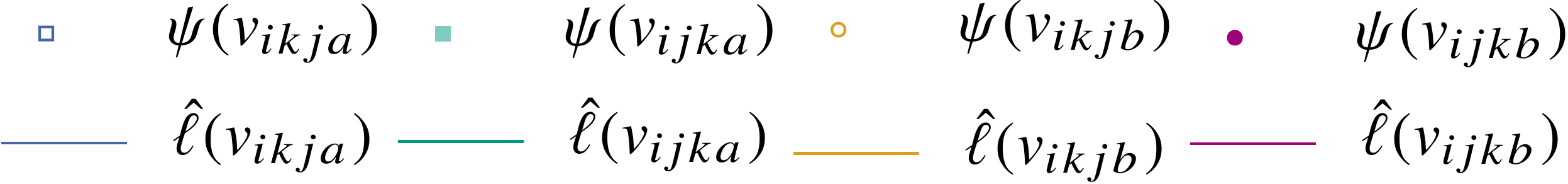}
   \caption{Under the assumption that all Apollonius vertices 
   $v_{ikja}, v_{ijka},v_{ikjb}$ and $v_{ijkb}$ exist on the trisector
   $\tri{ijk}$, we consider all possible orderings of these vertices. 
   As of Lemma~\ref{lemma:ordering_equivalence}, each of these 
   orderings is equivalent to a respective ordering of the 
   points $\arc{v_{ikja}}, \arc{v_{ijka}},\arc{v_{ikjb}}$ and 
   $\arc{v_{ijkb}}$ on the oriented arc 
   $(\arc{\eta},\arc{o},\arc{\theta})$ of \yspace. For every 
   possible ordering a possible location of $\yinv{S_a}$ and 
   $\yinv{S_b}$ is considered, such that the shadow regions 
   $\sh{S_n}$ for $n\in\{a,b\}$ is of type $(\chi,\phi)$ since
   we examine a classic configuration. 
   From top to bottom, Left:  \answer 1, 2, 3.
   From top to bottom, Right: \answer 4, 5, 6. }
   \label{fig:11a_11f}
  \end{figure}

   Any of these ordering on the trisector is 
   equivalent to the corresponding ordering of 
   $\arc{v_{ikja}}$, $\arc{v_{ikjb}}$, $\arc{v_{ijkb}}$ and $\arc{v_{ijka}}$
   on the arc $(\yinv{\eta}, \yinv{o}, \yinv{\theta})$ as stated in Lemma~\ref{lemma:ordering_equivalence}.
   
   We now study separately all these possible cases in \yspace 
   following the same approach. Firstly, we place the images of the 
   Apollonius vertices on the arc according to the \answer we are 
   examining. Then, we consider a possible location for each of the circles 
   $\yinv{S_a}$ and $\yinv{S_b}$ taking into consideration the remarks 
   made in the Section~\ref{sub:The_classic_configuration}. Lastly, we draw some conclusions 
   regarding the relative position of $\yinv{S_a}$ and $\yinv{S_b}$ 
   with the lines $\arc{v_{ikjb}}, \arc{v_{ijkb}}$ and 
   $\arc{v_{ijka}},\arc{v_{ikja}}$ respectively. The later observations
   are then translated as \insphere test's results based on 
   Lemma~\ref{lemma:yinv_insphere_equiv}. 
   
   Let us consider one case in detail, for example the \answer 2 configuration; a similar approach will be applied to each \answer. 
   In Figure~\ref{fig:11a_11f} (Left Column, 2nd Row), 
   we consider a 
   random\footnote{In Figure~\ref{fig:11a_11f}, the circles 
   $\yinv{S_a}$ and $\yinv{S_b}$ always appear to 
   be centered on the same side of the line going through 
   $\yinv{\mathcal{A}}$  and $\yinv{\OO}$. This was done for 
   reasons of consistency and does not always 
   correspond to reality, 
   since it would be equivalent to $\inv{C_a}$ and $\inv{C_b}$ 
   always lying on the same side of the plane going through the 
   points $\inv{C_i}, \inv{C_j}$ and $\inv{C_k}$. }  
   layout of the points  $\arc{v_{ikja}}$, $\arc{v_{ikjb}}$, 
   $\arc{v_{ijkb}}$ and $\arc{v_{ijka}}$ (and the respective 
   tangent planes at these points) that appear in the order 
   \answer 2 dictates. 
   In the same figure, we provide
   a possible location of $\yinv{S_n}$, for $n\in\{a,b\}$ with respect
   to the selected layout;  $\yinv{S_n}$ 
   must be tangent to both $\ill{v_{ikjn}}$ and $\ill{v_{ijkn}}$, and
   centered according to the analysis of Section~\ref{sub:The_classic_configuration}. 

   Finally, we inspect the relative position of $\yinv{S_a}$ 
   (resp. $\yinv{S_b}$) with the lines $\arc{v_{ikjb}}$ and 
   $\arc{v_{ikjb}}$ (resp. $\arc{v_{ikja}}$ and 
   $\arc{v_{ikja}}$). In any such random layout, it must hold that
   \begin{itemize}
    \item 
    $\yinv{S_a}$ intersects the negative side of $\arc{v_{ikjb}}$
    but does not intersect the negative side of $\arc{v_{ijkb}}$ and, 
    \item 
    $\yinv{S_b}$ intersects the negative side of $\arc{v_{ijka}}$
    but does not intersect the negative side of $\arc{v_{ijka}}$. 
   \end{itemize}
   
   Another way of proving this, is by looking at the shadow regions 
   of $S_a$ and $S_b$ on the arc. For example, in a
   \answer 2 configuration, $v_{ikja}\prec v_{ikjb}\prec v_{ijka}$ 
   and subsequently  $v_{ikjb}\in\sh{S_a}$, since $\sh{S_a}$ 
   consists of all points $p\in\tri{ijk}$ with 
   $v_{ikja}\prec p\prec v_{ijka}$. As a result the sphere 
   $\tts{v_{ikjb}}$ must intersect the sphere $S_a$ \ie, 
   $\yinv{S_a}$ intersects the negative side of $\ill{v_{ikjb}}$. 

   Lastly, we translate the obtained relative positions of circles 
   and lines of \yspace to \insphere tests outcomes. For example, if 
   $\yinv{S_a}$ intersects the negative side of $\ill{v_{ikjb}}$, 
   we conclude that \insphere$(S_i,S_k,S_j,S_b,S_a)$ is negative, as
   an immediate result of Lemma~\ref{fig:zspace_yspace_equivalency}. 
   In conclusion, we get that if the Apollonius vertices we seek to 
   order appear as in \answer 2, then 
   \begin{itemize}
    \item  
    \insphere$(S_i,S_k,S_j,S_b,S_a)=-$ and 
    \insphere$(S_i,S_j,S_k,S_b,S_a)=+$,
    \item  
    \insphere$(S_i,S_k,S_j,S_a,S_b)=+$ and 
    \insphere$(S_i,S_j,S_k,S_a,S_b)=-$.
   \end{itemize}
   
   \begin{table}[tbp]
   \begin{center}
   \begin{tabular}{|c||c|c|c|c|c|c|}
   \hline
   \answer: &  \: 1 \: & \: 2 \: & \: 3 \: & \: 4 \: & \: 5 \: & \: 6 \: \\
   \hline\hline
   $\text{\insphere}(S_i,S_k,S_j,S_b;S_a)$&$+$&$-$&$+$&$+$&$+$&$-$\\ \hline
   $\text{\insphere}(S_i,S_j,S_k,S_b;S_a)$&$+$&$+$&$+$&$-$&$+$&$-$\\ \hline
   $\text{\insphere}(S_i,S_k,S_j,S_a;S_b)$&$+$&$+$&$-$&$-$&$+$&$+$\\ \hline
   $\text{\insphere}(S_i,S_j,S_k,S_a;S_b)$&$+$&$-$&$-$&$+$&$+$&$+$\\ \hline
   \end{tabular}
   \end{center}
   \caption{Case A: Signs of all possible \insphere tests 
   that follow from the analysis of each \answer. 
   Notice that only 
   the rows that correspond to \answer 1 and \answer 5 are identical 
   and therefore we only need the signs of these \insphere predicates 
   to determine the \answer most of the time. If all predicates return 
   positive, we require some auxiliary tests to distinguish between the 
   two cases.}
   \label{tab:2VS2}
   \end{table}

   Ultimately, we create a table of the four possible \insphere 
   outcomes that hold in each of the \answer's 1 to 6 
   (see Table~\ref{tab:2VS2}). A simple way of distinguishing 
   the ordering of the Apollonius vertices becomes clear now, due 
   to tuple of outcomes being so different in most \answer's. 
   Indeed, if $\mathcal{Q}=(Q_1,Q_2,Q_2,Q_4)$ 
   denotes the ordered tuple of the \insphere predicate outcomes, where
   \begin{align*}
   Q_1 &= \text{\insphere}(S_i,S_k,S_j,S_b,S_a), \ \ 
   Q_2 = \text{\insphere}(S_i,S_j,S_k,S_b,S_a),\\
   Q_3 &= \text{\insphere}(S_i,S_k,S_j,S_a,S_b), \ \ 
   Q_4 = \text{\insphere}(S_i,S_j,S_k,S_a,S_b),
   \end{align*} 

   \noindent
   then the \order predicate returns :
   \begin{itemize}
    \item 
     the ordering of \answer 2, if $\mathcal{Q}=(-,+,+,-)$ or,
    \item 
     the ordering of \answer 3, if $\mathcal{Q}=(+,+,-,-)$ or,
    \item 
     the ordering of \answer 4, if $\mathcal{Q}=(+,-,-,+)$ or,
    \item 
     the ordering of \answer 6, if $\mathcal{Q}=(-,-,+,+)$.
   \end{itemize}

\begin{figure}[htbp]
    \centering
    \includegraphics[width=0.95\textwidth]{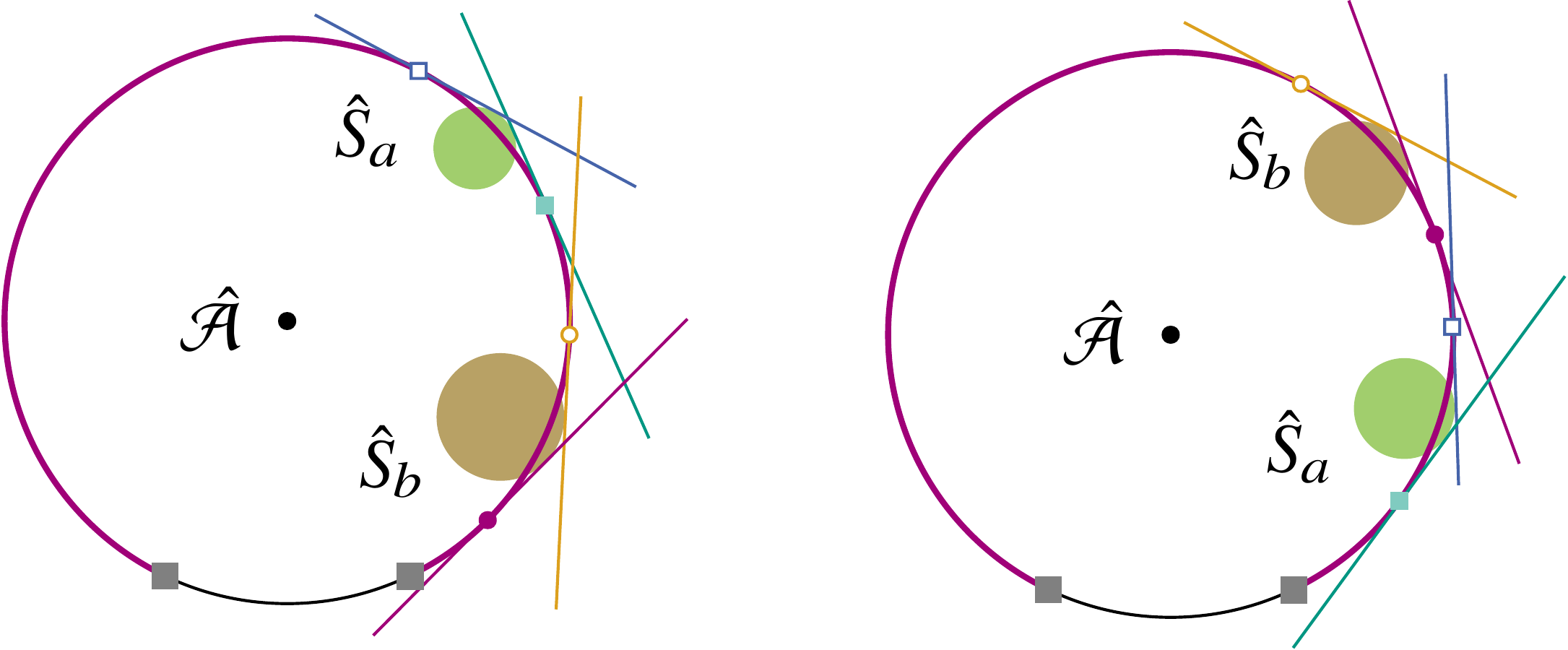}
    \caption{If $\mathcal{Q}=(+,+,+,+)$ then we must determine 
     if the ordering of the Apollonius vertices correspond to \answer 1 
     (Left) or 5 (Right). It is apparent that we are in the first 
     case if and only if the ray $(\mathcal{A},t)$ ``meets'' 
     $\yinv{C_a}$ first as $t$ traverses the arc 
     $(\arc{\eta},\arc{o},\arc{\theta})$.}
    \label{fig:2VS2_N1_vs_N5}
   \end{figure}

   Finally, if $\mathcal{Q}=(+,+,+,+)$ then either \answer 1 or 
   \answer 5 is the correct ordering of the vertices
   (see Figure~\ref{fig:2VS2_N1_vs_N5}). To resolve this 
   dilemma, we distinguish cases depending on the ordering 
   of the midpoints $M_a$ and $M_b$ of the arcs 
   $(\arc{v_{ikja}},\arc{v_{ijka}})$ 
   and $(\arc{v_{ikjb}},\arc{v_{ijkb}})$ respectively. Since it must 
   either hold that 
   $\{v_{ikja}\prec v_{ijka}\}\prec \{v_{ikjb}\prec v_{ijkb}\}$ 
   (\answer 1) or 
   $\{v_{ikjb}\prec v_{ijkb}\}\prec \{v_{ikja}\prec v_{ijka}\}$ 
   (\answer 5), then we are obviously in the former case if 
   $M_a\prec M_b$ or in the latter if $M_b\prec M_a$. 

   To determine the ordering of $M_a$ and $M_b$ on the arc
   $(\yinv{\eta},\yinv{o},\yinv{\theta})$, we shall use the auxiliary
   point $\yinv{o}$. Initially, we reflect on the fact that, 
   for $n\in\{a,b\}$, $\yinv{C_n}$ is known to lie on the open ray
   from $\mathcal{A}$ towards $M_n$. 
   It is also apparent that the points $\OO$, $\mathcal{A}$ 
   and $\yinv{o}$ are collinear and appear in this order on the 
   line $\yinv{\ell}$ they define. 

   Based on the definition of \yspace and the remarks of 
   Section~\ref{ssub:the_yspace_analysis}, the midpoint $M_n$, 
   for $n\in\{a,b\}$, satisfies

   \begin{itemize}
   \item 
   $M_n\prec\yinv{o}$ if and only if
   $\text{\orient}(\inv{C_n},\inv{C_i},\inv{C_j},\OO)<0$,
   \item
   $\yinv{o}\prec M_n$ if and only if
   $\text{\orient}(\inv{C_n},\inv{C_i},\inv{C_j},\OO)>0$,
   \item
   $M_n\equiv\yinv{o}$ if and only if
   $\text{\orient}(\inv{C_n},\inv{C_i},\inv{C_j},\OO)=0$.
   \end{itemize}
   
   Lastly, we notice that 
   $\text{\orient}(\inv{C_b},\inv{C_i},\inv{C_j},\inv{C_a})<0$ is 
   equivalent to $\yinv{C_b}$ lying on the ``right side'' of the 
   oriented line going from $\mathcal{A}$ to $\yinv{C_a}$.
   
   Ultimately, we determine the relative position of $M_a$ and $M_b$ 
   by combining all the information extracted of the \orient 
   predicates mentioned, using the following algorithm. 

\begin{figure}[htbp]
    \centering
    \includegraphics[width=0.95\textwidth]{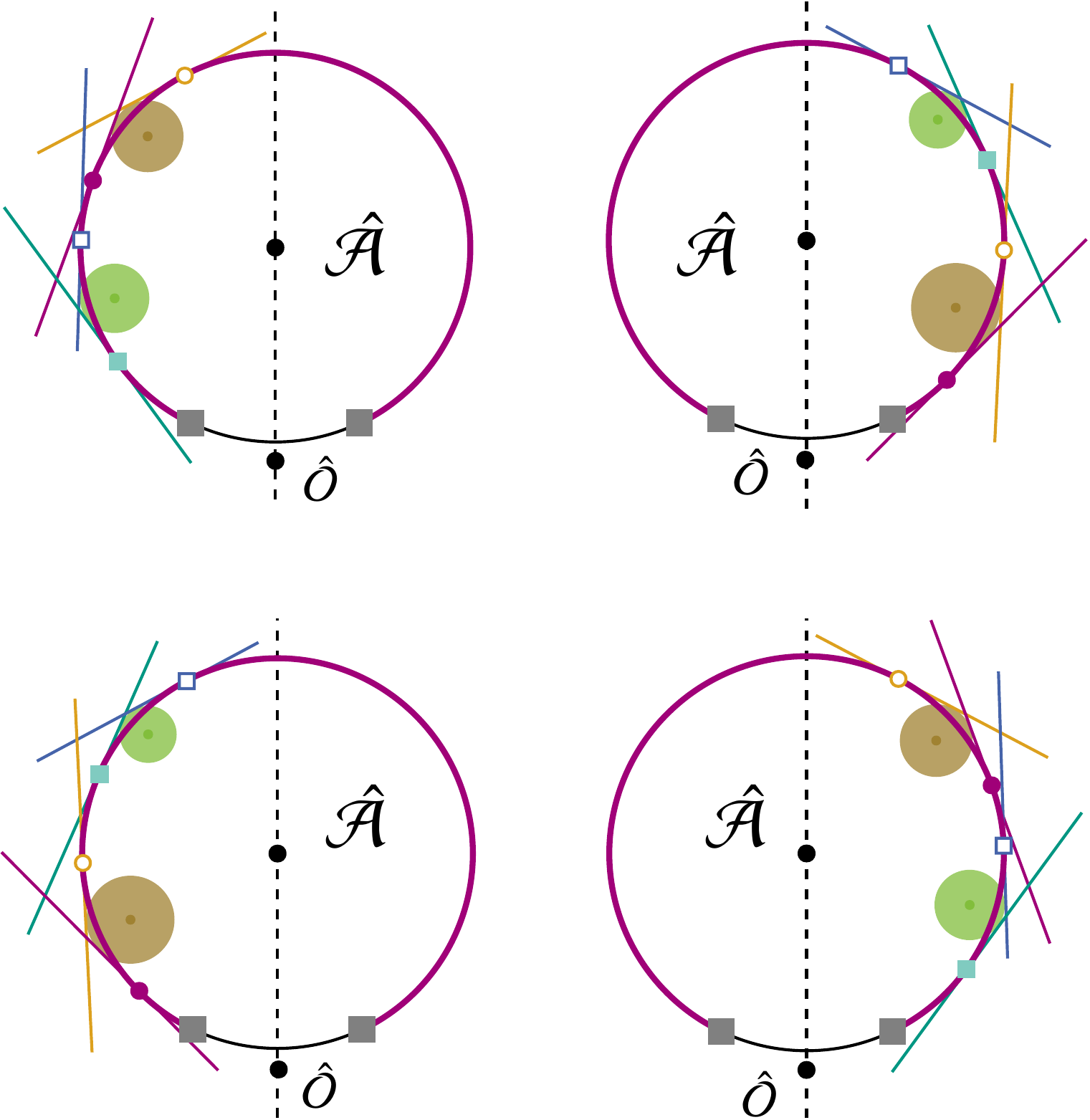}
    \caption{If $o_1\cdot o_2>0$, the 
    centers $\yinv{C_a}$ and $\yinv{C_b}$ must lie on the same side of 
    the line that goes through $\yinv{\mathcal{A}}$ and $\yinv{\OO}$.
    No matter which side the centers lie on, if 
    $o_3<0$ 
    or equivalently $\yinv{C_a}$ lies on the left side of the 
    oriented line that goes from $\yinv{\mathcal{A}}$ to  $\yinv{C_b}$, 
    (Top 2 Figures) then we obtain the ordering described 
    in \answer 1. Otherwise, we obtain the ordering described 
    in \answer 5 (Bottom 2 Figures).}
    \label{fig:2vs2_step2a}
   \end{figure}

\begin{figure}[htbp]
    \centering
    \includegraphics[width=0.95\textwidth]{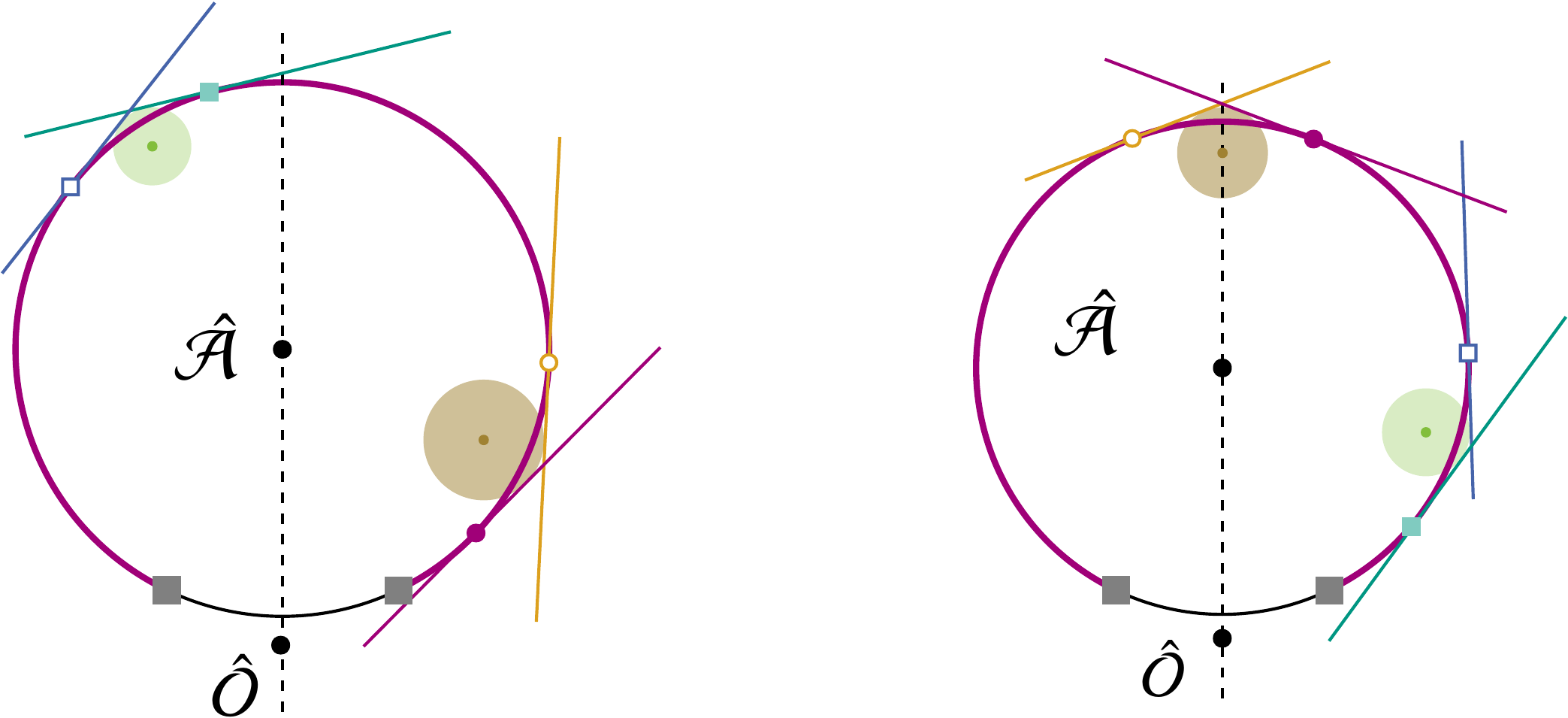}
    \caption{If $o_1\cdot o_2\leq 0$, the 
    centers $\yinv{C_a}$ and $\yinv{C_b}$ lie on different sides of 
    the line that goes through $\yinv{\mathcal{A}}$ and $\yinv{\OO}$ 
    (Left) or only one of them lies on the line (since we are in 
    either \answer 1 or 5).
    No matter which side the centers lie on, if 
    $o_1< o_2$
    we obtain the ordering described in \answer 1. Otherwise, 
    we obtain the ordering described in \answer 5 (Bottom 2 Figures).
    }
    \label{fig:2vs2_step2b}
   \end{figure}

   \begin{description}
    \item[Step 1.] 
    We evaluate 
    $\Pi=o_1\cdot o_2 $, where 
    $o_1=\text{\orient}(\inv{C_a},\inv{C_i},\inv{C_j},\OO)$ and
    $o_2=\text{\orient}(\inv{C_b},\inv{C_i},\inv{C_j},\OO)$. 
    If $\Pi>0$ go to Step 2a, otherwise go to Step 2b.
    \item[Step 2a.] 
    Either $M_a,M_b\prec \yinv{o}$ or $\yinv{o}\prec M_a,M_b$. 
    In either case, we evaluate
    $o_3=\text{\orient}(\inv{C_b}, \inv{C_i}, \inv{C_j}, \inv{C_a})$. 
    If $o_3<0$ then $M_a\prec M_b$, and the \order predicate returns the 
    ordering of \answer 1. Otherwise, $M_b\prec M_a$ and the ordering of
    \answer 5 is returned.
    (see Figure~\ref{fig:2vs2_step2a}).
    \item[Step 2b.] 
    Either $\yinv{o}$ lies in-between $M_a$ and $M_b$ or 
    is identical with one of them.
    In both cases, if 
    $o_1<o_2$ then $M_a\prec M_b$ and the \order predicate return the ordering of \answer 1, otherwise, $M_b\prec M_a$ and the ordering 
    of \answer 5 is returned.
    (see Figure~\ref{fig:2vs2_step2b}).
   \end{description} 

   \paragraph*{\textbf{Analysis of Case B}}
   Given that $\sh{S_a}$ and $\sh{S_b}$ are both of the form 
   $(\chi,+\infty)$ and therefore only the Apollonius vertices
   $v_{ikja}$ and $v_{ikjb}$ exist on $\tri{ijk}$, the ordering
   of these vertices on $(\yinv{\eta}, \yinv{o}, \yinv{\theta})$ 
   is either
   \begin{description}
   \item[\answer 1.] $v_{ikja}\prec v_{ikjb}$ or,
   \item[\answer 2.] $v_{ikjb}\prec v_{ikja}$.
   \end{description}

   A similar analysis with the Case A is used to resolve the predicate 
   in Case B; we create a table regarding the possible outcomes 
   of the \insphere tests with inputs 
   $(S_i$,$S_k$,$S_j$,$S_a$,$S_b)$ and 
   $(S_i$,$S_k$,$S_j$,$S_b$,$S_a)$. Recall that the outcome of 
   $Q_1=\text{\insphere}(S_i,S_k,S_j,S_a,S_b)$ 
   (resp. $Q_2=\text{\insphere}(S_i,S_k,S_j,S_a,S_b)$) 
   is $-$, $0$ or $+$ if the circle $\yinv{S_b}$ (resp. $\yinv{S_a}$)
   intersects, is tangent to or does not intersect the negative side 
   of $\ill{v_{ikja}}$ (resp. $\ill{v_{ikjb}}$).

   Using a simpler approach, we observe that 
   \begin{itemize}
    \item 
    in \answer 1, $v_{ikja}$ does not belong to the shadow region 
    of the sphere $S_b$ on $\tri{ijk}$ and therefore $\tts{v_{ikja}}$
    does not intersect $S_b$ or equivalently $Q_1=+$. Moreover,
    in this case,  $v_{ikjb}$ belongs to the shadow region 
    of the sphere $S_a$ on $\tri{ijk}$ and therefore $\tts{v_{ikjb}}$  
    intersects  $S_a$ or equivalently $Q_2=-$. 
    \item 
    In \answer 2, $v_{ikja}$ belongs to the shadow region 
    of the sphere $S_b$ on $\tri{ijk}$ and therefore $\tts{v_{ikja}}$  
    intersects  $S_b$ or equivalently $Q_1=-$. Furthermore, 
    $v_{ikjb}$ does not belong to the shadow region 
    of the sphere $S_a$ on $\tri{ijk}$ and therefore $\tts{v_{ikjb}}$   
    does not intersects $S_a$ or equivalently $Q_2=+$. 
   \end{itemize}

   In conclusion we can answer the 
   \text{\order}$(S_i,S_j,S_k,S_a,S_b)$ predicate in case B by 
   evaluating $Q_1$; if $Q_1=+$ then return \answer 1 otherwise,
   if $Q_1=-$ return \answer 2. 
   Equivalently, we could evaluate $Q_2$ instead of $Q_1$; 
   if $Q_2=-$ then return \answer 1 otherwise
   if $Q_2=-$ return \answer 2. The following equivalencies are 
   depicted in Table~\ref{tab:1VS1} and this concludes the analysis of 
   Case B. 

   \begin{table}[bt]
   \begin{center}
   \begin{tabular}{|c||c|c|}
   \hline
   & OrderCase 1 & OrderCase 2 \\
   \hline\hline
   $\text{\insphere}(S_i,S_k,S_j,S_b;S_a)$ & $+$ & $-$ \\ \hline
   $\text{\insphere}(S_i,S_k,S_j,S_a;S_b)$ & $-$ & $+$ \\ 
   \hline
   \end{tabular}
   \end{center}
   \caption{Case B: Signs of all possible \insphere tests 
   that follow from the analysis of each \answer. Notice that each 
   column is distinct and therefore we can determine the  \answer
   after the outcomes of the  \insphere predicates.}
   \label{tab:1VS1}
   \end{table}

   \begin{figure}[tbp]
   \centering
   \includegraphics[width=0.4\textwidth]{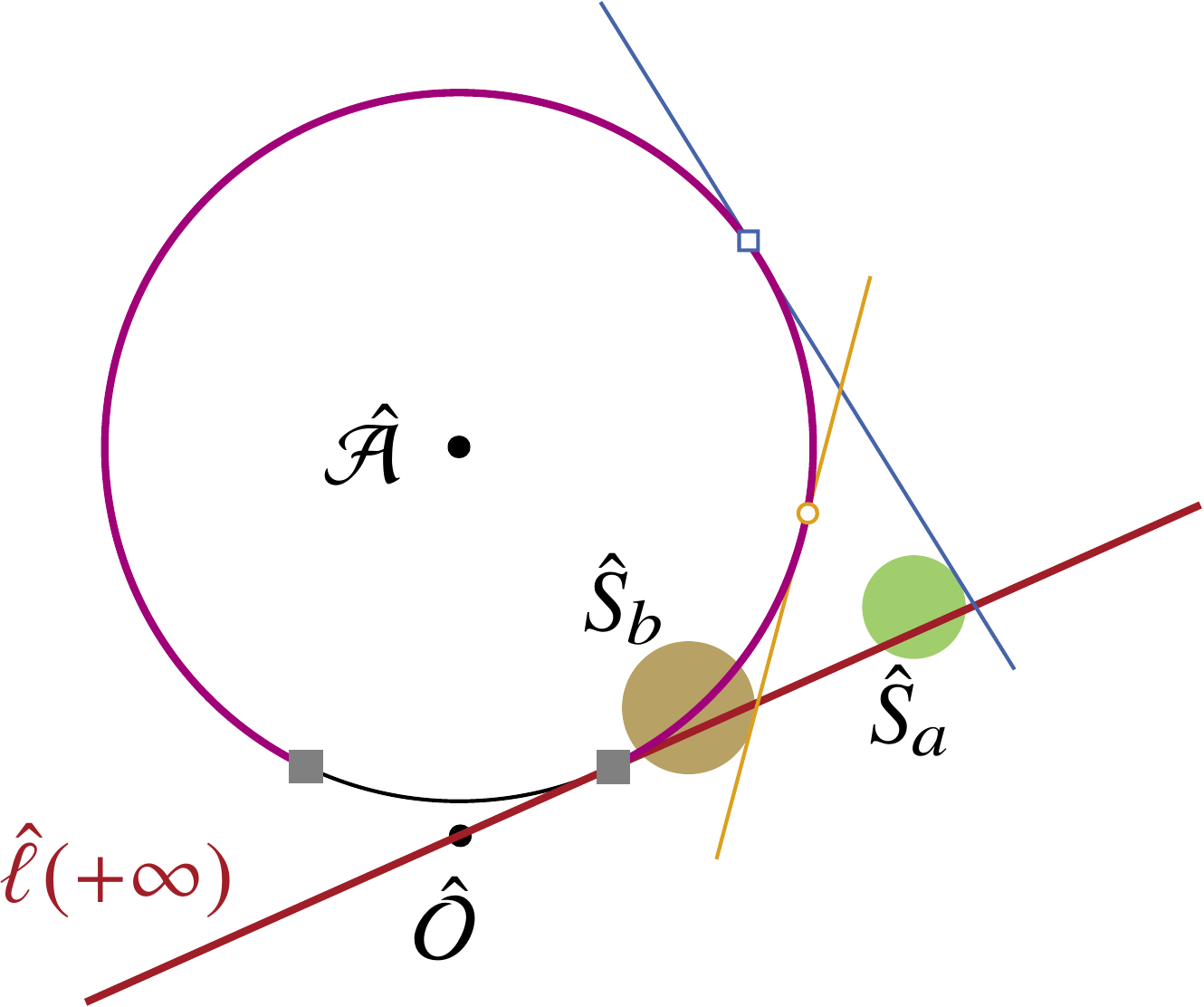}\hspace{1 cm}
   \includegraphics[width=0.4\textwidth]{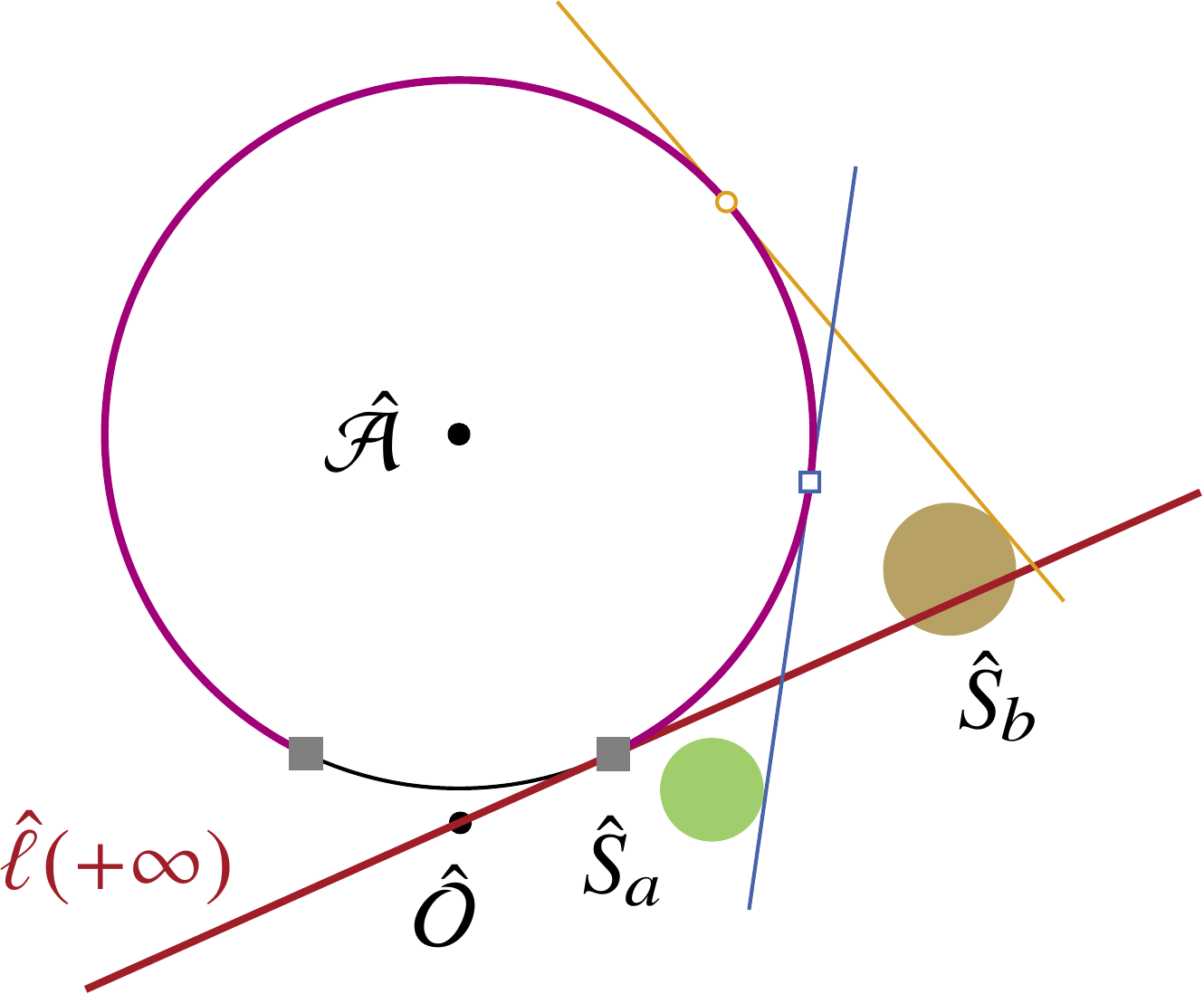}
   \caption{In Case B, it is assumed that only the Apollonius vertices 
   $v_{ikja}$ and $v_{ijkb}$ exist on the trisector
   $\tri{ijk}$. We consider the two possible orderings of these vertices:
   \answer 1 (Left) and \answer 2 (Right). Similar with Case A, 
   we consider the corresponding ordering of the points  
   $\arc{v_{ikja}}$ and $\arc{v_{ijkb}}$ on the arc 
   $(\arc{\eta},\arc{o},\arc{\theta})$. A possible location for 
   the circles $\yinv{S_a}$ and $\yinv{S_b}$ is drawn based on 
   the analysis of Section~\ref{sub:the_main_algorithm}.}
   \label{fig:12a_12b}
   \end{figure}

   \paragraph*{\textbf{Analysis of Case C}}

   In Case C, it is assumed that $\sh{S_a}=(\chi,\phi)$ hence
   $v_{ikja}\prec v_{ijka}$ while $\sh{S_b}=(\chi,+\infty)$ and 
   consequently only $v_{ikjb}$ exists on the arc 
   $(\yinv{\eta}, \yinv{o}, \yinv{\theta})$. All three possible 
   orderings of these three Apollonius vertices on the 
   arc are 
   \begin{description}
   \item[\answer 1.] $v_{ikja}\prec v_{ijka}\prec v_{ikjb}$,
   \item[\answer 2.] $v_{ikja}\prec v_{ikjb}\prec v_{ijka}$ and 
   \item[\answer 3.] $v_{ikjb}\prec v_{ikja}\prec v_{ijka}$.
   \end{description}

   The analysis of this Case uses the same tools and analysis 
   presented in the previous two cases with small adjustments, 
   since $\sh{S_a}=(\chi,\phi)$ and $\sh{S_b}=(\chi,+\infty)$ in 
   the case studied. 
   Let us denote by $Q_1,Q_2$ and $Q_3$ the results of the 
   \insphere predicates with inputs $(S_i,S_k,S_j,S_b,S_a)$, 
   $(S_i,S_k,S_j,S_a,\allowbreak S_b)$ and $(S_i,S_j,S_k,S_b,S_a)$ respectively.

   Notice now that
   \begin{itemize}
   \item 
   in \answer 1, $v_{ikjb}$,$ v_{ikja}$ and $v_{ijka}$ 
   do not belong to the shadow region of $S_a$, $S_b$ and $S_b$ 
   respectively and therefore it must stand that $Q_1=Q_2=Q_3=+$.
   \item 
   in \answer 2, $v_{ikjb}$ and $v_{ijka}$ 
   belong to the shadow region of $S_a$ and $S_b$ 
   respectively and for this reason $Q_1=-$ and $Q_3=-$.
   On the other hand, $v_{ikja}$ does not belong to 
   the shadow region of $S_b$ and therefore $Q_2=+$. Finally,
   \item 
   in \answer 3, both $v_{ikja}$ and $v_{ijka}$ 
   belong to the shadow region of $S_b$ and 
   consequently $Q_2=-$ and $Q_3=-$ whereas,
   $v_{ikjb}$ does not belong to the shadow region of $S_a$
   and therefore $Q_1=+$.
   \end{itemize}

   Since the tuple $\mathcal{Q}=(Q_1,Q_2,Q_3)$ is different 
   in each \answer 1 to 3, we can answer the predicate by 
   evaluating the three \insphere predicates hence $\mathcal{Q}$ 
   and correspond it the respective ordering (also see 
   Table~\ref{tab:1VS2}):
   \begin{itemize}
    \item if $\mathcal{Q}=(+,+,+)$, return the ordering of \answer 1 or,
    \item if $\mathcal{Q}=(-,+,-)$, return the ordering of \answer 2 otherwise,
    \item if $\mathcal{Q}=(+,-,-)$, return the ordering of \answer 3.
   \end{itemize}

   \begin{table}[tbp]
   \begin{center}
   \begin{tabular}{|c||c|c|c|}
   \hline
   & OrderCase 1 & OrderCase 2 & OrderCase 3 \\
   \hline\hline
   $\text{\insphere}(S_i,S_k,S_j,S_b;S_a)$ & $+$ & $-$ & $+$ \\ \hline
   $\text{\insphere}(S_i,S_k,S_j,S_a;S_b)$ & $+$ & $+$ & $-$ \\ \hline
   $\text{\insphere}(S_i,S_j,S_k,S_a;S_b)$ & $+$ & $-$ & $-$ \\ \hline
   \end{tabular}
   \end{center}
   \caption{Case C: Signs of all possible \insphere tests 
   that follow from the analysis of each \answer. Notice that each 
   column is distinct and therefore we can determine the  \answer
   after the outcomes of the \insphere predicates, as in Case B.}
   \label{tab:1VS2}
   \end{table}

   \begin{figure}[tbp]
   \centering
   \includegraphics[width=0.4\textwidth]{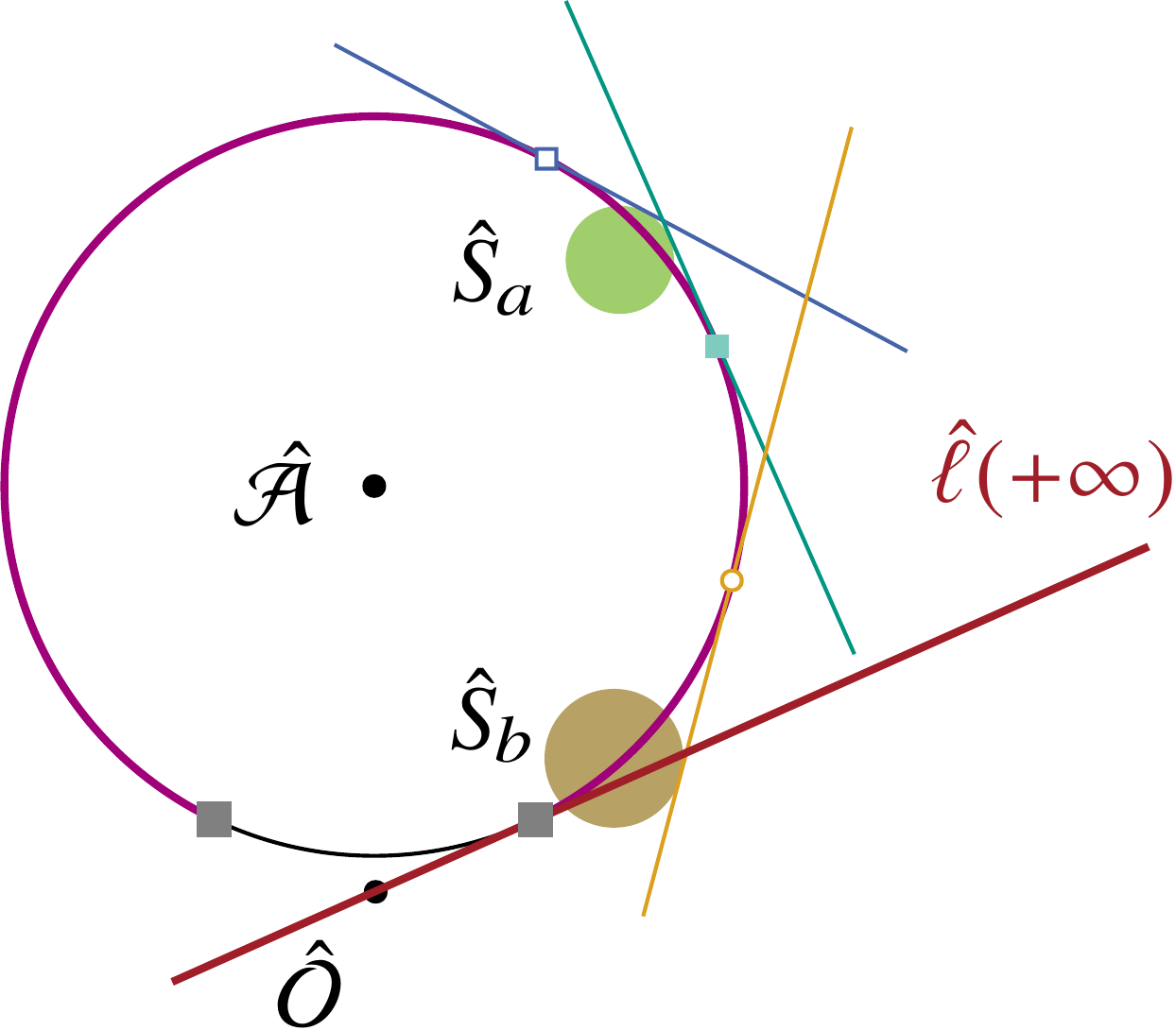}\hspace{1 cm}
   \includegraphics[width=0.4\textwidth]{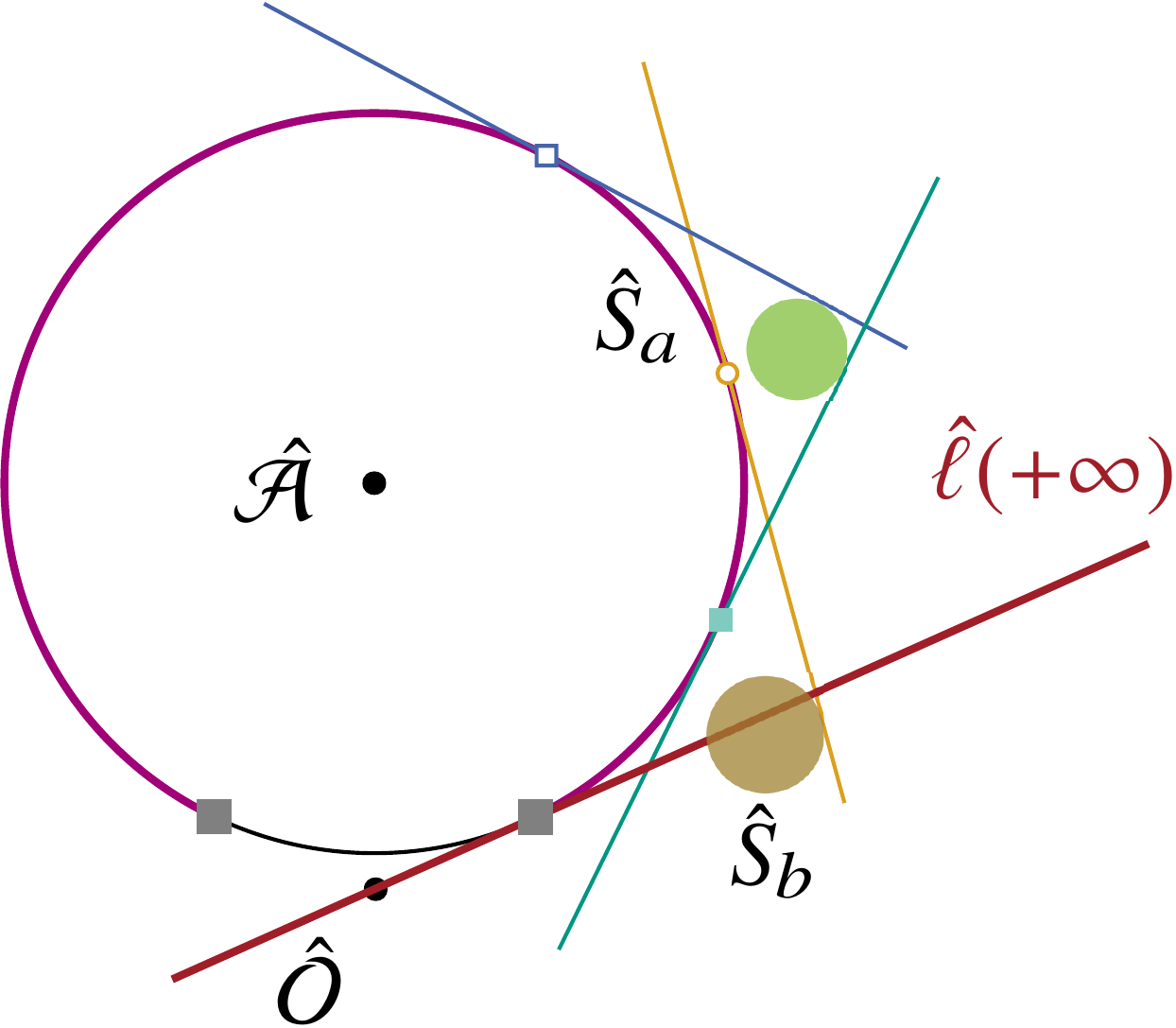}\\
   \includegraphics[width=0.4\textwidth]{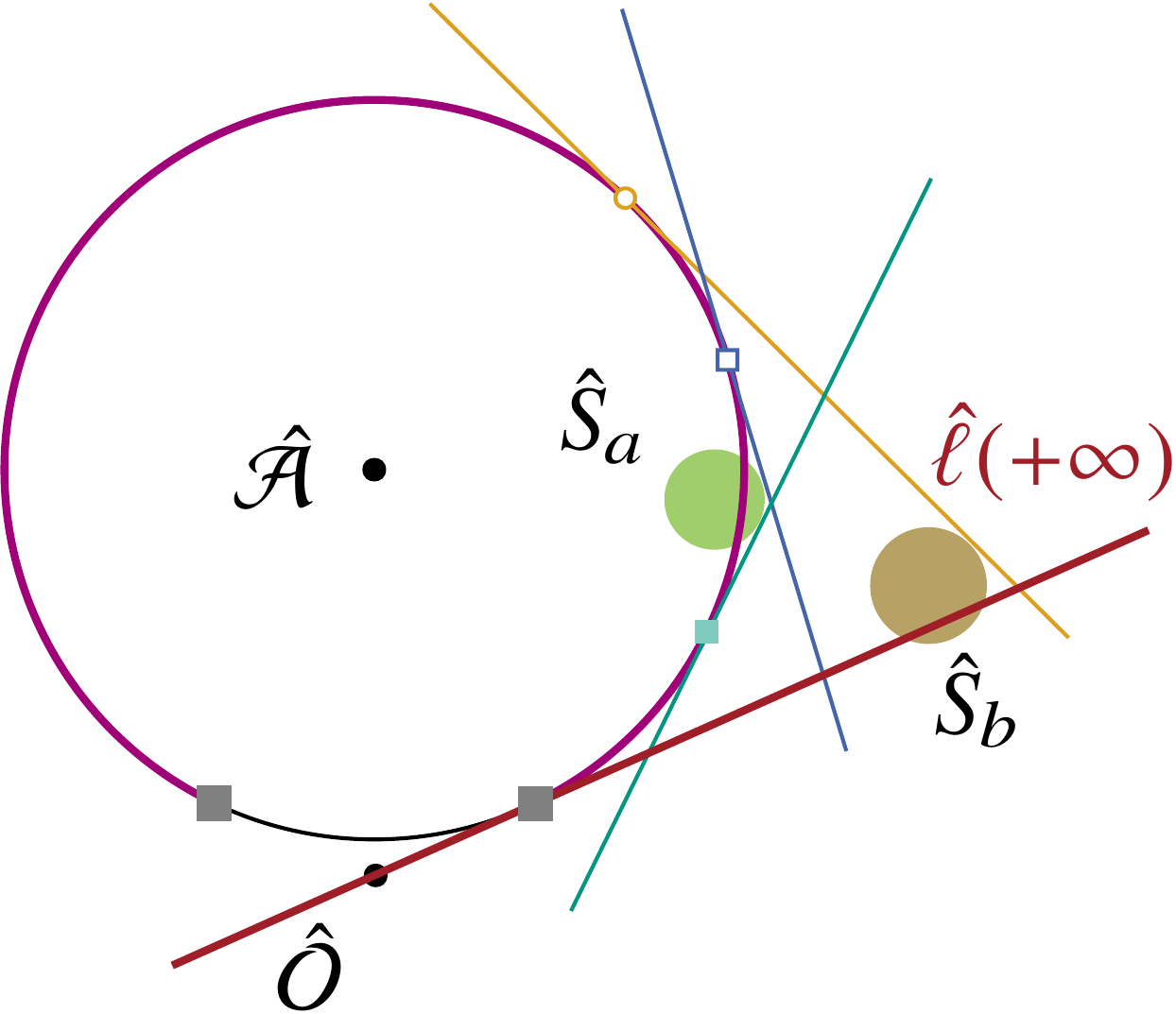}
   \caption{In Case C, it is assumed that only the Apollonius vertices 
   $v_{ikja}, v_{ijka}$ and $v_{ijkb}$ exist on the trisector
   $\tri{ijk}$. We consider the three possible orderings of these vertices:
   \answer 1 (Top Left), \answer 2 (Top Right) and \answer 3 (Bottom). 
   Similar with Case A and B, 
   we consider the corresponding ordering of the points  
   $\arc{v_{ikja}}$ and $\arc{v_{ijkb}}$ on the arc 
   $(\arc{\eta},\arc{o},\arc{\theta})$. A possible location for 
   the circles $\yinv{S_a}$ and $\yinv{S_b}$ is drawn based on 
   the analysis of Section~\ref{sub:the_main_algorithm}.}
   \label{fig:13a_13c}
   \end{figure}

   \paragraph*{\textbf{Algebraic Cost to resolve the Cases A, B or C}}
   The analysis of the Cases A, B and C showed that the answer 
   of the $\text{\order}(S_i,S_j,S_k,S_a,S_b)$ predicate
   in a classic configuration ultimately amounts to determining 
   the outcomes of up to four \insphere predicates and, if needed, 
   some auxiliary \orient tests.

   To answer any of the \insphere predicates that may require 
   evaluation, we must 
   perform operations of maximum algebraic degree 10 (in the input 
   quantities), as mentioned in Section~\ref{sub:the_insphere_predicate}. 

   Regarding the auxiliary \orient primitives, we observe that
   \begin{align*}
    \text{\orient}(\inv{C_b},\inv{C_i},\inv{C_j},\inv{C_a})
      &= \sgn(D^{uvw}_{bija})
      = \sgn(\inv{p_i}\inv{p_j}\inv{p_a}\inv{p_b})\sgn(E^{xyzp}_{bija})\\
      &= \sgn(E^{xyzp}_{bija}),  
   \end{align*}
   where the quantity $E^{xyzp}_{bija}$ and  is an expression of 
   algebraic degree 5 on the input quantities. The expression
   $\text{\orient}(\inv{C_n}, \inv{C_i}, \inv{C_j}, \OO)$, for $n\in\{a,b\}$ 
   can be evaluated as shown in Section~\ref{ssub:the_yspace_analysis},
   \begin{align*}
   \text{\orient}(\inv{C_n},\inv{C_i},\inv{C_j},\OO)&=
   \sgn(\inv{p_i}\inv{p_i}\inv{p_n}D^{uvw}_{nij})=
   \sgn(D^{xyz}_{nijk})\\
   &= \text{\orient}(C_n,C_i,C_j,C_k)
   \end{align*}

   \noindent and therefore its evaluation requires operations of
   algebraic degree 4 (in the input quantities). 

   In conclusion, since the evaluation of the \insphere predicates 
   is the most degree-demanding operation throughout the 
   evaluation of the \order predicate in a classic configuration, we 
   have proven the following lemma. 

   \begin{lemma}
   The \order predicate in a classic configuration can be evaluated 
   by determining the sign of quantities of algebraic 
   degree at most 10 (in the input quantities).
   \end{lemma}

   \subsubsection{Ordering the Apollonius vertices in 
   a non-classic configuration} 
   \label{sub:ordering_in_a_non_classic_configuration}
   
   In the previous section we presented a way to resolve the 
   \text{\order}$(S_i,S_j,S_k,S_a,S_b)$ predicate under the 
   assumption that $\sh{S_a}$ and $\sh{S_b}$ were 
   either $(\chi,+\infty)$ or $(\chi,\phi)$ (not necessary the same); 
   we called this \emph{a classic configuration}. 
   In this section, we will assume we are in a non-classic 
   configuration \ie, at least one of $\sh{S_a}$ or $ \sh{S_b}$  
   is $(-\infty,\phi)$ or $(-\infty,\phi)\cup(\chi,\phi)$. 
   For convenience, these last two forms of a shadow region are 
   labelled as non-classic whereas the classic forms are 
   $(\chi,+\infty)$ and $(\chi,\phi)$.

   If $\sh{S_n}$ has a non-classic type, for $n=a$ or $b$, then we claim 
   that there exist a sphere $S_N$, for $N=A$ or $B$ respectively,
   such that:
   \begin{itemize}
    \item 
    if $\sh{S_n}=(-\infty,\phi)$ then $\sh{S_N}=(\chi,+\infty)$ 
    and $v_{ijkn}\equiv v_{ikjN}$ or,
    \item 
    if $(-\infty,\phi)\cup(\chi,\phi)$ then $\sh{S_N}=(\chi,\phi)$
    and $v_{ijkn}\equiv v_{ikjN}$ as well as $v_{ikjn}\equiv v_{ijkN}$.
   \end{itemize}

   \noindent If these conditions hold, we will say that 
   $S_n$ and $S_N$ are \emph{equivalent spheres}. Notice 
   that if $\sh{S_n}$ has a non-classic type then the shadow region
   of its equivalent sphere has a classic type and vice versa. The utility 
   of this equivalency is that it enable us to make a connection 
   between a classic and a non-classic configuration in the following 
   way. 

   When the predicate $\text{\order}(S_i,S_j,S_k,S_a,S_b)$ is 
   called then 
   \begin{enumerate}
    \item 
    if $\sh{S_a}$ and $\sh{S_b}$ have a classic type,
    we are in a classic and therefore, we resolve the predicate
    based on the analysis of 
    Section \ref{sub:ordering_in_a_classic_configuration}.
    \item 
    If $\sh{S_a}$ has a classic type and $\sh{S_b}$ does not, 
    then we call \order$(S_i,S_j,S_k,S_a,S_B)$. Since both
    $\sh{S_a}$ and $\sh{S_B}$ have a classic type, this predicate 
    can be evaluated using analysis of 
    Section~\ref{sub:ordering_in_a_classic_configuration} with 
    some adjustments. The predicate's outcome would be the 
    ordering of $v_{ikja}$, $v_{ikjB}$ and any of the existing
    $v_{ijka}$ or $v_{ijkB}$. Using the property of equivalent 
    spheres, we could answer the initial predicate by 
    substituting $v_{ikjB}$ with $v_{ijkb}$ and, if it exists, 
    $v_{ijkB}$ with $v_{ikjb}$.
    \item 
     If $\sh{S_b}$ has a classic type and $\sh{S_a}$ does not,
     then we follow a similar analysis with the previous case. 
     We evaluate \order$(S_i$, $S_j,S_k,S_A,S_b)$ and 
     in the resulting ordering of the Apollonius vertices 
     $v_{ikjA}$, $v_{ikjb}$ and any of the existing
     $v_{ijkA}$ or $v_{ijkb}$, we will substitute  
     $v_{ikjA}\equiv v_{ijka}$ and if necessary, 
     $v_{ijkA}\equiv v_{ikja}$, to obtain the answer to the 
     initial \order predicate. 
     \item 
     Finally, if both $\sh{S_a}$ and $\sh{S_b}$ do not have 
     a classic type we evaluate $\text{\order}(S_i$,$S_j$,$S_k$,$S_A$,$S_B)$.
     As before, we substitute $v_{ikjA}\equiv v_{ijka}$, 
     $v_{ikjB}\equiv v_{ijkb}$ and if necessary, 
     $v_{ijkA}\equiv v_{ikja}$ and/or $v_{ijkB}\equiv v_{ikjb}$, 
     and the acquired ordering is the answer of the initial 
     \order predicate.
   \end{enumerate}

   The evaluation of the \order predicate called in any of these 
   4 cases will eventually require determining \insphere or 
   \orient predicates with inputs that involve the 
   sites $S_i,S_j,S_k,S_A$ (or $S_a$) and $S_B$(or $S_b$). 
   The list of all possible predicates that must be 
   evaluated, in the worst case scenario and assuming a classic 
   configuration, would be:
    %
    %
    %
   \begin{itemize}
   \item \insphere$(S_i,S_k,S_k,S_a,S_b)$, 
   \item \insphere$(S_i,S_j,S_k,S_a,S_b)$, 
   \item \insphere$(S_i,S_k,S_j,S_b,S_a)$, 
   \item \insphere$(S_i,S_j,S_k,S_b,S_a)$, 
   \item \orient$(C_a,C_i,C_j,C_k)$,  
   \item \orient$(C_b,C_i,C_j,C_k)$ and 
   \item \orient$(\inv{C_a},\inv{C_i},\inv{C_j},\inv{C_b})$.
   \end{itemize}

   \noindent 
   It is apparent that we must be able to answer these predicates 
   when either one or both of $S_a$ and $S_b$ are substituted by 
   $S_A$ and $S$ respectively. 

   Firstly, we present a way of defining an equivalent sphere 
   $S_N$ when $\sh{S_n}$ has a non-classic type, 
   for $N=A$ or $B$ and $n=a$ or $b$ respectively. 
   Since $\yinv{C_n}$ cannot coincide with
   $\yinv{\mathcal{A}}$ (because there are either 1 or 2  
   cotangent lines to $\wcone$ and $\yinv{S_n}$), these points 
   define a line $\yinv{\ell_n}$. If a random point $\yinv{C_N}$ 
   is selected on $\yinv{\ell_n}$ such that $\yinv{\mathcal{A}}$ 
   lies in-between $\yinv{C_N}$ and $\yinv{C_n}$, then we 
   may choose an appropriate radius such that a circle $\yinv{S_n}$,
   centered at $\yinv{C_N}$, is tangent to any of the existing 
   lines $\ill{v_{ikjn}}$ and $\ill{v_ijkn}$. 

\begin{figure}[htbp]
    \centering
    \includegraphics[width=0.7\textwidth]{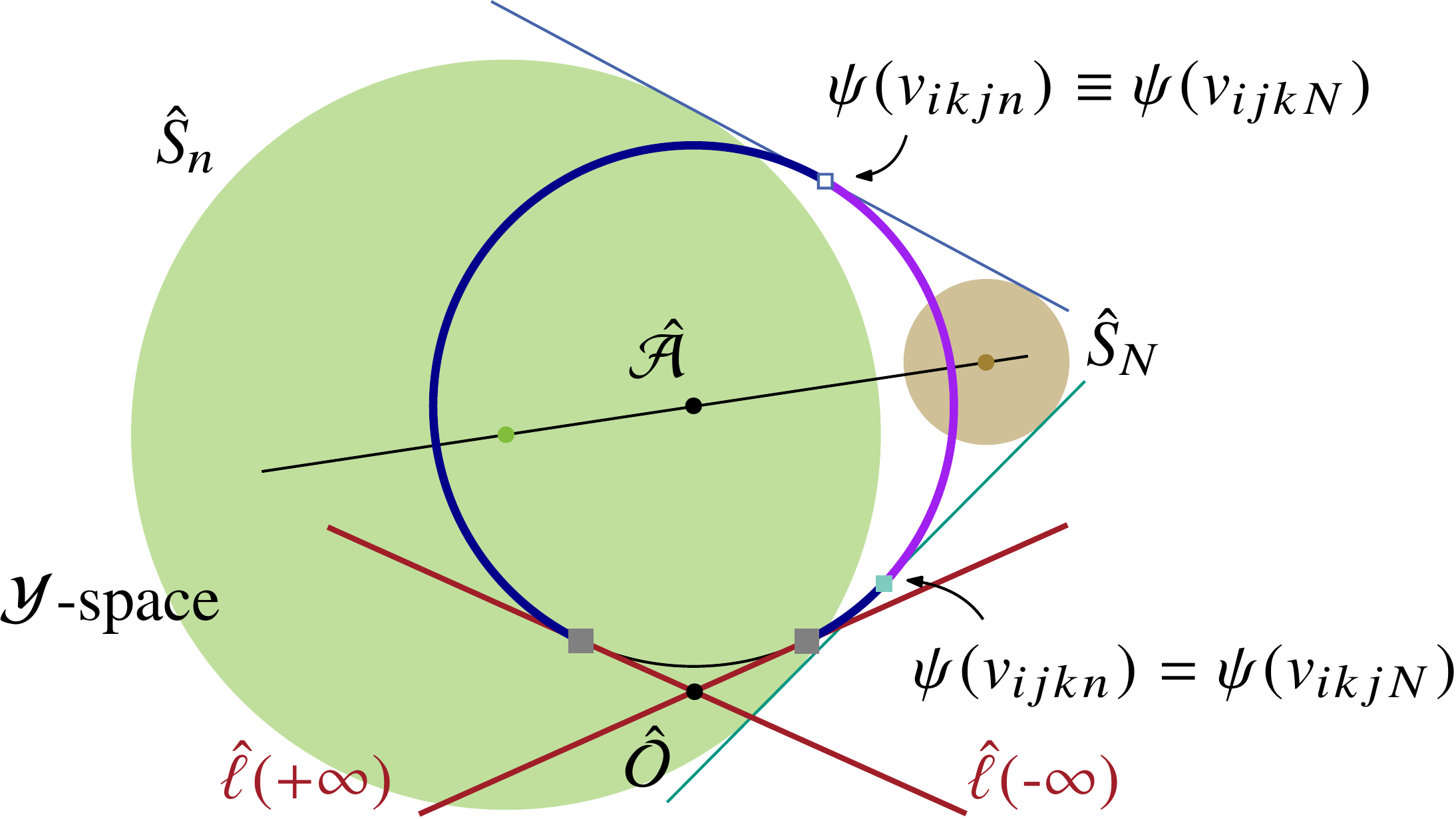}
    \caption{The shadow region of $S_n$ is 
    $(-\infty,\phi)\cup(\chi,+\infty)$ as its image in \yspace is 
    the blue area of the arc. Notice that the 
    respective image of $\sh{S_N}$ is the purple area and therefore 
    $\sh{S_n}$ must equal $(\chi,\phi)$. Since the endpoints of the 
    two shadow regions coincide and based on Lemma~\ref{lemma:phi_chi}, 
    it must hold that $v_{ijkN}\equiv v_{ikjn}$ and 
    $v_{ikjN}\equiv v_{ijkn}$. Therefore, $\yinv{S_N}$ and $\yinv{S_n}$ 
    are equivalent.}
    \label{fig:14a}
   \end{figure}

   Notice that any sphere $S_N$ of \wspace whose corresponding image 
   in \yspace is the circle $\yinv{S_N}$ has the desired properties 
   of an equivalent sphere of $S_n$. Indeed, if $\sh{S_n}$ is 
   $(\chi,\phi)$ then it must stand that 
   $\sh{S_N}=(-\infty,\phi)\cup(\chi,+\infty)$ and 
   specifically the actual endpoints of these shadow regions on 
   the trisector $\tri{ijk}$ coincide. To prove this argument, 
   we only need observe in \yspace that the circle $\yinv{C_n}$ 
   intersects the negative side of a line $\ill{p}$ only for 
   $v_{ikjn}\prec v_{ijkn}$ whereas, these are the only family of 
   lines $\ill{p}$ for $p\in\tri{ijk}$ that do not intersect 
   $\yinv{S_N}$. As a conclusion the shadow region of $S_n$ and 
   $S_N$ must be complementary \ie, 
   $\sh{S_N}=(-\infty,\phi)\cup(\chi,+\infty)$. From 
   Lemma~\ref{lemma:phi_chi}, we deduce that $v_{ijkN}\prec v_{ikjN}$ and 
   since these endpoints coincide with the endpoints of 
   $\sh{S_n}$ it must hold that $v_{ijkN}\equiv v_{ikjn}$ and 
   $v_{ikjN}\equiv v_{ijkn}$, since $v_{ikjn}\prec v_{ijkn}$ (see Figure~\ref{fig:14a}). 

\begin{figure}[htbp]
    \centering
    \includegraphics[width=0.7\textwidth]{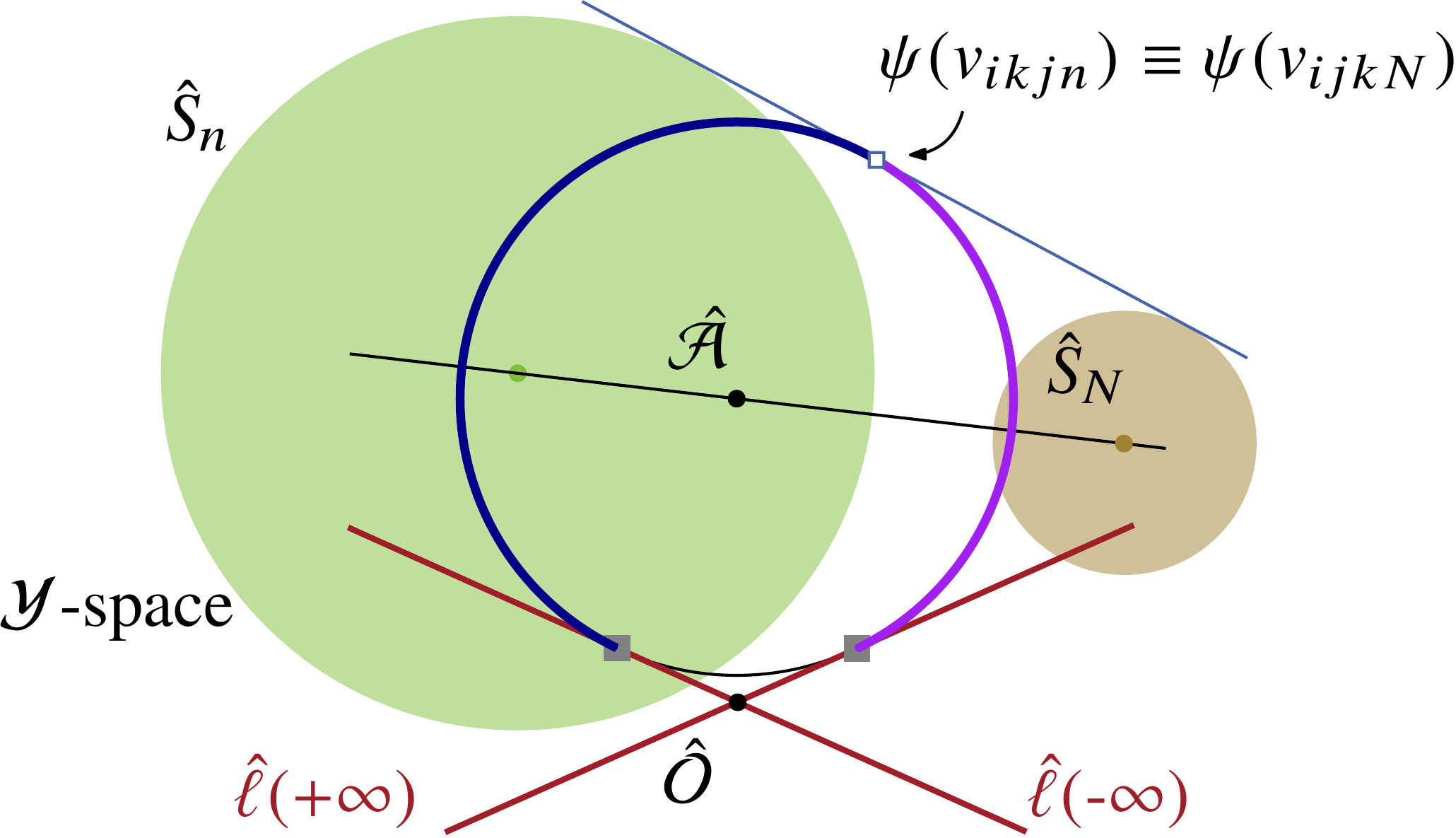}
    \caption{The shadow region of $S_n$ is 
    $(-\infty,\phi)$ as its image in \yspace is 
    the blue area of the arc. Notice that the 
    respective image of $\sh{S_N}$ is the purple area and therefore 
    $\sh{S_n}$ must equal $(\chi,+\infty)$. Since the endpoints of the 
    two shadow regions coincide and based on Lemma~\ref{lemma:phi_chi}, 
    it must hold that $v_{ijkN}\equiv v_{ikjn}$. Therefore, $\yinv{S_N}$ 
    and $\yinv{S_n}$ are equivalent.}
    \label{fig:14b}
   \end{figure}

   Using a similar analysis, one can consider an equivalent 
   sphere $S_N$ of $S_n$, when $\sh{S_n}$ is assumed to be 
   $(-\infty,\phi)$. The center of the respective circle 
   $\yinv{C_N}$ is selected in the same way as above, and the radius 
   of $\yinv{S_N}$ is chosen such that the circle is tangent 
   to $\ill{v_{ijkn}}$. Again, we can conclude that 
   $\sh{S_N}$ and $\sh{S_n}$ are complementary since the 
   family of lines $\ill{p}$ for $\yinv{\eta}\prec p$ 
   are the locus of lines $\ill{p}$, with 
   $\yinv{p}\in (\yinv{\eta},\yinv{o},\yinv{\theta})$, whose negative
   side is intersected by $\yinv{S_n}$ and simultaneously, 
   whose negative side is not intersected by $\yinv{S_N}$ (see Figure~\ref{fig:14b}).

   An interesting observation is that $S_N$ is not uniquely defined 
   in the sense that we do not provide its exact coordinates expressed 
   as a function of the input quantities. This is a consequence of the  fact
   that there are infinite spheres $S_n$ that all share the same Apollonius vertices $v_{ikjn}$ and $v_{ijkn}$.

\begin{figure}[htbp]
    \centering
    \includegraphics[width=0.7\textwidth]{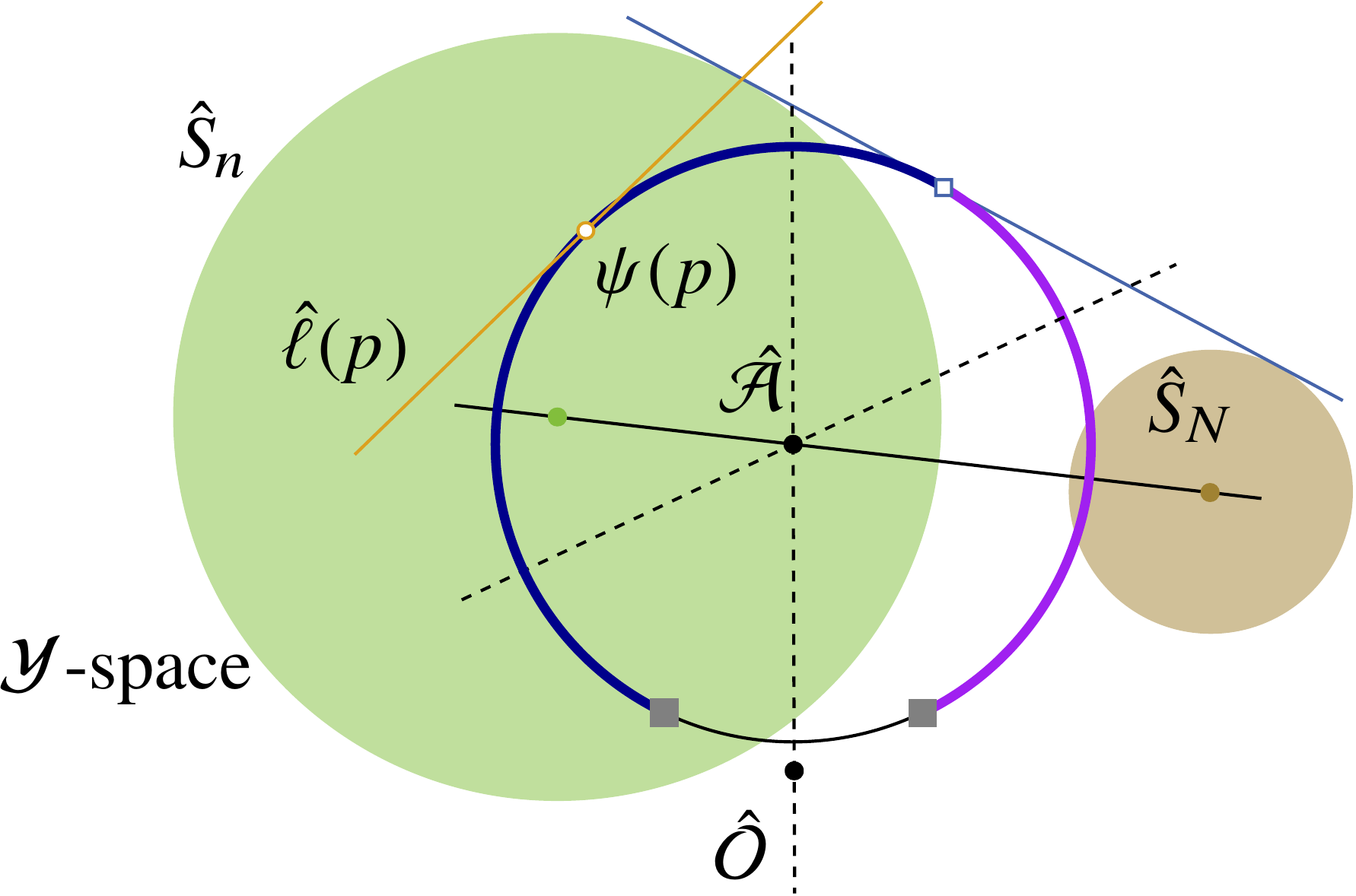}
    \caption{ If $S_N$ is an equivalent sphere of $S_n$, then 
    it must hold that the centers $\yinv{C_n}$, $\yinv{C_N}$ and 
    $\yinv{\mathcal{A}}$ are collinear and the former two points 
    lie on opposite sides with respect to the latter. Observe that 
    they also lie on opposite sides with respect to any line that 
    goes through $\yinv{\mathcal{A}}$. Lastly, it is apparent that 
    a point $\arc{p}$ on the arc $(\arc{\eta},\arc{o},\arc{\theta})$
    must lie on the image of the shadow region of either $S_n$ or 
    $S_N$.}
    \label{fig:15}
   \end{figure}

   Resuming the analysis of the properties of the equivalent sphere, 
   we notice that if a point $p\in\tri{ijk}$ lies on the shadow region 
   of $S_n$ then it must not lie on the shadow region 
   of $S_N$ and vice versa. An equivalent statement would be that 
   a sphere $\tts{p}$, for $p\in\tri{ijk}$, intersects $S_n$ 
   if and only if it does not intersect $S_N$ (see Figure~\ref{fig:15}). 
   If $p$ is chosen to 
   be either $v_{ikjm}$ or $v_{ijkm}$, where 
   $m\in\{a,b\}\backslash\{n\}$, we get the following relations
   \begin{align*}
   \text{\insphere}(S_i,S_j,S_k,S_m,S_N) &=
     -\text{\insphere}(S_i,S_j,S_k,S_m,S_n), \\
   \text{\insphere}(S_i,S_k,S_j,S_m,S_N) &=
     -\text{\insphere}(S_i,S_k,S_j,S_m,S_n).
   \end{align*}
   
   Moreover, if $S_m$ has a non-classic type and 
   $S_M$ is an equivalent sphere, where
   $M=B$ if $m=b$ or $M=A$ if $m=a$, it is known that 
   $v_{ikjM}\equiv v_{ijkm}$ and, if $v_{ikjm}$ also exists, 
   then $v_{ijkM}\equiv v_{ikjm}$. Therefore, using the previous 
   observation for $p=v_{ijkM}$ or $v_{ikjM}\}$, 
   we obtain the following expressions,
   \begin{align*}
   \text{\insphere}(S_i,S_j,S_k,S_M,S_N) 
    &= -\text{\insphere}(S_i,S_j,S_k,S_M,S_n) \\
     &=-\text{\insphere}(S_i,S_k,S_j,S_m,S_n), \\
   \text{\insphere}(S_i,S_k,S_j,S_M,S_N) 
     &=-\text{\insphere}(S_i,S_k,S_j,S_M,S_n) \\
     &=-\text{\insphere}(S_i,S_j,S_k,S_m,S_n).
   \end{align*}

   These last four equalities can be used to evaluate any \insphere 
   predicate that arises during the evaluation of the \order 
   predicate in the case of a non-classic configuration. 

   Regarding 
   the respective \orient predicates that may have to be evaluated, 
   we consider the fact that $\yinv{\mathcal{A}}$, $\yinv{C_n}$
   and $\yinv{C_N}$ are collinear and the latter two lie on 
   opposite sides of $\yinv{\mathcal{A}}$. Subsequently, 
   it is also true that $\yinv{C_n}$
   and $\yinv{C_N}$ must lie on opposite sides with respect to any line
   $\yinv{\lambda}$ that goes through $\yinv{\mathcal{A}}$ (see 
   Figure~\ref{fig:15}).

   If we choose $\yinv{\lambda}$  to be the line $\yinv{\ell}$ 
   that goes through $\yinv{\OO}$ and bear in mind that the
   position of a point of \yspace with respect to this line 
   corresponds to the position of its pre-image in \zspace against 
   the plane $\Pi_{ijk}$, we infer that $C_n$ and $C_N$ lie on 
   different sides of $\Pi_{ijk}$ and therefore
   \[\text{\orient}(C_N,C_i,C_j,C_k)=-\text{\orient}(C_n,C_i,C_j,C_k).\]

   If  $\yinv{\lambda}$ is chosen to be the line that goes 
   through $\yinv{C_m}$ for $m\in\{a,b\}\backslash\{n\}$ then 
   it must hold in \yspace that $\inv{C_n}$ and  $\inv{C_N}$ 
   lie on different sides with respect to the plane that goes through
   $\inv{C_i}$, $\inv{C_j}$ and $\inv{C_m}$, which is equivalent to 
   \begin{align*}
   \text{\orient}(\inv{C_N},\inv{C_i},\inv{C_j},\inv{C_m}) &=
     -\text{\orient}(\inv{C_n},\inv{C_i},\inv{C_j},\inv{C_m}),\\
   \text{\orient}(\inv{C_m},\inv{C_i},\inv{C_j},\inv{C_N}) &=
   -\text{\orient}(\inv{C_m},\inv{C_i},\inv{C_j},\inv{C_n}) .
   \end{align*}
   \noindent Finally, combining the last two equations, we obtain that 
   \begin{align*}
   \text{\orient}(\inv{C_N},\inv{C_i},\inv{C_j},\inv{C_M}) &=
   - \text{\orient}(\inv{C_n},\inv{C_i},\inv{C_j},\inv{C_M}) \\
   &=\text{\orient}(\inv{C_n},\inv{C_i},\inv{C_j},\inv{C_m}).
   \end{align*}

   In conclusion, we have shown that 
   the evaluation of all 7 \insphere or \orient 
   predicates, that may involve one or two equivalent spheres, 
   can be amounted to the evaluation of respective predicates 
   that contain only the original spheres $S_a$ and $S_b$ instead.
   Ultimately, we proved that the algebraic cost of the 
   \order predicate in a non-classic configuration is the same 
   as in a classic configuration, yielding the following lemmas.

   \begin{lemma}
   The \order predicate in a non-classic configuration can be 
   evaluated by determining the sign of quantities of algebraic 
   degree at most 10 (in the input quantities).
   \end{lemma}

   \begin{lemma}
   The \order predicate can be 
   evaluated by determining the sign of quantities of algebraic 
   degree at most 10 (in the input quantities).
   \end{lemma}

\section{Conclusion and Future Work} 
\label{sec:conclusion_and_future_progress}
  In this paper, we presented a clever way of combining 
  various subpredicates in order to answer the \emph{master} 
  \conflict predicate. 
  The design of all predicates and primitives was made in such a way 
  such that the maximum algebraic cost of answering them would not 
  exceed 10 (on the input quantities). Based on current bibliography, 
  this is a quite small bound if compared with the respective 2D 
  version of the \conflict predicate (16 as shown in \cite{Emiris2006Predicates} and 6 as shown in \cite{millman2007degeneracy}).
  It is also remarkable that both the {\sc{VertexConflict}} (equivalent 
  to \insphere in non degenerate configurations) and the \conflict 
  predicates share the same algebraic degree. 
  
  Through our attempt to answer the master predicate, various 
  useful primitives were also developed. These tools can also 
  be used in the context of an incremental algorithm that evaluates 
  the Apollonius diagram of a set of spheres.

  A natural extension of the work presented in this paper involves 
  answering the \conflict predicate in the case where the 
  trisector of the first three input sphere is an ellipse 
  (or a circle) or a parabola. One can follow a similar analysis 
  with the hyperbolic case presented here, but several
  modifications have to be made for the analysis to be complete. 

  Ultimately, we would like to resolve the \conflict predicate even in 
  degenerate configurations, \ie if one or more sub-predicates return a 
  degenerate answer. This task can be handled in various ways; 
  our intention is to resolve any degeneracies that may arise using a 
  qualitative perturbation scheme, in accordance to the one 
  presented in \cite{devillers2012qualitative}. 
  



\newpage

\bibliographystyle{plain}
\bibliography{ms}

\end{document}